\documentclass{aa}

\usepackage{euclid}
\usepackage{graphicx}
\usepackage{natbib}
\usepackage{scalerel}
\usepackage{verbatim}
\usepackage[table]{xcolor}

\usepackage{amsmath,amsfonts,amssymb}
\usepackage{txfonts}
\usepackage{yfonts}
\usepackage{longtable}

\usepackage{soul,color,lscape}

\usepackage[pdfencoding=auto,psdextra]{hyperref}
\hypersetup{
    colorlinks=true,
    linkcolor=blue,
    filecolor=magenta,      
    urlcolor=blue,
    citecolor=blue
}
\makeatletter
\renewcommand*\aa@pageof{, page \thepage{} of \pageref*{LastPage}}
\makeatother

\usepackage[utf8]{inputenc}

\usepackage[switch, modulo]{lineno}
\usepackage{physics}

\newcommand{\done}[1]{}

\usepackage[dvipsnames]{xcolor}

\usepackage[normalem]{ulem}

\begin{document}

\title{Comparing explicit likelihood and likelihood-free simulation-based inference for weak lensing cosmic shear}

\titlerunning{PISCO Paper}
\authorrunning{S. Vinciguerra, N. Martinet and M. Gatti}

\newcommand{\orcid}[1]{}	
\author{
S.~Vinciguerra\orcid{0009-0005-4018-3184}\thanks{\email{simone.vinciguerra@lam.fr}}\inst{\ref{aff1}}
\and N.~Martinet\orcid{0000-0003-2786-7790}\inst{\ref{aff1}}
\and M.~Gatti\orcid{0000-0001-6134-8797}\inst{\ref{aff2}}
}

%%%% please do not edit the affiliation list -- contact ECEB Bureau for changes
\institute{
Aix-Marseille Universit\'e, CNRS, CNES, LAM, Marseille, France\label{aff1}
\and
Institute of Space Sciences (ICE, CSIC), Campus UAB, Carrer de Can Magrans, s/n, 08193 Barcelona, Spain\label{aff2}
}

\date{}

\setcounter{page}{1}

\abstract{

Simulation-based inference (SBI) has become a major tool for extracting cosmological information from weak-lensing (WL) surveys, particularly from non-Gaussian observables. We compare its two main paradigms: explicit likelihood inference (ELI), based on a Gaussian likelihood built from an emulator and covariance matrix, and likelihood-free inference (LFI), which learns the likelihood directly from simulations using neural density estimators. Using Gaussian random field mocks representative of the non-tomographic final Euclid data release, we analyse shear two-point correlation functions (shear-2PCFs), compressed with linear or non-linear methods, together with a fundamentally different map-level convolutional neural network (CNN) statistic, focusing on $\Omega_{\rm m}$ and $S_8$. We deploy posterior calibration diagnostics developed for LFI, including the test of accuracy with random points (TARP), revealing that ELI becomes strongly miscalibrated in the presence of emulation inaccuracies or likelihood non-Gaussianity, whereas LFI remains consistently well calibrated. These effects lead to substantial disagreement between ELI and LFI constraints, which largely disappears once they are addressed. We further show that the choice of compression scheme can significantly degrade ELI while leaving LFI largely unaffected. Although shear-2PCFs should capture all the information in Gaussian fields, finite compression and non-Gaussian likelihoods cause ELI constraints to differ by up to a factor of two from those inferred with the CNN, while the discrepancy is reduced to $\approx 30\%$ for LFI, underscoring the robustness of the deep-learning probe. Overall, our results indicate that in our simple setup which in particular neglects systematic biases, LFI provides a more robust and better-calibrated framework, while highlighting accurate non-Gaussian likelihood modelling and posterior calibration diagnostics as essential for future ELI analyses.

}


\maketitle

% ========================================================= % 
\section{Introduction}

Stage-IV weak-lensing (WL, \citealp{Schneider_1996}) surveys, such as Euclid (\citealp{Laureijs_2011, Mellier_2025}) and the Vera C. Rubin Observatory Legacy Survey of Space and Time (LSST; \citealp{Abell_2009}), will provide an unprecedented view of the large-scale structure (LSS) of the Universe and enable precision measurements of key cosmological parameters, including the matter density $\Omega_{\rm m}$, the clustering amplitude $S_8 \equiv \sigma_8\sqrt{\Omega_{\rm m}/0.3}$, and the dark-energy equation-of-state parameters (e.g., $w_0$ and $w_a$ in dynamical models, \citealp{Linder_2003}). By mapping the coherent WL distortions of galaxy images due to the intervening LSS along the line-of-sight, these surveys will probe the growth of cosmic structures over a large fraction of cosmic history, offering powerful tests of the standard $\Lambda$CDM paradigm and shedding light on the nature of the late-time accelerated expansion of the Universe (e.g. \citealp{Amendola_2018}).

Fully exploiting the statistical power of these surveys requires inference pipelines capable of extracting information beyond the traditional analysis framework. Current WL analyses predominantly rely on two-point statistics, such as cosmic shear correlation functions (shear-2PCFs), combined with an approximate Gaussian likelihood model to describe the connection between data and cosmology (e.g. \citealp{Li_2023, Wright_2025, Abbott_2026}). Although this approach has provided robust cosmological constraints, it compresses the lensing field into second-order correlations and does not fully exploit the information generated by the non-linear evolution of matter. In particular, gravitational clustering introduces non-Gaussian features in the matter distribution that become increasingly important at small angular scales, where a significant fraction of the cosmological information is expected to reside. Higher-order statistics (HOS, e.g. \citealp{Ajani2023, Vinciguerra_2026}) and field-level approaches (e.g. \citealp{Perraudin_2019, Ribli_2019, Porqueres_2021}) have therefore emerged as promising alternatives to recover information contained beyond the two-point level. 

However, including non-Gaussian probes alone does not solve the full inference problem. As increasingly informative and non-linear summaries of the lensing field are considered, the Gaussian likelihood approximation becomes progressively less justified (e.g. \citealp{Carron_2013}). The distributions of these summary statistics can significantly deviate from multivariate Gaussianity, potentially leading to biased or suboptimal parameter constraints if the likelihood shape is incorrectly modelled (e.g. \citealp{Oehl_2026}). Considerable effort has therefore been devoted to developing more accurate analytical descriptions of non-Gaussian likelihoods, although such approaches remain relatively uncommon and often require additional modelling assumptions. In parallel, neural density estimators (NDEs, e.g. \citealp{Papamakarios_2019}) have emerged as an increasingly popular approach to bypass the need for an explicit likelihood model by learning the statistical relationship between observables and cosmological parameters directly from simulations.

Hence, two approaches have become central in this context. Explicit likelihood inference (ELI; e.g. \citealp{Kilbinger_2015} for a review) follows the conventional statistical framework by combining the signal modelling of the summary statistics, often obtained through an emulator, with an analytically specified likelihood and an estimated covariance matrix. This framework benefits from interpretability and a clear connection between the data model and the inferred parameters. Nevertheless, accurate emulation of the cosmological dependence of the observables and reliable covariance estimation from a finite number of simulations remain significant challenges, particularly as the dimensionality of the data-vector (DV) increases (\citealp{Hartlap_2006}).

Likelihood-free inference (LFI; e.g. \citealp{Cranmer_2020} for a review)\footnote{LFI is also referred to in the literature as simulation-based inference (SBI, e.g. \citealp{Cranmer_2020}) or implicit likelihood inference (ILI, e.g. \citealp{Alsing_2018}).} provides an alternative strategy by directly learning the posterior distribution or likelihood from datasets consisting of pairs of cosmological summary statistics and their corresponding ground-truth parameters, using flexible NDE. By avoiding an explicit likelihood model, LFI can in principle capture complex non-Gaussian distributions and scale naturally to high-dimensional summaries. However, this flexibility comes at the cost of reduced interpretability and introduces additional requirements for validation, calibration, and robustness testing (e.g. \citealp{Talts_2020, Lemos_2023}). In particular, the absence of an explicit likelihood model makes it essential to assess whether the inferred posterior accurately represents the information contained in the dataset.

Both ELI and LFI have been successfully applied to HOS cosmological parameter inference, including recent applications to WL observations and simulated survey analyses (e.g. \citealp{Martinet_2018, Gatti_2022, Deraps_2024, Armijo_2025, Gomes_2025} for ELI, and \citealp{Fluri_2022, Jeffrey_2024, Novaes_2024} for LFI). Nevertheless, a systematic comparison between ELI and LFI performed on the same simulation suite, using identical observables and matched compression strategies, remains only marginally explored. Such a comparison is crucial in order to determine whether differences between the two frameworks originate from their statistical assumptions, their numerical implementation, or the information content of the adopted summary statistics.

Therefore, in this work, we perform a side-to-side analysis of ELI and LFI for WL Stage-IV cosmological inference using a common suite of Euclid-like simulations tailored on the final data release (DR3) in terms of redshift distribution and noise properties. We sample the cosmological parameter space generating full-sky Gaussian shear maps with the \texttt{GLASS} framework (\citealp{Tessore_2023}) across multiple independent realisations per node, together with dedicated fiducial realisations for covariance estimation and parameter-shifted maps for numerical derivatives. Flat-sky Cartesian patches are then extracted and used to measure shear-2PCFs, compressed to the parameter-space dimensionality via either MOPED (\citealp{Heavens_2017}), a Fisher-optimal linear compression, or a non-linear MLP (NN) compressor (\citealp{Rumelhart_1986}), and map-level CNN estimator (\citealp{Lecun_1998}). ELI relies on gaussian process emulators (GP) to predict the cosmological signal and combines it with the estimated covariance matrix under a Gaussian likelihood approximation, while LFI trains NDE directly on simulated data to learn the full likelihood shape without any parametric assumption. Beyond comparing final parameter constraints, we assess posterior calibration, robustness to simulation noise, and the impact of different compression strategies, providing both a validation framework and guidance for future cosmological applications of ELI and LFI.

The paper is organised as follows. Section~\ref{sec:simulations} defines the WL simulations and the generation of Euclid-like shears maps used in Section~\ref{sec:statistics} to measure the summary statistics adopted in this work. Section~\ref{sec:compression} describes how the probes are compressed with different techniques and how the covariance estimation is performed. These key-ingredients represent the input dataset of ELI and LFI pipelines introduced in Sections~\ref{sec:eli}  and \ref{sec:lfi}, respectively. Section~\ref{sec:eli_vs_lfi_shear2pcfs} presents the comparison between ELI and LFI using shear-2PCFs, while Section~\ref{sec:null_test} compares this same statistic against a map-level CNN estimator. Finally, Section~\ref{sec:conclusions} summarises our conclusions and discusses the implications for Stage-IV WL surveys.
% ============================================================
\section{Simulations}
\label{sec:simulations}
% ============================================================

This section describes the simulation framework used to generate the WL maps that underpin all subsequent analyses. We present the cosmological parameter space and sampling strategy in Sect.~\ref{sec:lhs}, the forward model for full-sky convergence and shear map generation in Sect.~\ref{sec:glass}, the shape noise injection strategy in Sect.~\ref{sec:shape_noise}, and the flat-sky patch extraction procedure in Sect.~\ref{sec:patches}.

% ----------
\subsection{Cosmological parameter sampling}
\label{sec:lhs}
% ----------

We sample the cosmological parameter space using Latin Hypercube Sampling (LHS), varying $\Omega_{\rm m}$ and $\sigma_8$ across flat priors based on the cosmo-SLICS N-body suite ($\Omega_{\rm m} \in [0.10, 0.55]$, $\sigma_8 \in [0.44, 1.56]$, \citealp{Deraps_2019}). All remaining cosmological parameters are held fixed at a fiducial cosmology ($h = 0.6898$, $n_s = 0.969$, $w_0 = -1$, $\Omega_b = 0.0473$). Two independent LHS ($\rm LHS_{0}$, $\rm LHS_{1}$) of 5,100 cosmological nodes each are generated: one for deep learning and NN compression ($\rm LHS_{0}$), and one for training emulators of our observables for the ELI and to perform the LFI analysis ($\rm LHS_{1}$). We refer to these $\rm LHS$ as our model since these are used to learn the impact of cosmology on our statistics within the parameter space. 

A dedicated fiducial-cosmology configuration, set to the cosmo-SLICS fiducial values ($\Omega_{\rm m}^{\rm fid}=0.2905$, $\sigma_8^{\rm fid}=0.8364$), is also used for covariance estimation used in ELI and MOPED compression, for the latter we also numerically compute derivatives via symmetric finite differences. In that case, each of the two varied parameters ($\Omega_{\rm m}$, $\sigma_8$) is shifted independently by $\pm 16\%$ around its fiducial value while all other parameters are held fixed, yielding four additional simulation nodes per derivative. Following \citet{Ajani2023}, the $16\%$ step size is chosen to be large enough that the finite-difference estimate is not dominated by numerical noise, yet small enough that the linear approximation underlying the finite-difference formula remains valid.

% ----------
\subsection{Forward model: GLASS weak lensing maps}
\label{sec:glass}
% ----------

\begin{figure}[h]
  \centering
  \includegraphics[width=\columnwidth, trim=0 0 0 55, clip]{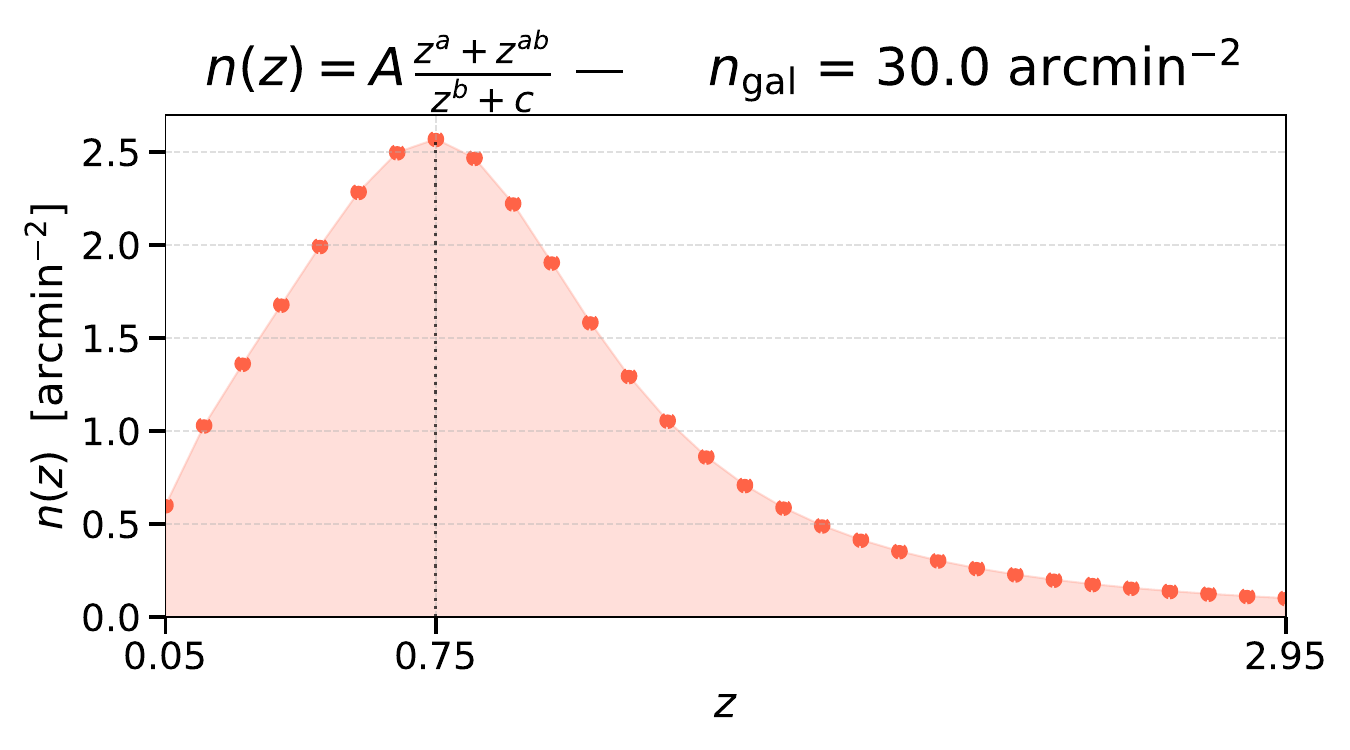}
  \caption{Source redshift distribution $n(z)$ used in this work (Euclid DR3-like, single non-tomographic bin,  \citealp{Ajani2023}). The solid curve shows the parametric form of Eq.~(\ref{eq:nz}) (see text for details).}
  \label{fig:nz}
\end{figure}

Full-sky WL Gaussian convergence and shear maps with $n_{\rm side} = 512$ are generated using the \texttt{GLASS}\footnote{\url{https://glass.readthedocs.io/}} simulation framework (\citealp{Tessore_2023}). For each cosmological node, \texttt{CAMB}\footnote{\url{https://camb.readthedocs.io}} (\citealp{Lewis_2000}) computes angular matter power spectra with Halofit non-linear corrections up to $\ell_{\max} = 2048 \, (4 \times \rm n_{\rm side})$. Starting from the 5,100 nodes of each LHS, we discard nodes for which \texttt{CAMB} fails to provide reliable predictions, which occurs for $\Omega_{\rm m} \lesssim 0.15$ near the lower boundary of the prior. This yields effective node counts of 4,645 and 4,610 for $\mathrm{LHS}_0$ and $\mathrm{LHS}_1$ respectively, which are used for all subsequent analyses. \texttt{GLASS} then generates correlated Gaussian random fields on the sphere in radial shells of 200~Mpc comoving spacing and sampled via a fixed random seed. For each shell $i$, a convergence plane $\kappa_i$ is accumulated via multi-plane lensing, and the corresponding shear components $(\gamma_{1,i}, \gamma_{2,i})$ are obtained through inverse spherical Kaiser--Squires inversion (\citealp{Kaiser_1993}). The n(z)-weighted convergence and shear fields are then computed as
\begin{equation}
    \kappa = \frac{\sum_i n_{\rm gal}(z_i)\,\kappa_i}{\sum_i n_{\rm gal}(z_i)}\,, \qquad
    \gamma_{1,2} = \frac{\sum_i n_{\rm gal}(z_i)\,\gamma_{1,2\,,i}}{\sum_i n_{\rm gal}(z_i)}\,,
    \label{eq:kappa_gamma}
\end{equation}
where the galaxy number density per shell is $n_{\rm gal}(z_i) = \int_{z_i}^{z_{i+1}} n(z)\,\mathrm{d}z$, and $n(z)$ is the source redshift distribution mimicking a Euclid DR3-like single non-tomographic bin extending to $z_{\rm max} = 3$. We stress here that we opted for a non-tomographic configuration as a good trade-off between constraining power and computational resources, while reserving the tomographic analysis for a subsequent paper. Specifically, we adopt the parametric form of \citet{Fu_2008} with Euclid-like parameters fitted to COSMOS (\citealp{Massey_2007b}) data in \citet{Ajani2023} (their Eq.~1, Table~2),
\begin{equation}
  n(z) = A\,\frac{z^{a} + z^{ab}}{z^{b} + c}\,,
  \label{eq:nz}
\end{equation}
with $A=1.7865$, $a=0.471$, $b=5.1843$, $c=0.7259$, normalised to $n_{\rm gal} = 30\,\mathrm{arcmin}^{-2}$ (Fig.~\ref{fig:nz}). Maps are pixellised on the \texttt{HEALPix} RING grid \citep{Gorski2005} at resolution $n_{\rm side} = 512$ (pixel scale $\approx 6'.9$).
By varying the random seed used for Gaussian random fields generation, we produced 1, 10, and 200 independent full-sky realisations per cosmological node for the model, derivatives, and covariance estimation respectively.
% ----------
% ----------
\subsection{Shape-noise injection}
\label{sec:shape_noise}
% ----------

In order to imprint the effect of shape noise in our mocks, we approximate it as independent Gaussian noise that is propagated at the pixel level to full-sky shears \texttt{HEALPix} maps. Specifically, we mimic a non-tomographic Euclid-like setup expected for the DR3 but accessing only $100 \, \rm deg^2$ of the area (about $1\, \%$ of the total). We reserve a tomographic analysis and larger footprint coverage for a future more realistic framework. The noise standard deviation per pixel of intrinsic ellipticity maps follows from the intrinsic ellipticity dispersion $\sigma_\epsilon = 0.26$ per component (\citealp{Martinet2019}) and the effective galaxy density $n_{\rm gal} = 30\,\mathrm{arcmin}^{-2}$:
\begin{equation}
    \sigma_{\rm e_{i}^{\rm int}} = \frac{\sigma_\epsilon}{\sqrt{n_{\rm gal}\,A_{\rm pix}}}\,,
    \label{eq:sigma_kappa}
\end{equation}
where $A_{\rm pix}$ is the \texttt{HEALPix} pixel area in arcmin$^2$. We assume the intrinsic ellipticity dispersion to be identical for both components, so that $\rm e_1^{\rm int}$ and $\rm e_2^{\rm int}$ share the same Gaussian width $\sigma_{\rm e_{i}^{\rm int}}$. For each pixel, the two components are nonetheless drawn from independent realisations of $\mathcal{N}(0,\,\sigma_{\rm e_{i}^{\rm int}})$, forming the complex intrinsic ellipticity $\rm e^{\rm int} = \rm e_1^{\rm int} + \mathrm{i}\,\rm e_2^{\rm int}$, which is then injected into the corresponding pixel of the noise-free shear map via the reduced-shear composition
\begin{equation}
    \rm e^{\rm obs} = \frac{\rm e^{\rm int} + \gamma}{1 + \bar\gamma\,\rm e^{\rm int}}\,,
    \label{eq:reduced_shear}
\end{equation}
where $\bar\gamma$ denotes complex conjugation. In our fiducial configuration we adopt a \emph{varying-phase-only} draw strategy for $\rm e^{\rm int}$: writing $\rm e^{\rm int} = |\rm e^{\rm int}|\,\mathrm{e}^{2\mathrm{i}\phi}$, we fix the amplitude to $|\rm e^{\rm int}| = \sigma_{\rm e_{i}^{\rm int}}\sqrt{2}$, where $\sigma_{\rm e_{i}^{\rm int}} = \sigma_\epsilon/\sqrt{n_{\rm gal}\,A_{\rm pix}}$, and draw only the phase $\phi$ uniformly from $[0, \pi)$. The fixed amplitude is chosen so that the per-component RMS matches the fully Gaussian case, that is the expectation values $\mathbb{E}[|\rm e_1^{\rm int}|^2] + \mathbb{E}[|\rm e_2^{\rm int}|^2] = 2\sigma_{\rm e_{i}^{\rm int}}^2$, while eliminating amplitude fluctuations that would otherwise introduce additional variance in the noise realisation. Although this is our nominal strategy, we checked our results to be consistent when drawing noise varying both the phase and the amplitude at the same time. 

Noise seeds are assigned according to the role of each simulation set. For both $\rm LHS_{0}$ and $\rm LHS_{1}$, each independent spherical realisation of a given cosmology receives a statistically independent noise draw, and different cosmological nodes additionally receive independent seeds, that is, the noise seed varies both across realisations of the same cosmology and across cosmological nodes. Crucially, the two LHS receive completely independent set of noise seeds. This overall strategy prevents overfitting in both the CNN and LFI trainings by ensuring that the networks cannot memorise the noise pattern associated with any particular cosmological node. We note that this strategy is in principle suboptimal for ELI: by varying the noise seed across cosmological nodes, additional variance is introduced in the training targets of the GP emulator, potentially making the emulation of the cosmological signal more challenging. In our low-resolution, non-tomographic setup, however, we verified that the resulting ELI posteriors are insensitive to this choice, with no significant difference observed when fixing the noise seed across nodes instead. This effect is expected to become more impactful at higher resolution, where shape noise dominates at smaller angular scales and the per-node noise variance is larger relative to the cosmological signal.

While for the covariance estimation the noise seed varies realisation by realisation, for the derivative maps, the same set of noise seeds is reused identically across all shifted cosmologies ($\pm\Omega_{\rm m}$, $\pm\sigma_8$): each of the realisations per shift direction shares its noise draw with the corresponding realisation of every other shift. This deliberate design choice maximises the cosmological sensitivity of the finite-difference derivative by cancelling the common noise contribution, so that the derivative signal is not diluted by noise fluctuations between the two shift directions.

% ----------
\subsection{Flat-sky patch extraction}
\label{sec:patches}
% ----------
Once shape noise has been injected at the full-sky \texttt{HEALPix} level, flat-sky patches are extracted from the noisy convergence and shear maps via Cartesian projection (\texttt{\texttt{HEALPix} CartesianProj}). Each patch covers $10^\circ \times 10^\circ$ ($64 \times 64$ pixels, $\approx 9.4'$ pixel resolution). Central pointings are selected at low latitudes ($|\delta| < 20^\circ$) with $3^\circ$ separation in both longitude and latitude, yielding 102 non-overlapping quasi-independent patches per full-sky realisation; this strategy minimises projection distortions and patch overlap. For the single full-sky realisation available per node for signal modelling, only the first 50 out of the 102 available patches are retained. These 50 tiles span the full declination range $|\delta| < 20^\circ$ across roughly half the right-ascension coverage, ensuring that all projection distortion effects, which depend exclusively on declination, are fully captured in this subset. This is sufficient to average over shape noise and cosmic variance fluctuations while halving the computational cost relative to the full set. By contrast, the covariance and derivative maps make use of all 102 patches per realisation, since an accurate estimate of the DV inverse covariance matrix requires sufficient independent realisations to suppress numerical noise in its entries (\citealp{Hartlap_2006, Percival_2021}). The full set of 102 projected patches is also used for the numerical derivatives in order to assess their convergence. Therefore, with 1, 10, and 200 independent full-sky realisations produced for the model, derivatives, and covariance respectively, the total number of flat-patch based DVs available at each stage is then $n_{\rm flat}^{\rm model} = 50 / \rm node$, $n_{\rm flat}^{\rm der} = 1\,020$, and $n_{\rm flat}^{\rm cov} = 20\,400$.   

% ============================================================
\section{Summary statistics}
\label{sec:statistics}
% ============================================================

We employ two summary statistics to compress WL maps into DVs, both measured on flat-sky patches for consistency: the shear two-point correlation functions ($\xi_\pm$, Sect.~\ref{sec:2pcfs}), a Gaussian probe, and a map-level estimator based on convolutional neural networks (CNN, Sect.~\ref{sec:cnn}), a non-linear probe. Since our simulations are based on Gaussian random fields, $\xi_\pm$ alone is expected to capture the full available information up to numerical effects such as any information loss introduced by the adopted binning. We nevertheless include the CNN as a second fiducial estimator to demonstrate the robustness of a map-level, non-linear summary against the well-understood two-point reference.

% ----------
\subsection{Shear two-point correlation functions}
\label{sec:2pcfs}
% ----------
The shear two-point correlation functions $\xi_\pm(\theta)$ quantify the correlation of the complex observed ellipticity $\rm e^{\rm obs} = \rm e^{\rm obs}_1 + \mathrm{i}\rm e^{\rm obs}_2$ between pairs of pixels separated by angular scale $\theta$. Decomposing the ellipticity into tangential ($\rm e^{\rm obs}_t$) and cross ($\rm e^{\rm obs}_\times$) components with respect to the line joining each pair, they are defined as
\begin{equation}
    \xi_\pm(\theta) = \langle \rm e^{\rm obs}_t\,\rm e^{\rm obs}_t \rangle(\theta) \pm \langle \rm e^{\rm obs}_\times\,\rm e^{\rm obs}_\times \rangle(\theta)\,,
    \label{eq:xi_pm}
\end{equation}
where angle brackets denote averaging over all pixel pairs within an angular separation bin. In terms of the lensing E- and B-mode angular power spectra $C_\ell^{EE}$ and $C_\ell^{BB}$, the two functions are related via
\begin{equation}
    \xi_\pm(\theta) = \int_0^\infty \frac{\mathrm{d}\ell\,\ell}{2\pi}\,\left(C_\ell^{EE} \pm C_\ell^{BB}\right) J_{0/4}(\ell\theta)\,,
    \label{eq:xi_pm_cl}
\end{equation}
where $J_0$ and $J_4$ are zeroth- and fourth-order Bessel functions of the first kind for $\xi_+$ and $\xi_-$ respectively. Both $\xi_+$ and $\xi_-$ receive contributions from E- and B-modes, with opposite sign for the B-mode term. In the absence of B-modes, both reduce to integrals over $C_\ell^{EE}$ weighted by different Bessel kernels, making them sensitive to partially different combinations of angular scales and thus carrying partially independent cosmological information. Both functions are measured on each flat-sky patch using \textsc{TreeCorr} \citep{jarvis2015}, with galaxy positions assigned to pixel centres and coordinates expressed in arcminutes. We adopt 8 logarithmically spaced angular bins ranging from the pixel scale ($\approx 9.4'$) to the patch diagonal ($10^\circ \times \sqrt{2} \approx 14^\circ \approx 840'$), and the $\xi_+$ and $\xi_-$ vectors are concatenated into a single DV of 16 elements.

% ----------
\subsection{CNN-based statistics}
\label{sec:cnn}
% ----------
While traditional WL statistics compress the field information into a one-dimensional DV, each capturing a specific aspect of the lensing signal (e.g. the matter dispersion at a given angular scale for $\xi_{\pm}$), convolutional neural networks (CNNs) take a complementary approach by learning the mapping between the two-dimensional map and the cosmological parameters, exploiting its full spatially structured information through learnable convolutional filters \citep{LeCunEtAl1998}. Specifically, a convolutional layer computes
\begin{equation}
    h^{(l)}_{c,ij} = \sigma\!\left(
    \sum_{c'}\sum_{p,q} W^{(l)}_{c,c',p,q}\,
    h^{(l-1)}_{c',i+p,\,j+q} + b^{(l)}_c
    \right),
    \label{eq:conv}
\end{equation}
where $c$ and $c'$ index output and input channels, $(i,j)$ are spatial positions, $(p,q)$ run over the filter extent, $W^{(l)}$ and $b^{(l)}$ are the learnable weights and biases of layer $l$, and $\sigma$ is a non-linear activation function. Successive convolutional layers, separated by non-linear activations and downsampling operations, build a hierarchy of spatial features: early layers detect local edges and orientations, while deeper layers capture large-scale morphological patterns of the lensing field. We adopt the ResNet-18 architecture (\citealp{He_2016}), in which residual skip connections allow gradients to propagate through the network without vanishing, and the final fully connected layer is replaced by a regression head mapping the learned spatial representation to point estimates of $(\Omega_{\rm m}, \sigma_8)$. We tested deeper ResNet architectures finding that the complexity of our inference problem was well characterized by 18 convolutional layers. 

We consider a CNN that takes two-channel ellipticity maps $(\rm e^{\rm obs}_1, \rm e^{\rm obs}_2)$ as input. We also test an alternative version using single-channel convergence maps $\kappa$, allowing us to assess whether Kaiser–-Squires inversion leads to any measurable loss of cosmological information. The first convolutional layer is adapted to accept the appropriate number of input channels, while the rest of the architecture is kept identical. Training minimises the mean squared error (MSE) between predicted and true cosmological parameters using AdamW optimisation (\citealp{Loshchilov_2019}). We consider the following hyperparameters and priors: learning rate ($10^{-6}$--$10^{-4}$, log-uniform), weight decay ($10^{-4}$--$5\times10^{-2}$, log-uniform), batch size (\{4, 8, 16, 32, 64, 128\}), learning rate scheduler type (cosine annealing, step, plateau), dropout rate (0.3--0.6), activation function (LeakyReLU, ReLU, ELU), convolutional channel widths, kernel sizes, and fully connected hidden-layer dimensions, are optimised via \textsc{Optuna} \citep{akiba2019} over 150 random trials, each evaluated by minimum validation MSE loss. See e.g. \citealt{Goodfellow_2016} for a comprehensive overview of these architectural choices and their role in deep learning. A warm-up phase of 15 epochs precedes early stopping, which is triggered after 8 epochs without improvement (5\% of the 150 maximum training epochs). Cosmological parameters are normalised to $[0, 1]$ for training and rescaled to physical units at inference time. Input ellipticity maps are instead fed to the network at their native pixel values without additional standardisation: since the maps are noise-dominated, the shape noise contribution ensures a homogeneous pixel value distribution across all cosmological nodes, making per-channel normalisation unnecessary.

The networks are trained and validated on $\mathrm{LHS}_0$ using an 80\%/20\% train/validation split of 4,645 nodes (3,716 training and 929 validation nodes), each equipped with $n_{\rm flat}=50$ patches, yielding 185,800 training and 46,450 validation flat-patch realisations. Performance is evaluated on all 4,610 nodes of $\mathrm{LHS}_1$ (230,500 flat-patch realisations).
% ============================================================
\section{Data compression}
\label{sec:compression}
% ============================================================
Compressing the high-dimensional DVs produced by each estimator to the dimensionality of the parameter space ($n_{\rm p}=2$) is useful for tractable likelihood evaluation and essential for density estimation. We employ two complementary strategies. The Massive data compression for parameter-dependent covariance matrices (MOPED, Sect.~\ref{sec:moped}, \citealp{Heavens_2000}) that is an analytic, Fisher-optimal compression that projects the DV onto a fixed set of vectors constructed from the covariance matrix and DV derivatives with respect to the cosmological parameters; it is interpretable, lossless under Gaussian assumptions, and requires no training data beyond a covariance matrix and derivatives with respect to the cosmological parameters. Neural network compression (Sect.~\ref{sec:nn_compression}) instead learns a non-linear mapping from DV to compressed summaries by training a residual Multilayer Perceptron (MLP, \citealp{Rumelhart_1986}) on the full simulation suite; it requires no explicit derivative computation but relies on sufficient training data and introduces an additional optimisation step. Both methods reduce the DV to $n_{\rm p}=2$ summaries and are applied consistently to all statistics and their associated covariance matrices.

% ----------
\subsection{Covariance and derivative estimation}
\label{sec:covder}
% ----------
The DV covariance matrix $\hat{\mathsf{C}}$ is estimated from $200 \times 102 = 20\,400$ independent \texttt{GLASS} flat-patch realisations at the fiducial cosmology. Sub-samples at 25\%, 50\%, 75\%, and 100\% of the available realisations are used to assess convergence of the covariance estimate. Covariance matrices are regularised and condition numbers are monitored throughout. The considerable number of covariance realizations compared to the typical length of our input DVs makes unnecessary any correction factor (e.g. \citealp{Hartlap_2006}). Numerical derivatives $\partial\boldsymbol{d}/\partial\theta_i$ are computed via symmetric finite differences,
\begin{equation}
    \frac{\partial\boldsymbol{d}}{\partial\theta_i} \approx \frac{\boldsymbol{d}(\theta_i^{\rm fid} + \Delta\theta_i) - \boldsymbol{d}(\theta_i^{\rm fid} - \Delta\theta_i)}{2\,\Delta\theta_i}\,,
    \label{eq:finite_diff}
\end{equation}
with $\Delta\theta_i = 0.16\,\theta_i^{\rm fid}$ (i.e.\ $\pm 16\%$ of the fiducial value), each parameter varied independently while all others are held fixed. Each shifted DV is averaged over $10 \times 102 = 1\,020$ independent flat-patch realisations per shift direction. The absolute shifts are therefore $\Delta\Omega_{\rm m} = 0.0465$ and $\Delta\sigma_8 = 0.134$. Convergence of the derivative estimate with respect to the number of realisations per shift direction is verified by comparing results at 25\%, 50\%, 75\%, and 100\% of the available realisations as done for the covariance. Both the covariance and derivative estimates are found to be well converged at the full sample size, with no significant variation in the resulting posteriors across the four sub-sample fractions.

% ----------
\subsection{MOPED compression}
\label{sec:moped}
% ----------
The MOPED algorithm \citep{Heavens_2000} compresses the DV $\boldsymbol{d}$ to a set of $n_{\rm p}$ scalars $\{y_m = \boldsymbol{b}_m^\top \boldsymbol{d}\}$, one per parameter, by constructing compression vectors $\{\boldsymbol{b}_m\}$ that are Fisher-optimal and mutually uncorrelated under the data covariance. For the $m$-th parameter, the compression vector is
\begin{equation}
    \boldsymbol{b}_m =
    \frac{
        \hat{\mathsf{C}}^{-1}\boldsymbol{\mu}_{,m}
        - \displaystyle\sum_{q=1}^{m-1}
          \left(\boldsymbol{\mu}_{,m}^\top\boldsymbol{b}_q\right)\boldsymbol{b}_q
    }{
        \sqrt{
            \boldsymbol{\mu}_{,m}^\top\hat{\mathsf{C}}^{-1}\boldsymbol{\mu}_{,m}
            - \displaystyle\sum_{q=1}^{m-1}
              \left(\boldsymbol{\mu}_{,m}^\top\boldsymbol{b}_q\right)^2
        }
    }\,,
    \label{eq:moped}
\end{equation}
where $\boldsymbol{\mu}_{,m} \equiv \partial\boldsymbol{d}/\partial\theta_m$ is the derivative of the mean DV with respect to the $m$-th parameter (Eq.~\ref{eq:finite_diff}), $\hat{\mathsf{C}}$ is the estimated covariance matrix, and the Gram--Schmidt sum enforces that $y_m$ is uncorrelated with all previously formed compressed statistics $\{y_1,\ldots,y_{m-1}\}$. For the first parameter the sum is empty and Eq.~(\ref{eq:moped}) reduces to $\boldsymbol{b}_1 = \hat{\mathsf{C}}^{-1}\boldsymbol{\mu}_{,1} / \sqrt{\boldsymbol{\mu}_{,1}^\top\hat{\mathsf{C}}^{-1}\boldsymbol{\mu}_{,1}}$. For a Gaussian likelihood with parameter-independent covariance, the compressed statistics $\{y_m\}$ retain all the Fisher information about $\boldsymbol{\theta}$, making the compression exactly lossless \citep{Heavens_2000}. In practice, the compression matrix $\mathsf{B} = [\boldsymbol{b}_1, \ldots, \boldsymbol{b}_{n_{\rm p}}]^\top$ reduces our DVs from $\mathcal{O}(10^1)$--$\mathcal{O}(10^2)$ elements to $n_{\rm p}=2$ compressed summaries, with the compressed covariance given by $\mathsf{C}_{\rm comp} = \mathsf{B}\,\hat{\mathsf{C}}\,\mathsf{B}^\top$. This compression is applied independently to each estimator, using the covariance matrix and numerical derivatives computed for that estimator's DV as described in Sect.~\ref{sec:covder}.

% ----------
\subsection{Neural network compression}
\label{sec:nn_compression}
% ----------
As a flexible, data-driven alternative to MOPED, independent from the likelihood property and covariance estimation, we train a MLP to learn a non-linear compression mapping $\boldsymbol{t} = f_{\rm MLP}(\boldsymbol{d};\,\boldsymbol{w})$ from the full DV $\boldsymbol{d}$ to $n_{\rm p}=2$ compressed summaries. An MLP applies successive layers of the form
\begin{equation}
    \boldsymbol{h}^{(l)} = \sigma\!\left(\mathsf{W}^{(l)}\,\boldsymbol{h}^{(l-1)} + \boldsymbol{b}^{(l)}\right),
    \label{eq:mlp}
\end{equation}
where $\mathsf{W}^{(l)}$ and $\boldsymbol{b}^{(l)}$ are the learnable weight matrix and bias of layer $l$, and $\sigma$ is a non-linear activation function. Our compressor adds residual skip connections between hidden layers to improve gradient flow, and uses LeakyReLU activations and dropout regularisation to prevent overfitting. The network is trained as a regressor on simulated DV--parameter pairs $\{(\boldsymbol{d}_i, \boldsymbol{\theta}_i)\}$ from $\mathrm{LHS}_0$, minimising the MSE between the predicted compressed summaries and the true parameters using AdamW optimisation with a cosine annealing learning rate schedule.

Hyperparameters (hidden-layer width within 128, 256, 512, or 1024 units, dropout rate, learning rate, weight decay, batch size, and scheduler type) are optimised via a random grid search over up to 50 trials, each evaluated by minimum validation MSE loss. The search uses 4,645 nodes from $\mathrm{LHS}_0$, split into 80\% for training (3,716 nodes) and 20\% for validation (929 nodes), each node equipped with $n_{\rm flat}=50$ patches; training runs for up to 150 epochs with early stopping (patience of 8 epochs, after which the trial is rejected). The best-performing architecture is selected and the trained compressor is applied to all summary statistics of $\mathrm{LHS}_1$. Compressed covariance matrices are then computed by applying the trained network to the $20\,400$ fiducial flat-patch realisations.
% ============================================================
% ============================================================
\section{Explicit Likelihood Inference}
\label{sec:eli}
% ============================================================

We present the ELI framework in the following subsections. We describe the emulator architecture and likelihood modelling in Sect.~\ref{sec:gp_and_likelihood}, the validity of likelihood Gaussian approximation in Sect.~\ref{sec:gaussianity}, and the MCMC sampling strategy in Sect.~\ref{sec:eli_mcmc}.

% ------
\subsection{Emulator and likelihood}
\label{sec:gp_and_likelihood}
% ------

In the ELI framework, we model the mean DV as a function of cosmological parameters $\boldsymbol{\theta} = (\Omega_{\rm m}, \sigma_8)$ using a GP emulator, and assume a multivariate Gaussian likelihood. For each summary statistic and compression type, a set of independent GP, one per compressed DV bin, is trained on a randomly selected subset of 250 cosmological nodes from $\mathrm{LHS}_1$ (out of the 4,609 nodes available for inference, after reserving one node as the mock observation for the MCMC chain), validated and diagnosed on the remaining nodes, and used to predict the mean DV at arbitrary $\boldsymbol{\theta}$ within the priors indicated in Sect. \ref{sec:lhs}. We verified that increasing the number of training nodes beyond 250 yields consistent posterior constraints, confirming that this number is sufficient to capture the cosmological dependence of all considered statistics while minimising computational cost. The GP implementation uses the \texttt{GaussianProcessRegressor} from \texttt{scikit-learn} \citep{Pedregosa_2011}. Each GP uses a kernel composed of a constant amplitude factor $C$, which sets the overall variance scale of the covariance function, multiplied by a Mat\'ern-5/2 covariance function $k_{\rm Mat\acute{e}rn}(\boldsymbol{x}, \boldsymbol{x}'; \{\ell_d\})$, where $\boldsymbol{x} = (\Omega_{\rm m}, \sigma_8)$ and $\boldsymbol{x}' = (\Omega_{\rm m}', \sigma_8')$ are two points in cosmological parameter space and $\ell_d$ is an independent length scale for each parameter dimension $d \in \{\Omega_{\rm m}, \sigma_8\}$ controlling how rapidly the emulated quantity varies with that parameter. Having one $\ell_d$ per dimension, rather than a single shared length scale, constitutes automatic relevance determination (ARD): a large $\ell_d$ indicates that the emulated bin is nearly insensitive to $\theta_d$, effectively down-weighting its contribution. An additive white-noise term with amplitude $\sigma_n^2$ absorbs independently and identically distributed Gaussian observation noise in the training targets. The full kernel evaluated on the training set is
\begin{equation}
    k(\boldsymbol{x}, \boldsymbol{x}') = C \cdot k_{\rm Mat\acute{e}rn}(\boldsymbol{x}, \boldsymbol{x}'; \{\ell_d\}) + \sigma_n^2\,\mathbf{I}\,,
    \label{eq:gp_kernel}
\end{equation}
where $\sigma_n^2\,\mathbf{I}$ is the noise contribution added to the diagonal of the training covariance matrix. All kernel hyperparameters ($C$, $\{\ell_d\}$, $\sigma_n^2$) are jointly optimised by maximising the log marginal likelihood via the L-BFGS-B algorithm, with 10 random restarts to reduce sensitivity to local optima. Tests with other kernels and hyperparameters did not show relevant improvements in the emulation process. Inputs (cosmological parameters) are standardised to zero mean and unit variance; outputs (DV bins) are normalised internally by the GP during training and automatically de-normalised to physical units at prediction time. Extensive emulator quality diagnostics are applied to each trained GP ensemble on the test nodes of $\mathrm{LHS}_1$ and are presented in Appendix~\ref{app:gp_diagnostics}. As a cross-check of the GP emulator, we also trained an MLP and obtained highly consistent results, indicating that the observed emulation errors are not driven by the chosen architecture. Rather, they arise because the second MOPED component carries very little cosmological information, making its underlying cosmological dependence difficult to learn from the available simulations. Increasing the number of realizations per cosmological node, thereby reducing the impact of simulation noise, would likely improve the emulation accuracy.
Assuming a Gaussian likelihood with cosmological parameter-independent covariance, the log-likelihood is
\begin{equation}
\ln\mathcal{L}(\boldsymbol{\theta}) = -\frac{1}{2}\,
[\mathbf{d}_{\rm obs} - \boldsymbol{\mu}_{\rm GP}(\boldsymbol{\theta})]^{\rm T}\,
\hat{\mathsf{C}}^{-1}\,
[\mathbf{d}_{\rm obs} - \boldsymbol{\mu}_{\rm GP}(\boldsymbol{\theta})]\,,
\label{eq:log_likelihood}
\end{equation}
where $\mathbf{d}_{\rm obs}$ is the observed compressed DV, $\boldsymbol{\mu}_{\rm GP}(\boldsymbol{\theta})$ is the GP-predicted mean at parameters $\boldsymbol{\theta}$, and $\hat{\mathsf{C}}^{-1}$ is the inverse of the estimated DV covariance matrix (Sect.~\ref{sec:covder}).
% ------
\subsection{Likelihood Gaussianity validation}
\label{sec:gaussianity}
\begin{figure}[h]
    \centering
    \text{MOPED-shear-2PCFs Residuals}\\[0.5em]
    \includegraphics[width=0.8\columnwidth, trim=0 0 0 50, clip]{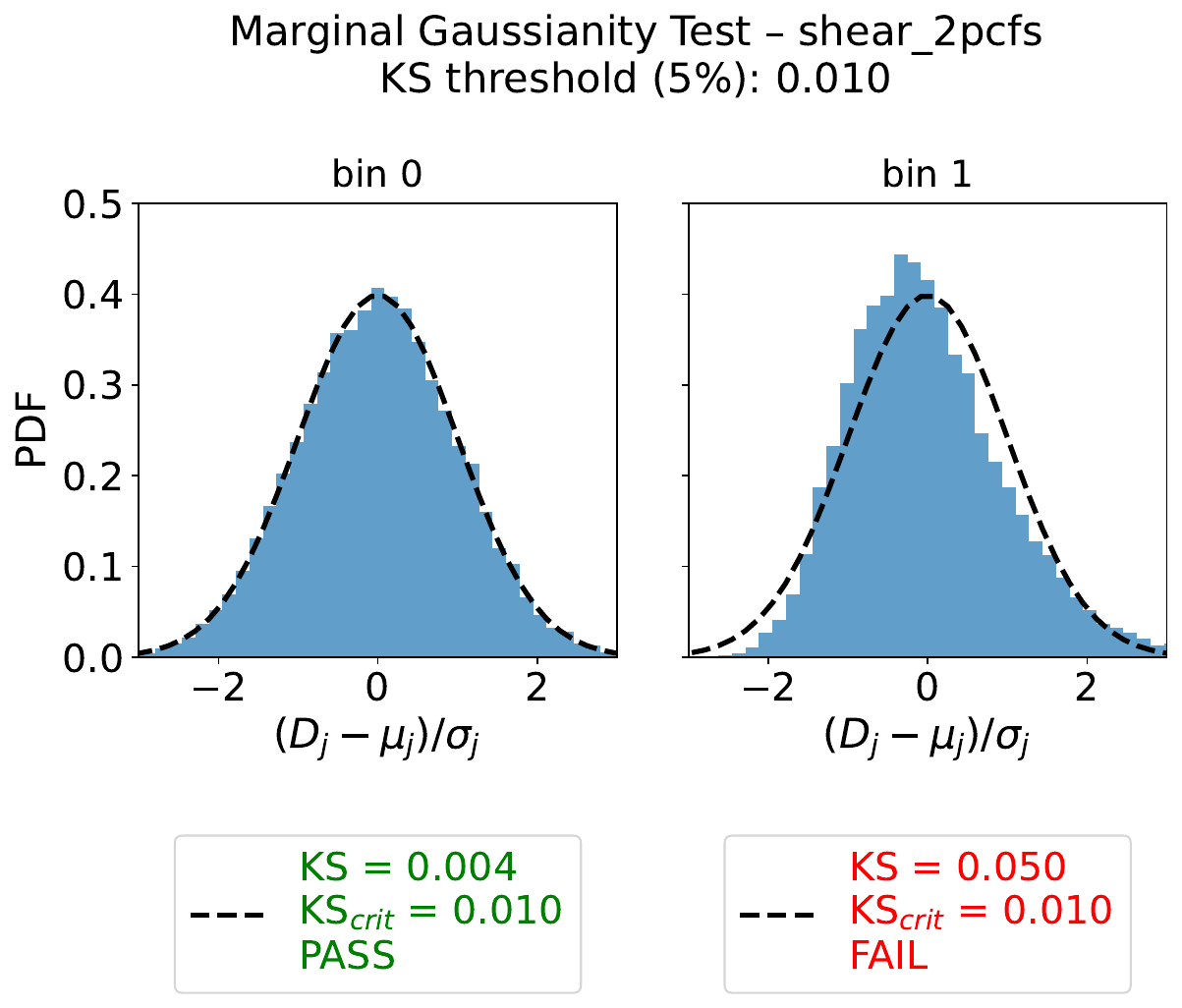}\\[1.2em]
    \text{NN-shear-2PCFs Residuals}\\[0.5em]
    \includegraphics[width=0.8\columnwidth, trim=0 0 0 50, clip]{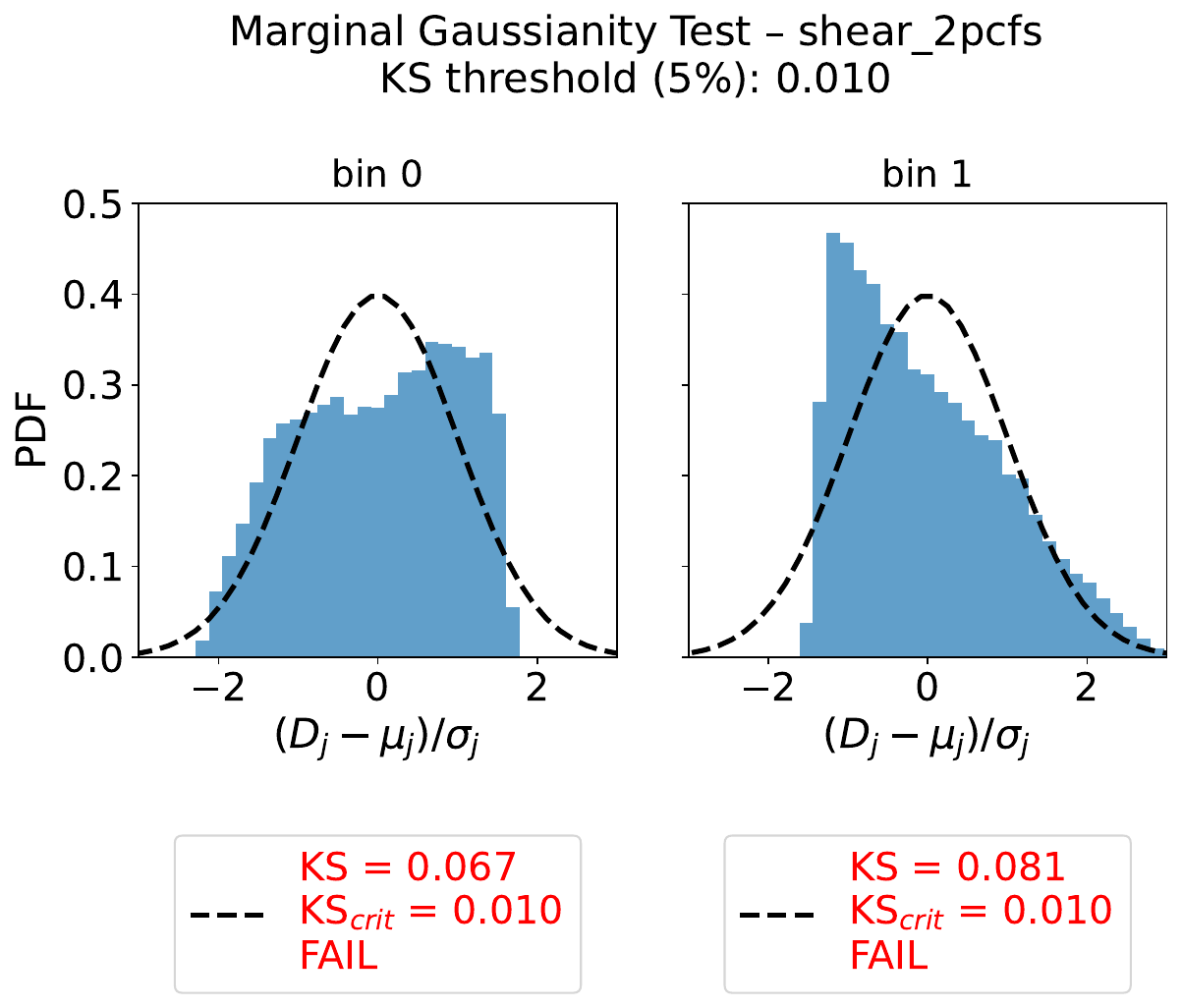}\\[1.2em]
    \text{CNN Residuals}\\[0.5em]
    \includegraphics[width=0.8\columnwidth, trim=0 0 0 50, clip]{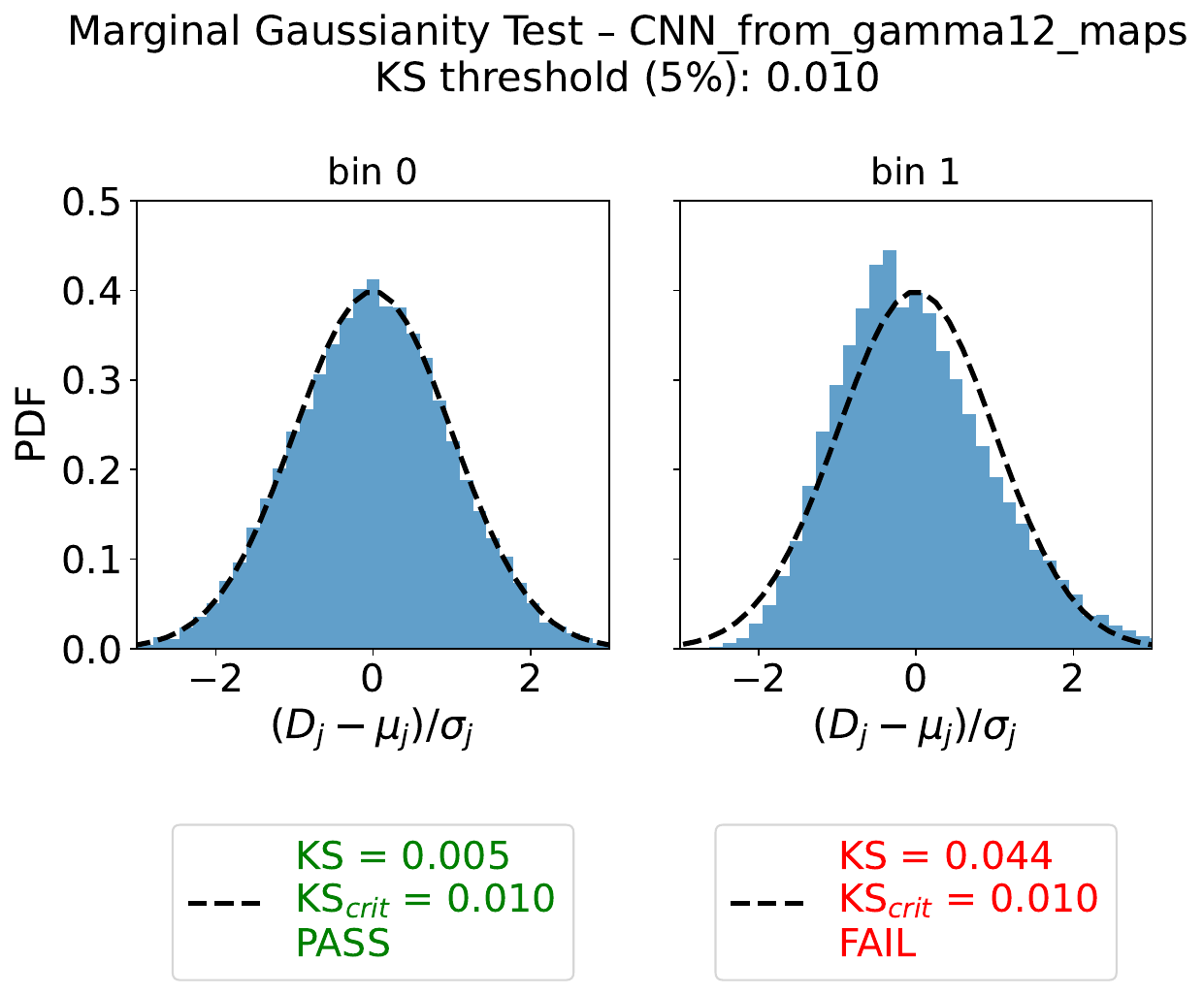}
    \caption{Marginal Gaussianity test for the two compressed summaries of shear-2PCFs under MOPED (top) and NN (middle) compression, and for the two CNN summaries (bottom). Each sub-panel shows the normalised distribution $(d_j - \mu_j)/\sigma_j$ of one compressed bin across the $20\,400$ fiducial flat-patch realisations, compared to the standard normal (dashed curve). The KS statistic and critical threshold ($D_{\rm crit}$ at 5\% significance) are reported; bins failing (passing) the test are labelled \textcolor{red}{FAIL} (\textcolor{green}{PASS}).}
    \label{fig:gaussianity_compression}
\end{figure}

\begin{figure}[h]
    \centering
    \includegraphics[width=1.0\columnwidth]{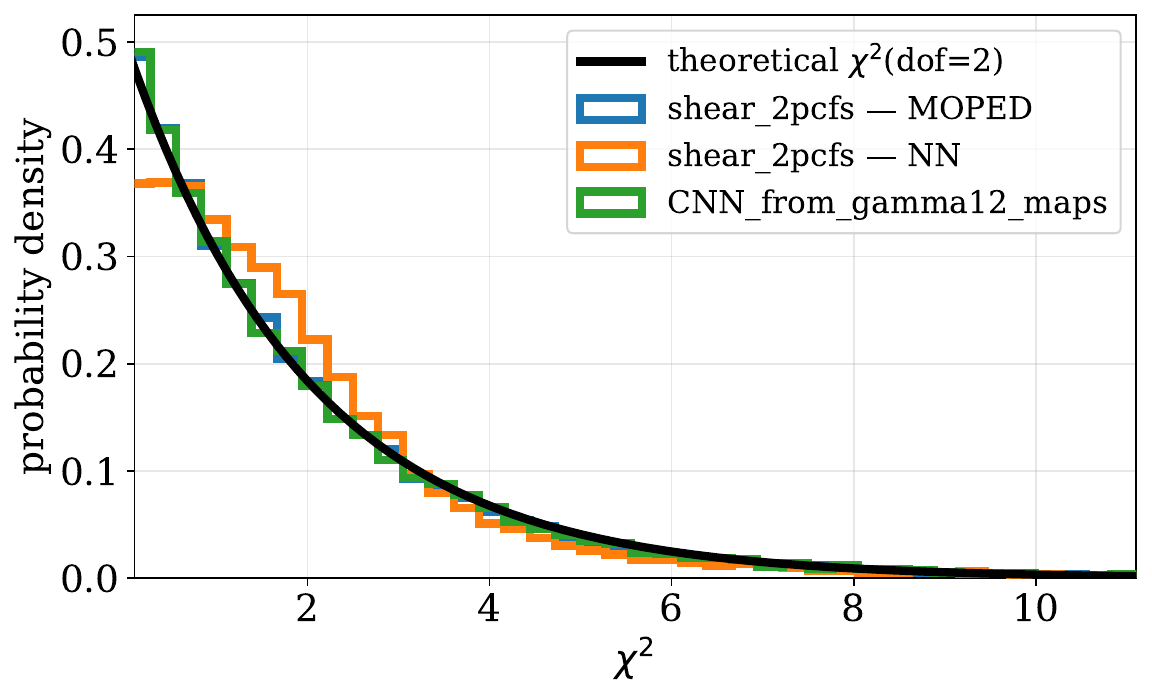}
    \caption{Global $\chi^2$ goodness-of-fit test comparing the empirical $\chi^2$ distribution (coloured histograms) across the $20\,400$ fiducial flat-patch realisations to the theoretical $\chi^2$ distribution with $n_{\rm p}=2$ degrees of freedom (solid black curve), for MOPED-shear-2PCFs (blue), NN-shear-2PCFs (orange), and CNN (green) compressed summaries.}
    \label{fig:chi2_compression}
\end{figure}

% ------
The Gaussian likelihood assumption of Eq.~(\ref{eq:log_likelihood}) requires that the DV distribution at fixed cosmology is well approximated by a multivariate Gaussian. We validate this at the fiducial cosmology using the $20\,400$ fiducial flat-patch realisations available.
In detail, a per-bin marginal Kolmogorov--Smirnov (KS) test compares the empirical distribution of each bin $d_i$ (standardised to zero mean and unit variance) to the standard normal $\mathcal{N}(0,1)$. The KS statistic is $D_N = \sup_x |F_N(x) - \Phi(x)|$, where $\sup_x$ denotes the supremum (largest value) over all $x$, $F_N$ is the empirical cumulative distribution function (CDF) and $\Phi$ is the standard normal CDF. A bin passes at the 5\% significance level when $D_N < D_{\rm crit} = 1.36 / \sqrt{N_{\rm real}}$, where $N_{\rm real}$ is the number of realisations.
Fig.~\ref{fig:gaussianity_compression} shows the per-bin marginal distributions for the two MOPED- and NN-compressed summaries of shear-2PCFs. Under MOPED compression, bin~0 passes the KS test (KS $= 0.004 < $ KS$_{\rm crit} = 0.010$) and bin~1 shows a mild failure (KS $= 0.050$), indicating that the MOPED-compressed summaries are close to Gaussian, as expected from their construction via linear Fisher-optimal projections of the data covariance. Under NN compression, however, both bins fail the KS test at a much higher level (KS $= 0.067$ and $0.081$ respectively), with histograms showing pronounced asymmetry and heavy tails relative to the best-fit Gaussian: the NN summaries are strongly non-Gaussian. We stress that the observed non-Gaussianity does not necessarily arise solely from the original DVs, but also reflects the properties of the non-linear compression performed by the NN. As such, the distribution of the compressed summaries is expected to encode both the non-Gaussian features of the input data and the characteristics of the compression itself. For completeness, the bottom row of Fig.~\ref{fig:gaussianity_compression} shows the CNN-compressed summaries: bin~0 passes the KS test (KS $= 0.005$) while bin~1 shows a mild failure (KS $= 0.044$), a behaviour qualitatively similar to the MOPED case. The overall picture is confirmed by the global $\chi^2$ goodness-of-fit test shown in Fig.~\ref{fig:chi2_compression}, which compares the empirical distribution of $\chi^2_i = (\mathbf{d}_i - \langle\mathbf{d}\rangle)^{\rm T}\,\hat{\mathsf{C}}^{-1}\,(\mathbf{d}_i - \langle\mathbf{d}\rangle)$ across the $20\,400$ covariance fiducial realisations to the theoretical $\chi^2$ distribution with $n_{\rm p}=2$ degrees of freedom. While MOPED-shear-2PCFs and CNN summaries closely follow the theoretical distribution, the NN-shear-2PCFs summaries show a clear departure from it, with the empirical distribution peaking away from zero, consistent with the strong non-Gaussianity already revealed by the marginal KS test. This difference has direct consequences for the validity of the Gaussian likelihood assumption in ELI, and for the relative performance of ELI and LFI under the two compression schemes, as discussed in Sect.~\ref{sec:eli_vs_lfi_shear2pcfs}.
We stress that no single Gaussianity diagnostic is individually sufficient to establish that a distribution is Gaussian: passing one test is not proof of Gaussianity, whereas failing any test is sufficient to probe its departure. This asymmetry is illustrated by comparing Fig.~\ref{fig:chi2_compression} with the per-bin marginal test of Fig.~\ref{fig:gaussianity_compression}. The global $\chi^2$ statistic, being an inverse covariance-weighted combination of all DV bins, is passed by both MOPED-shear-2PCFs and CNN, closely tracking the theoretical distribution. However, the per-bin marginal KS test reveals that both MOPED-shear-2PCFs and CNN mildly fail Gaussianity in their second bin, a feature the aggregate $\chi^2$ test is not sensitive enough to detect, particularly in presence of lengthier DVs.
% ------
\subsection{MCMC sampling from explicit posterior}
\label{sec:eli_mcmc}
% ------
Once the GP emulator is trained and validated, we sample the posterior distribution $p(\boldsymbol{\theta} \mid \mathbf{d}_{\rm obs}) \propto \mathcal{L}(\boldsymbol{\theta})\,p(\boldsymbol{\theta})$ by evaluating the log-likelihood of Eq.~(\ref{eq:log_likelihood}) at each step of a Markov Chain Monte Carlo (MCMC) exploration of the parameter space. Posterior samples are drawn using the \texttt{emcee} affine-invariant ensemble sampler \citep{ForemanMackey2013} with $n_{\rm walkers} = n_{\rm dim}^2 = 4$ walkers, 1\,000 burn-in steps, and 10\,000 production steps, using the flat uniform priors defined in Sect.~\ref{sec:lhs}. Chain convergence is verified on an initial subset of chains via the Gelman--Rubin diagnostic \citep{Gelman_Rubin_1992}, confirming that the chosen number of burn-in and production steps is sufficient. The inference is performed directly in $(\Omega_{\rm m}, \sigma_8)$ space; the derived parameter $S_8 \equiv \sigma_8\sqrt{\Omega_{\rm m}/0.3}$ is obtained by applying this transformation point-by-point to the posterior samples. The posterior median is adopted as the best-fit parameter value, and result figures display joint and marginal posteriors in the $(\Omega_{\rm m}, S_8)$ plane with the $S_8$ axis restricted to $[0.60, 0.95]$, which comfortably brackets the fiducial value $S_8^{\rm fid} = 0.823$.

% ----------
\section{Likelihood-Free Inference}
\label{sec:lfi}
% ----------

We present the LFI framework in the following subsections. We describe the NDE and inference mode in Sect.~\ref{sec:lfi_NDE}, the posterior quality assessment diagnostics in Sect.~\ref{sec:lfi_posterior_validation} and the MCMC sampling in Sect.~\ref{sec:lfi_mcmc}.

% ----------
\subsection{Neural Density Estimators} 
\label{sec:lfi_NDE}
% ----------
In the LFI framework, we train NDE to approximate the likelihood or posterior directly from simulated data, without assuming any parametric form for the distribution and without requiring a covariance matrix. We implement LFI using the \texttt{ltu-ili}\footnote{\url{https://github.com/maho3/ltu-ili}} package (\citealp{Ho_2024}), which provides a unified interface to the \texttt{sbi} library (\citealp{TejeroCantero_2020}) for simulation-based inference.
Two inference modes are distinguished. Neural Posterior Estimation (NPE) trains a network to approximate the posterior $p(\boldsymbol{\theta} \mid \mathbf{d})$ directly, by learning to map DVs to distributions over parameters. Neural Likelihood Estimation (NLE) instead trains a network to approximate the conditional likelihood $p(\mathbf{d} \mid \boldsymbol{\theta})$, from which the posterior is recovered via Bayes' theorem combined with the prior, consistently with the ELI framework. The key difference is that NPE is prior-dependent, the posterior is learned jointly with the prior geometry, while NLE is prior-independent: the learned likelihood can be combined with any prior at inference time without retraining, making it more flexible for prior sensitivity studies. In this work, we use NPE during the automated architecture grid search (Appendix~\ref{app:lfi_nde_grid}), where its ability to produce posterior samples directly without MCMC reduces the per-trial evaluation cost by orders of magnitude, while NLE is adopted for the final best-net inference and posterior validation (Sects.~\ref{sec:lfi_posterior_validation} and~\ref{sec:lfi_mcmc}), where prior independence and consistency with the ELI likelihood framework are essential.

Three NDE architectures are considered. The first two, Neural Spline Flow (NSF, \citealp{durkan2019neuralsplineflows}) and Masked Autoregressive Flow (MAF, \citealp{papamakarios2018maskedautoregressiveflowdensity}), belong to the family of normalising flows: they model the target distribution by learning a sequence of invertible transformations that warp a simple base distribution, here a standard Gaussian, into the complex, potentially multi-modal shape of the target distribution. The two architectures differ in how each transformation is parametrised. MAF uses simple linear rescalings of the input conditioned on previous iterations; this makes the transforms fast to evaluate and stable to train, at the cost of limited flexibility per layer. NSF replaces these linear rescalings with smooth, learnable curved mappings that can bend and stretch the probability density more freely, capturing sharper features and asymmetries in the likelihood with fewer layers. The third architecture, the Mixture Density Network (MDN, \citealp{bishop1994mdn}), takes a different approach: rather than transforming a base distribution, it directly parametrises the shape as a weighted sum of Gaussian components whose means, variances, and weights are predicted by a neural network as a function of the DV. MDN is therefore more interpretable but assumes that the target distribution can be well approximated by a finite mixture of Gaussians.

The key hyperparameters controlling each architecture are: the number of hidden units per layer, which determines how many intermediate values the network can use internally to represent complex patterns in the data; the number of successive transformations for NSF and MAF, which sets how many sequential steps are stacked to build up the final distribution from the base Gaussian where more steps allow more complex shapes but increase training cost, and the number of Gaussian components for MDN, which controls how many distinct peaks or modes the mixture can represent. All architectures additionally share two optimisation hyperparameters: the learning rate, which controls the step size during gradient-based training, and the batch size, which sets how many training examples are processed simultaneously at each update step. Each NDE is trained for up to 200 epochs which value is high enough to achieve full convergence of training and validation.

In order to train and validate the NDE, the compressed DVs (MOPED or NN) from $\mathrm{LHS}_1$ are split into training (80\%), validation (5\%), and test (15\%) subsets. One node of $\mathrm{LHS}_1$ is reserved as the mock observation for the MCMC chain, so the split is performed over the remaining 4,609 nodes, yielding 3,687 training, 230 validation, and 692 test nodes. With each node equipped with $n_{\rm flat}=50$ flat-patch measurements, this corresponds to 184,350 training, 11,500 validation, and 34,600 test flat-patch DV realisations.

% ----------
\subsection{Posterior quality assessment}
\label{sec:lfi_posterior_validation}
% ----------

Posterior quality of each NDE can be assessed using NLE via four complementary diagnostics reviewed in \cite{Ho_2024} and applied to 692 test cosmological nodes (15\% of total dataset), which the NDE has never seen during either training or architecture selection. To enable a direct comparison between frameworks, we mirror these same diagnostics for the ELI posterior, as described below and presented alongside the LFI results in Appendix~\ref{app:lfi_validation}.

\emph{Stability test.} To assess robustness to training stochasticity, the architecture is retrained independently $N_{\rm r}=3$ times from scratch with different random initialisations, each producing an independent ensemble posterior $\hat{p}_r(\boldsymbol{\theta} \mid \mathbf{d})$ for $r=0,1,2$. The spread across retrainings quantifies residual sensitivity to the NDE optimisation landscape.

\emph{Posterior predictive checks.} The posterior median of each parameter is compared against the true parameter value across all test nodes. A well-calibrated posterior should produce scatter consistent with the posterior width, with no systematic offset between predicted medians and true values.

\emph{PIT/rank histogram.} The Probability Integral Transform (PIT, \citealp{Czado_2009}) maps the true parameter value through the posterior cumulative distribution function, producing a variable in [0,1] that is uniformly distributed for a correctly calibrated inference procedure. Deviations from uniformity diagnose systematic biases or misestimated posterior widths. In practice, for each test node, $N_{\rm samp}$ samples are drawn from the posterior and the true parameter value is ranked among them, that is, we count how many posterior samples fall below the true value. If the posterior is perfectly calibrated, the true value is equally likely to fall anywhere among the samples, so the ranks should be uniformly distributed between 1 and $N_{\rm samp}$. Collecting these ranks into a histogram across all test nodes therefore provides a visual calibration check: a flat histogram indicates a well-calibrated posterior, a U-shaped histogram (true values systematically near the extremes of the posterior) indicates overconfidence, meaning the posterior is too narrow and excludes the truth too often, while an inverted-U shape (true values clustering near the centre) indicates under-confidence, meaning the posterior is unnecessarily wide.

\emph{TARP test.} The Test of Accuracy with Random Points (TARP, \citealp{Lemos_2023}) is a joint calibration diagnostic that probes the full multi-dimensional posterior rather than each parameter margin independently. For each test node, we compute the fraction of posterior probability mass lying in regions of lower density than the true parameter value; for a well-calibrated posterior, this fraction should equal the confidence level $\alpha$ on average, so that the Expected Coverage Probability (ECP) traces the diagonal $\mathrm{ECP}(\alpha) = \alpha$. Curves above (below) the diagonal indicate under-confidence (overconfidence). The full ECP curve is computed over 500 linearly spaced $\alpha$ bins, with inner and outer shaded bands estimated from 100 bootstrap resamples of the test set corresponding to $\pm 1\sigma$ and $\pm 2\sigma$ of the bootstrap ECP distribution respectively.

In addition, a marginal coverage test is computed independently for each parameter, probing calibration in each dimension individually rather than jointly. A well-calibrated marginal posterior traces the same diagonal $\hat{C}_d(\alpha) = \alpha$ as the TARP test, and deviations from it complement the joint diagnostic.

However, TARP is computationally demanding as it requires running one MCMC chain per test node to sample the NLE posterior; we therefore cap the evaluation at 50 test nodes (randomly sampled across the 692 available) and verified that results are consistent when this number is increased to 200 nodes for MOPED-compressed shear-2PCFs. 

\begin{figure}[h!]
    \centering

    % --- Top row ---
    \begin{minipage}{0.46\columnwidth}
        \centering
        \text{ELI MOPED}\\[1.em]
        \includegraphics[
            width=\linewidth,
            height=0.175\textheight,
            keepaspectratio
        ]{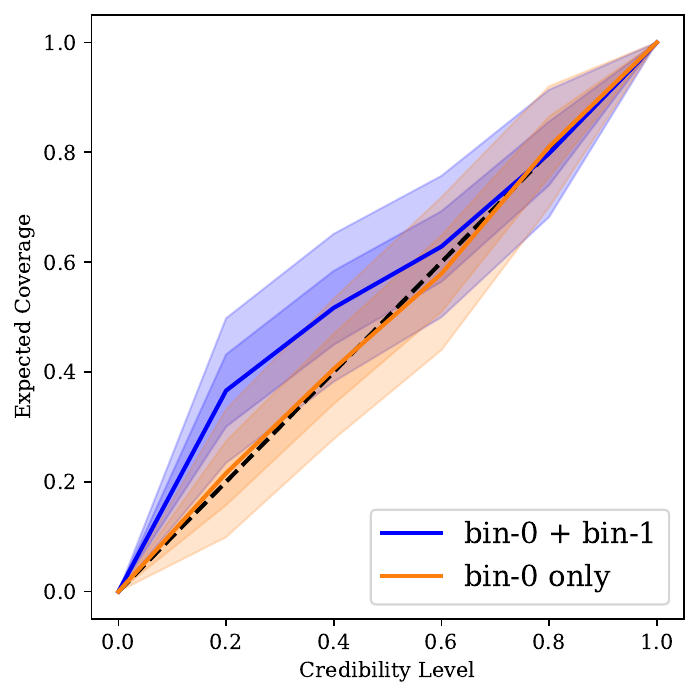}
    \end{minipage}
    \hfill
    \begin{minipage}{0.46\columnwidth}
        \centering
        \text{ELI NN}\\[1.em]
        \includegraphics[
            width=\linewidth,
            height=0.175\textheight,
            keepaspectratio,
            trim=0.3cm 0 0 0,
            clip
        ]{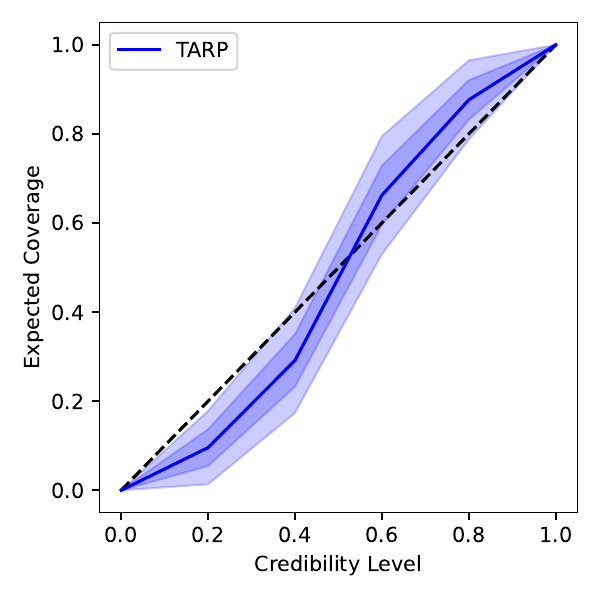}
    \end{minipage}

    \vspace{0.8em}

    % --- Bottom row ---
    \begin{minipage}{0.49\columnwidth}
        \centering
        \text{LFI MOPED}\\[0.2em]
        \includegraphics[
            width=\linewidth,
            height=0.33\textheight,
            keepaspectratio,
            trim=2cm 0 0 0,
            clip
        ]{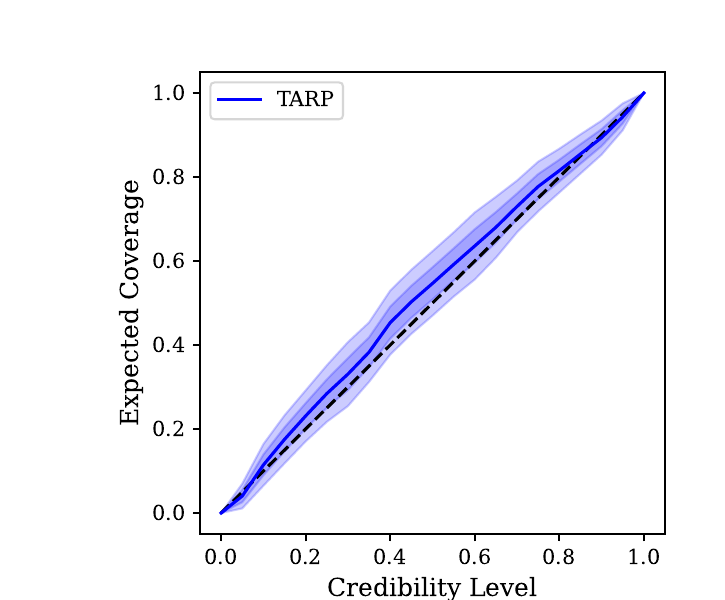}
    \end{minipage}
    \hfill
    \begin{minipage}{0.49\columnwidth}
        \centering
        \text{LFI NN}\\[0.2em]
        \includegraphics[
            width=\linewidth,
            height=0.33\textheight,
            keepaspectratio,
            trim=2cm 0 0 0,
            clip
        ]{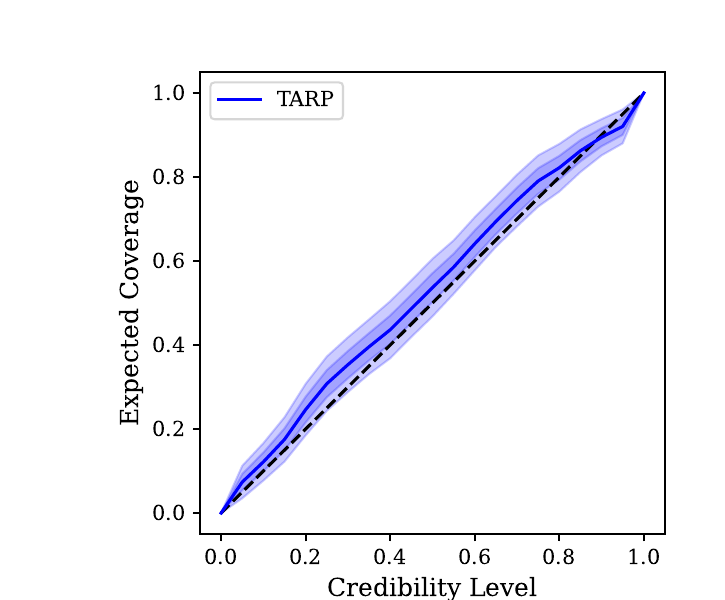}
    \end{minipage}

    \caption{TARP calibration test for shear-2PCFs, comparing ELI (top row) and LFI best-net (bottom row) under MOPED (left column) and NN (right column) compression. For ELI MOPED, the nominal both-bins (blue) and the bin-0-only (orange) configurations are shown. The expected coverage probability is plotted against the credibility level. The dashed diagonal indicates perfect calibration, and the shaded bands reflect the $1\sigma$ and $2\sigma$ scatter obtained via bootstrap resampling.}
    \label{fig:tarp_compression}
\end{figure}

In order to better understand the two inference techniques, we report in Fig.~\ref{fig:tarp_compression} the TARP curve for shear-2PCFs, comparing ELI (top row) and LFI from the best-net optimally selected as described in Appendix~\ref{app:lfi_nde_grid} (bottom row) under MOPED (left) and NN (right) compression. For ELI under MOPED compression, we additionally overlay the result obtained using only the well-emulated, near-Gaussian bin~0 (orange), alongside the nominal both-bins result (blue). The LFI panels lie close to the diagonal in both compression cases, confirming a well-calibrated joint posterior. The ELI panels, by contrast, show a mild but systematic departure from the diagonal mainly at low confidence levels for both the MOPED both-bins case and NN, indicating a degree of miscalibration that is absent in the LFI case; restricting to bin~0 alone in MOPED yields a markedly better-calibrated ELI posterior, closely tracking the diagonal and approaching the LFI level of calibration. This behaviour is consistent with the mild non-Gaussianity of the compressed likelihood identified by the marginal KS test of Sect.~\ref{sec:gaussianity} (Fig.~\ref{fig:gaussianity_compression}) and with the degraded GP emulation of bin~1 discussed in Appendix~\ref{app:gp_diagnostics}: since ELI enforces a Gaussian likelihood model onto a mildly non-Gaussian distribution and relies on a poorly emulated bin, its both-bins posterior cannot be perfectly calibrated, whereas LFI, which learns the likelihood shape directly from the simulations without any parametric assumption, recovers a well-calibrated posterior by construction. This effect is even more pronounced in the marginal coverage and PIT test (Appendix~\ref{app:lfi_validation}, Figs.~\ref{fig:lfi_coverage_compression}-\ref{fig:lfi_histogram_compression}), where ELI configurations show a markedly stronger departure from the diagonal than in the TARP test and inhomogeneous distribution of the ranks in the PIT, indicating that the per-parameter metrics are more sensitive diagnostic to this miscalibration than the TARP test alone.

% ------
\subsection{MCMC sampling from learned posterior}
\label{sec:lfi_mcmc}
% ------
Consistently with Sect.~\ref{sec:eli_mcmc}, posterior samples are drawn using the \texttt{emcee} ensemble sampler. Since we adopt NLE as the final inference mode, the sampler evaluates the learned conditional likelihood $\ln p(\mathbf{d}_{\rm obs} \mid \boldsymbol{\theta})$ at each step, which is combined with the prior via Bayes' theorem to obtain the log-posterior $\ln p(\boldsymbol{\theta} \mid \mathbf{d}_{\rm obs})$. The walker configuration, burn-in, and production steps are identical to those used in the ELI pipeline (Sect.~\ref{sec:eli_mcmc}).
% % ============================================================
\section{ELI vs.\ LFI with MOPED shear-2PCFs}
\label{sec:eli_vs_lfi_shear2pcfs}
% ============================================================

In this section, we present the main inference in our Euclid-like DR3 non-tomographic setup based on $100 \, \rm deg^2$ of simulated area using shear-2PCFs ($\xi_\pm$) as the summary statistic, applied to a single noise realisation of a fiducial mock observation ($\Omega_{\rm m}^{\rm fid} = 0.2905$, $S_8^{\rm fid} = 0.823$) extracted from $\rm LHS_{1}$. We compare the posteriors from ELI and LFI under MOPED compression in Sect.~\ref{sec:eli_vs_lfi}, compare MOPED and NN compression within each framework in Sect.~\ref{sec:compression_comparison}, and assess the sensitivity of the inferred posteriors to the shape noise and cosmic variance in Sect.~\ref{sec:bias_distribution}.

% ----------
\subsection{ELI vs.\ LFI}
\label{sec:eli_vs_lfi}
% ----------

\begin{figure}[h]
    \centering
    \includegraphics[width=1.0\columnwidth]{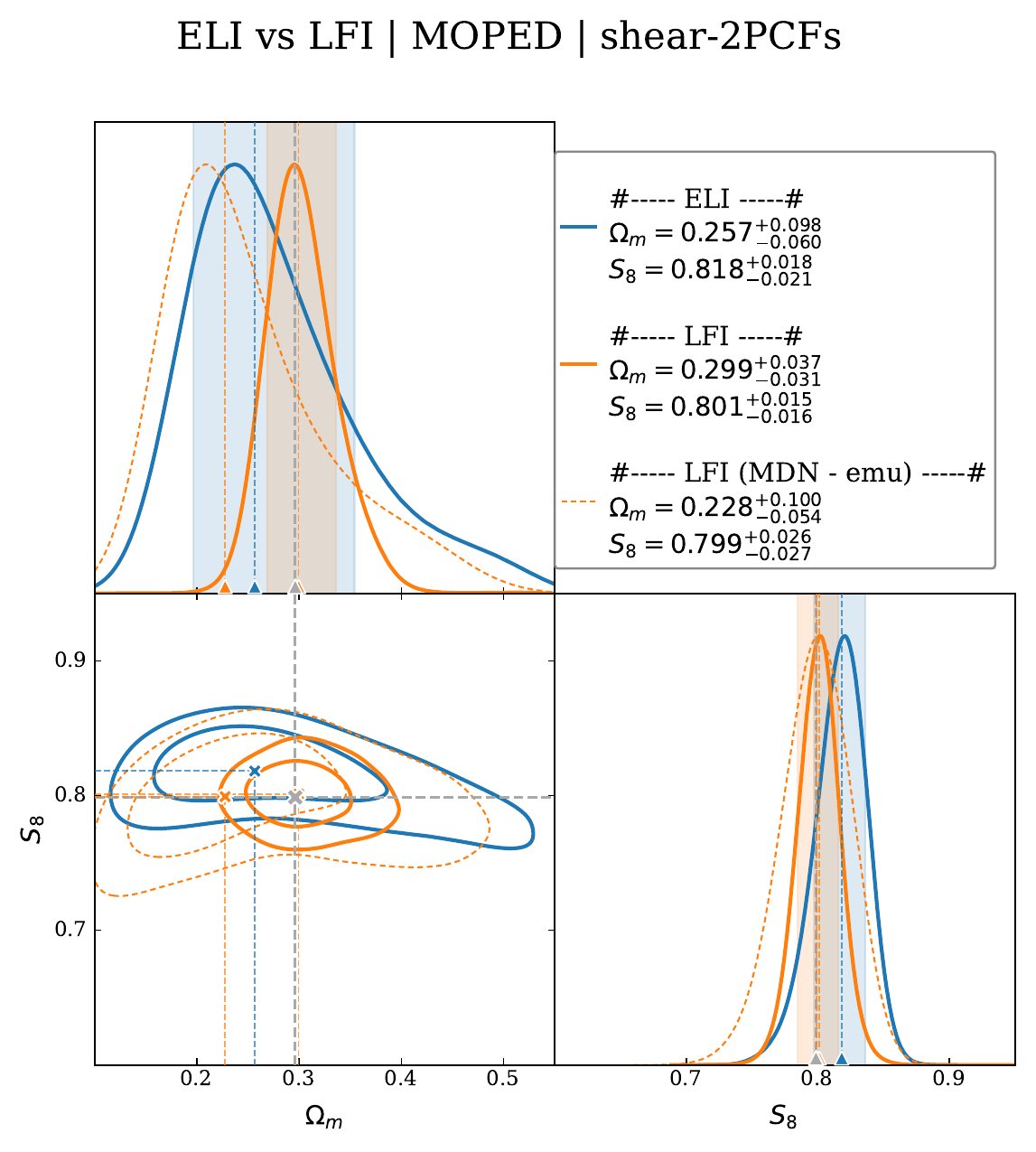}
    \caption{Posterior constraints in the $(\Omega_{\rm m}, S_8)$ plane from ELI (solid blue), LFI best-net (solid orange) and LFI with a single-component MDN from noisy emulator mean predictions (orange dashed), all with MOPED compression applied to shear-2PCFs. Contours show the $1\sigma$ and $2\sigma$ credible regions. The grey dashed lines and grey cross indicate the true target parameters ($\Omega_{\rm m}^{\rm fid} = 0.2905$, $S_8^{\rm fid} = 0.823$); coloured dashed lines and crosses mark the posterior medians. The diagonal panels show the marginal posteriors with $1\sigma$ shaded bands, the latter for the fiducial ELI \& LFI  posteriors only.}
    \label{fig:eli_vs_lfi}
\end{figure}

\begin{figure}[h]
    \centering
    \includegraphics[width=1.0\columnwidth]{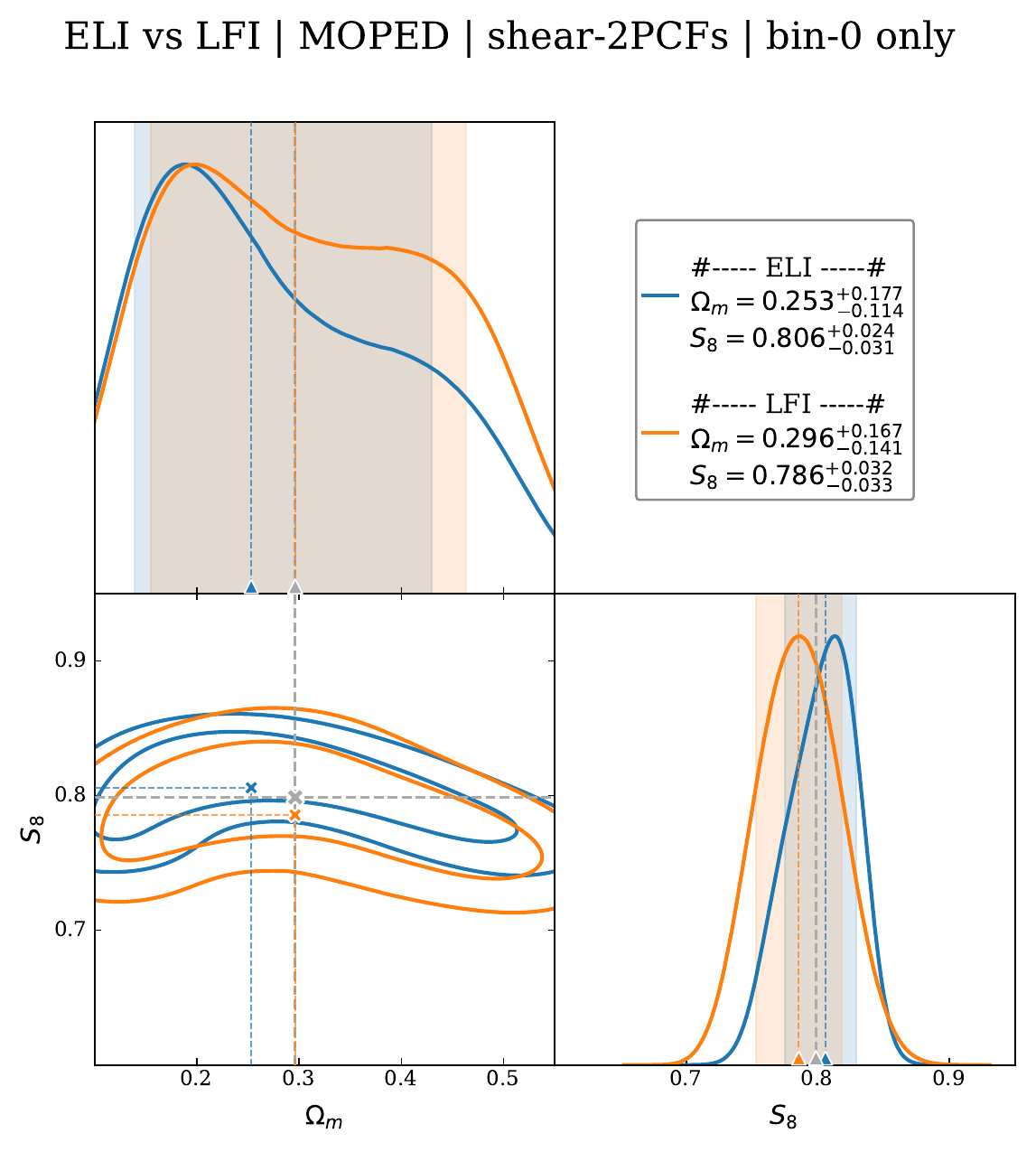}
    \caption{Same as Fig.~\ref{fig:eli_vs_lfi} but retaining only bin~0 of the MOPED-compressed DV.}
    \label{fig:eli_vs_lfi_bin0}
\end{figure}

Fig.~\ref{fig:eli_vs_lfi} compares the shear-2PCFs posteriors from ELI and LFI under MOPED compression, which provides the most direct framework-to-framework comparison by holding the compression scheme fixed.

Both posteriors are consistent with the target cosmology within their respective credible intervals. The ELI posterior has a median offset from the target ($\Omega_{\rm m}^{\rm ELI} = 0.257^{+0.098}_{-0.060}$, $S_8^{\rm ELI} = 0.818^{+0.018}_{-0.021}$), while the LFI posterior is better centred ($\Omega_{\rm m}^{\rm LFI} = 0.299^{+0.037}_{-0.031}$, $S_8^{\rm LFI} = 0.801^{+0.015}_{-0.016}$). Both frameworks well recover $(\Omega_{\rm m}^{\rm fid}, S_8^{\rm fid})$ within $1\sigma$, indicating no significant cosmological bias.

In terms of constraining power, ELI recovers larger ($\Omega_{\rm m}$) $S_{8}$ uncertainty up to a factor ($\approx 2$) $\approx 1.3$ compared to LFI. This striking difference, found to be even more pronounced using NN compression, can be explained by three caveats of the ELI framework: the cosmological dependence of the covariance matrix, neglected under our fixed-cosmology approximation; the non-Gaussianity of the likelihood, assumed Gaussian but non-negligible in bin~1 (Fig.~\ref{fig:gaussianity_compression}); and the emulator accuracy, degraded for bin~1 (Appendix~\ref{app:gp_diagnostics}).

We assess each caveat in turn. First, we tested the impact of the cosmological dependence of the covariance matrix using an approximate rescaling technique proposed in Eq.~(15) of \citet{Eifler_2009}. This is exact for a Gaussian random field assuming that the error-budget is dominated by the cosmic-variance contribution, a condition that we meet given the low-resolution and high galaxy density that we mimic in our setup. In this scenario, the covariance scales as $C(\theta) = A(\theta)\,C(\theta_{\rm fid})\,A(\theta)^\top$, with $A(\theta) = {\rm diag}[\boldsymbol{\mu}_{\rm GP}(\theta)/\boldsymbol{\mu}_{\rm GP}(\theta_{\rm fid})]$, so that the covariance is dynamically rescaled at every MCMC step rather than held fixed at its fiducial value. Repeating the ELI inference with this cosmology-dependent covariance yields posterior constraints consistent with the fiducial fixed-covariance analysis, confirming that this caveat has a negligible impact on our results and cannot explain the ELI--LFI discrepancy.

The remaining limitations of ELI, emulator accuracy and likelihood non-Gaussianity, are difficult to disentangle under MOPED compression because both affect bin~1, which carries most of the $\Omega_{\rm m}$ sensitivity, is the less accurately emulated of the two bins (Appendix~\ref{app:gp_diagnostics}), and mildly fails the Gaussianity test (Fig.~\ref{fig:gaussianity_compression}). We therefore perform two complementary consistency tests. First, in Fig.~\ref{fig:eli_vs_lfi}, we construct a synthetic LFI training set from the GP-predicted compressed DV means, adding Gaussian noise drawn from the MOPED covariance, and train a single-component MDN on these data. This reproduces within LFI the emulator mapping, covariance structure, and Gaussian-likelihood assumptions adopted by ELI. Under these matched assumptions, the resulting LFI (MDN -- emu) posterior closely reproduces the ELI posterior in both $\Omega_{\rm m}$ and $S_8$, showing that the two inference frameworks agree when applied to the same effective statistical model. Second, in Fig.~\ref{fig:eli_vs_lfi_bin0}, restricting both analyses to bin~0, which is accurately emulated and approximately Gaussian, likewise brings the ELI and LFI posteriors into close agreement. Together, these tests show that the observed discrepancy emerges when the assumptions underlying the ELI analysis are not satisfied, rather than from an intrinsic difference between the inference algorithms.

Together, these tests indicate that emulator inaccuracy and residual likelihood non-Gaussianity, rather than the covariance, are responsible for the discrepancy seen using MOPED compression, reflecting ELI-specific approximations rather than a fundamental incompatibility between the two paradigms. Similar conclusions hold for NN compression, where the healthy emulation of the compressed bins (Appendix~\ref{app:gp_diagnostics}) isolates likelihood non-Gaussianity (Figs.~\ref{fig:gaussianity_compression}-\ref{fig:chi2_compression}) as the sole remaining driver of the ELI--LFI discrepancy: once LFI is trained on the synthetic Gaussian dataset, the agreement is restored, confirming likelihood non-Gaussianity as its sole cause. These results also highlight the value of posterior validation metrics considered in this work as practical tools to identify such shortcomings in ELI analyses.

% ----------
% ----------
\subsection{Comparison of compression methods}
\label{sec:compression_comparison}
% ----------

We now compare MOPED (Sect.~\ref{sec:moped}) and NN (Sect.~\ref{sec:nn_compression}) compression within each inference framework separately. 

\begin{figure}[h]
    \centering
    \includegraphics[width=1.0\columnwidth]{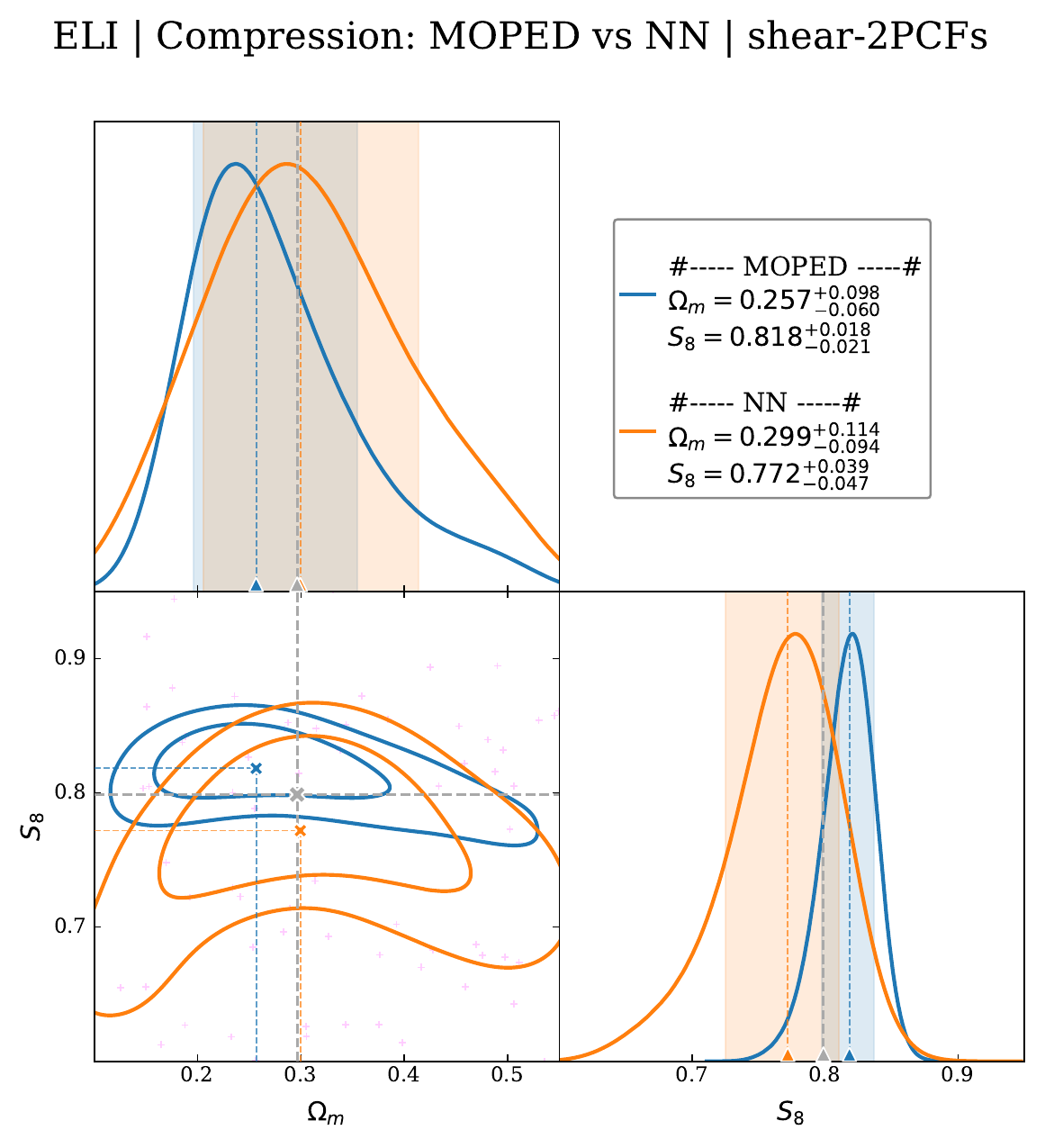}
    \caption{ELI posterior constraints in the $(\Omega_{\rm m}, S_8)$ plane comparing MOPED (blue) and NN (orange) compression for shear-2PCFs. Magenta $+$ symbols indicate the 250 $\mathrm{LHS}_1$ nodes randomly selected to train the GP emulator, shown after transformation to $(\Omega_{\rm m}, S_8)$ within the plotted $S_8$ range. Contours, markers, and shaded bands as in Fig.~\ref{fig:eli_vs_lfi}.}
    \label{fig:compression_eli}
\end{figure}

\begin{figure}[h]
    \centering
    \includegraphics[width=1.0\columnwidth]{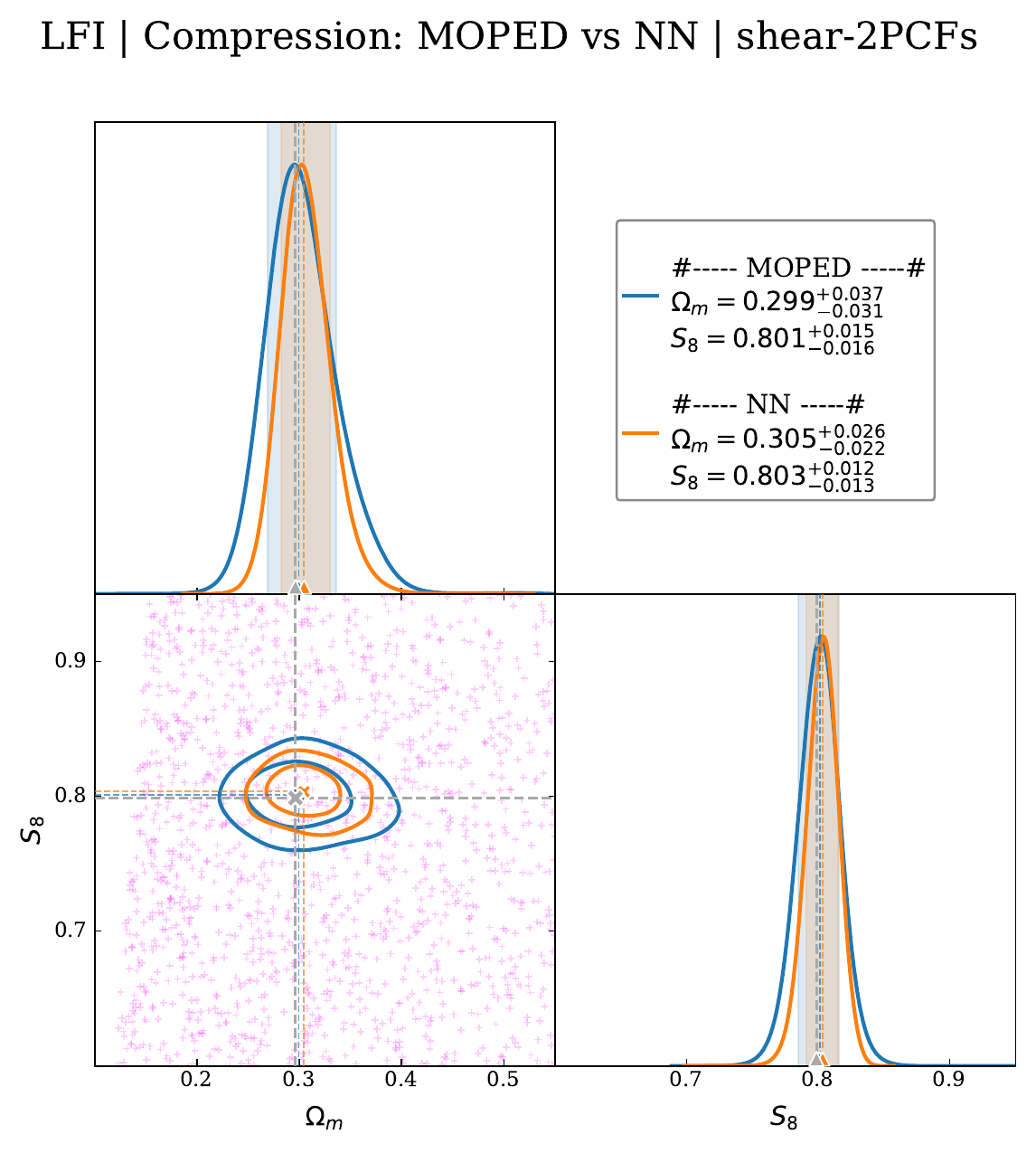}
    \caption{LFI posterior constraints in the $(\Omega_{\rm m}, S_8)$ plane comparing MOPED (blue) and NN (orange) compression for shear-2PCFs. Magenta $+$ symbols indicate the 3\,687 $\mathrm{LHS}_1$ nodes used to train the NDE, shown after transformation to $(\Omega_{\rm m}, S_8)$ within the plotted $S_8$ range. Contours, markers, and shaded bands as in Fig.~\ref{fig:eli_vs_lfi}.}
    \label{fig:compression_lfi}
\end{figure}

Fig.~\ref{fig:compression_eli} shows the ELI results: MOPED compression yields a well-constrained posterior ($\Omega_{\rm m}^{\rm MOPED} = 0.257^{+0.098}_{-0.060}$, $S_8^{\rm MOPED} = 0.818^{+0.018}_{-0.021}$), while NN compression produces markedly broader contours extending to the prior boundary ($\Omega_{\rm m}^{\rm NN} = 0.299^{+0.114}_{-0.094}$, $S_8^{\rm NN} = 0.772^{+0.039}_{-0.047}$), suggesting that the posterior has not retained the full information available under this configuration. Fig.~\ref{fig:compression_lfi} shows the LFI results: here both compressions yield tight and mutually consistent posteriors, with NN slightly outperforming MOPED ($\Omega_{\rm m}^{\rm NN} = 0.305^{+0.026}_{-0.022}$, $S_8^{\rm NN} = 0.803^{+0.012}_{-0.013}$ versus $\Omega_{\rm m}^{\rm MOPED} = 0.299^{+0.037}_{-0.031}$, $S_8^{\rm MOPED} = 0.801^{+0.015}_{-0.016}$).

Building on the posterior calibration results of Sect.~\ref{sec:lfi_posterior_validation} (Fig.~\ref{fig:tarp_compression}) and the interpretation developed in Sect.~\ref{sec:eli_vs_lfi}, we attribute the reversal in relative performance between ELI and LFI under NN compression to the strong non-Gaussianity of the NN-compressed likelihood (Figs.~\ref{fig:gaussianity_compression}-\ref{fig:chi2_compression}). MOPED, being a Fisher-optimal linear projection, is inherently better matched to the Gaussian likelihood assumption of ELI, while NN compression, being non-linear, produces summaries whose distribution departs more strongly from Gaussianity; this is deleterious for ELI but does not penalise LFI, which learns the likelihood shape directly from the simulations. Conversely, LFI benefits from NN compression, which yields modestly tighter constraints and can retain non-Gaussian features that are not fully exploited by the linear MOPED projection when its underlying assumptions are violated. Within our setup, we therefore recommend and use the MOPED compression for ELI and the NN compression for LFI.

\subsection{Bias distribution test}
\label{sec:bias_distribution}
% ----------

To assess the sensitivity of the inferred posteriors to the specific shape noise and cosmic variance realisation, we run independent MCMC chains on 20 different mock observations taken over different flat-projected patches carved from the same full-sky map of our target cosmology, using MOPED-compressed shear-2PCFs. Specifically, 1 of these 20 realizations corresponds to the one used for all posterior estimation shown so far.

For each realisation, we record the posterior median, mean, and maximum-a-posteriori (MAP) estimates of $\Omega_{\rm m}$ and $S_8$, together with the symmetric marginal $1\sigma$ posterior width $\sigma_\theta$. The relative bias for each realisation and estimator is defined as $(\hat{\theta} - \theta_{\rm true} - \widetilde{\Delta\theta})/\sigma_\theta$, where $\widetilde{\Delta\theta}$ is the median bias across the 20 realisations for that estimator; each point is therefore shifted by the common median, so that the plot isolates the realization-to-realization fluctuations around the overall trend rather than the absolute offset. Realisation ID~0 is highlighted by a coloured circle, corresponding to the specific noise realisation used for all triangle plots in Sects.~\ref{sec:eli_vs_lfi_shear2pcfs} and~\ref{sec:null_test}.

\begin{figure*}[t]
    \centering

    \begin{minipage}{0.49\textwidth}
        \centering
        \vspace{0.3em}
        \text{ELI}\\[-0.05em]
        \includegraphics[width=\linewidth, trim=0 0 0 1.5cm, clip]
        {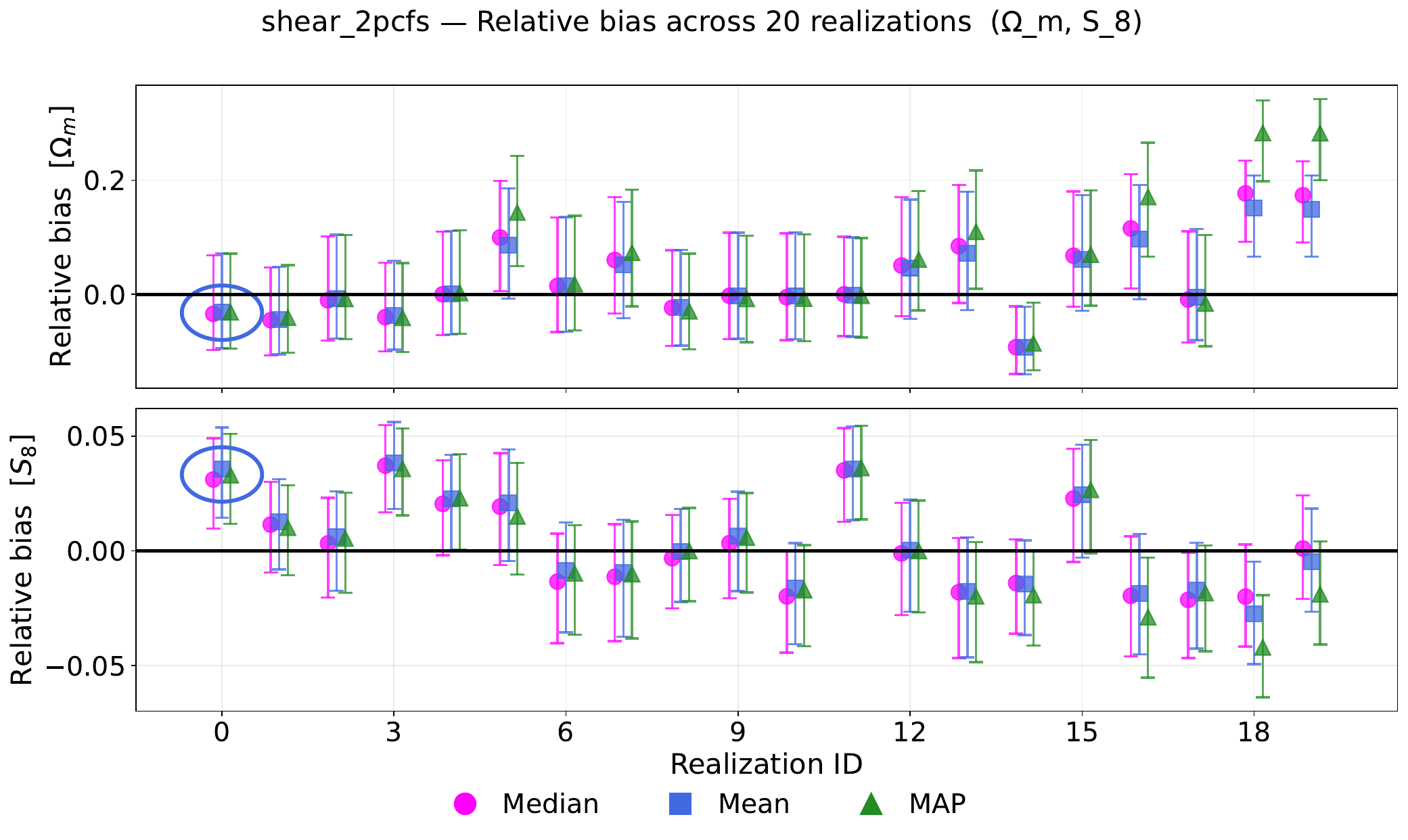}
    \end{minipage}
    \hfill
    \begin{minipage}{0.49\textwidth}
        \centering
        \vspace{0.3em}
        \text{LFI}\\[-0.05em]
        \includegraphics[width=\linewidth, trim=0 0 0 1.5cm, clip]
        {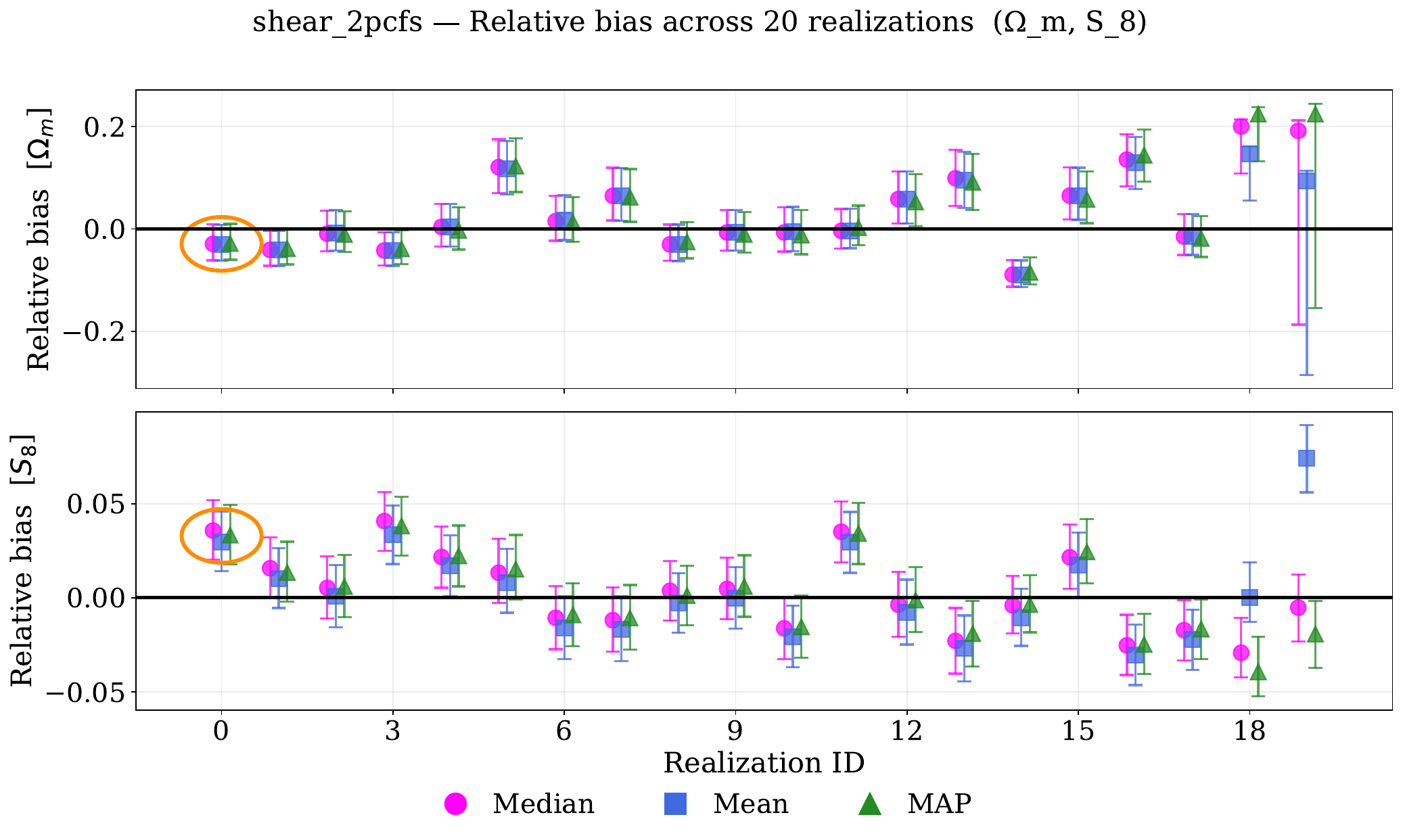}
    \end{minipage}

    \caption{Relative bias across 20 shape noise realisations of the fiducial cosmology for MOPED-compressed shear-2PCFs, for ELI (left) and LFI (right). Each point shows the relative bias $(\hat{\theta} - \theta_{\rm true} - \widetilde{\Delta\theta})/\sigma_\theta$ as a function of realisation ID for $\Omega_{\rm m}$ (top row) and $S_8$ (bottom row), where $\widetilde{\Delta\theta}$ is the median bias across the 20 realisations for that estimator and $\sigma_\theta$ is the per-realisation marginal $1\sigma$ posterior width. Each bias is therefore shifted by the common median, isolating realization-to-realization fluctuations around the overall trend. The three estimators are shown as separate markers: median (magenta circles), mean (blue squares), and MAP (green triangles), with a small horizontal jitter to avoid overlap. Errorbars show the asymmetric normalised $1\sigma$ interval. The highlighted circle at ID~0 marks the realisation used for all triangle plots in this paper.}
    \label{fig:bias_eli_lfi}
\end{figure*}

For ELI (Fig.~\ref{fig:bias_eli_lfi}, left), the three estimators (median, mean, MAP) show a consistent realization-to-realization trend for both $\Omega_{\rm m}$ and $S_8$, with fluctuations tightly correlated between the three estimators across all 20 realisations.

For LFI (Fig.~\ref{fig:bias_eli_lfi}, right), the same tight consistency between the three estimators is observed. The realization-to-realization trend is qualitatively similar to that of ELI, confirming that the fluctuations across realisations are driven by the shared noise input rather than by the inference framework.

The main conclusion from this diagnostic is that both ELI and LFI share a consistent realization-to-realization trend and that the three best-fit estimators are mutually consistent in both frameworks. This demonstrates that the inferred constraints are robust to the noise properties, and that the dominant source of variation across realisations is driven by shape noise and cosmic variance rather than any property of the likelihood model.

% % % ============================================================
\section{shear-2PCFs vs.\ CNN}
\label{sec:null_test}
% ============================================================

\begin{figure*}[t]
    \centering
    \begin{minipage}{0.49\textwidth}
        \centering
        \includegraphics[width=\linewidth]{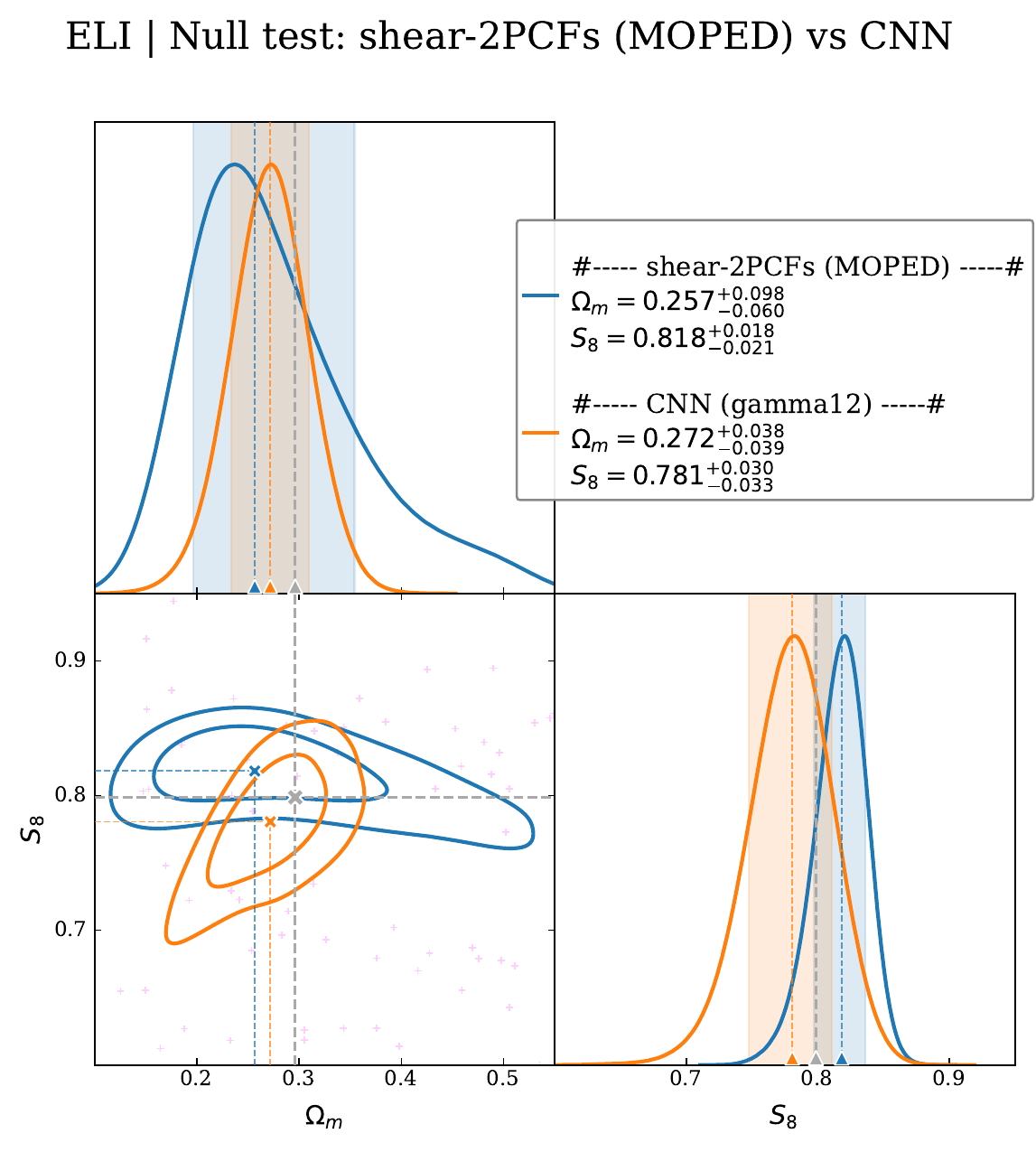}
    \end{minipage}
    \hfill
    \begin{minipage}{0.49\textwidth}
        \centering
        \includegraphics[width=\linewidth]{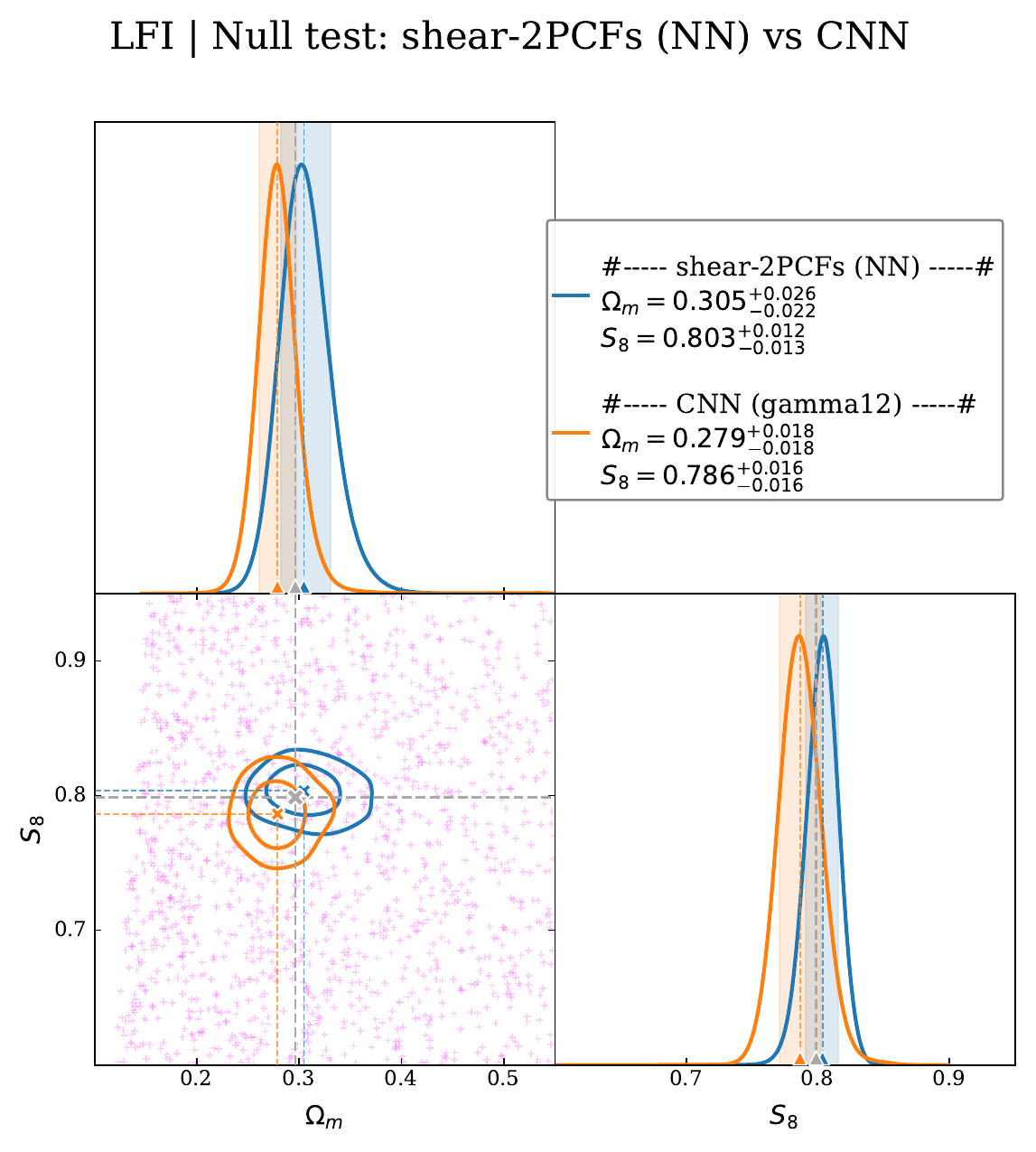}
    \end{minipage}
    \caption{Posterior constraints in the $(\Omega_{\rm m}, S_8)$ plane comparing shear-2PCFs (blue) and the CNN estimator (orange), for ELI (left, MOPED-compressed shear-2PCFs) and LFI (right, NN-compressed shear-2PCFs). Contours, markers, and shaded bands as in Fig.~\ref{fig:eli_vs_lfi}.}
    \label{fig:nulltest}
\end{figure*}

From a purely theoretical standpoint, in the idealised limit where both binning and compression are lossless, summary statistics applied to Gaussian random fields should extract no more cosmological information than a maximally informative Gaussian probe, here represented by the shear-2PCFs. In this section we test whether this expectation holds in practice for shear-2PCFs and the map-level CNN estimator in our simplified Gaussian-random-field simulated setting (Sect.~\ref{sec:glass}).

Fig.~\ref{fig:nulltest} shows the ELI (left) and LFI (right) posteriors from shear-2PCFs (blue) and the CNN estimator (orange), using the same mock observation as in Sect.~\ref{sec:eli_vs_lfi_shear2pcfs}. Following the conclusions of Sect.~\ref{sec:compression_comparison}, we adopt for each framework the compression scheme found to perform best: MOPED-compressed shear-2PCFs for ELI, and NN-compressed shear-2PCFs for LFI.

For ELI, the shear-2PCFs posterior is well-behaved ($\Omega_{\rm m}^{\rm 2PCF} = 0.257^{+0.098}_{-0.060}$, $S_8^{\rm 2PCF} = 0.818^{+0.018}_{-0.021}$). The CNN posterior, by contrast, shows a tighter $\Omega_{\rm m}$ but a systematically lower $S_8$ ($\Omega_{\rm m}^{\rm CNN} = 0.272^{+0.038}_{-0.039}$, $S_8^{\rm CNN} = 0.781^{+0.030}_{-0.033}$). This underestimation of $S_8$ is analogous to the behaviour observed for NN compression in Sect.~\ref{sec:compression_comparison}: the mild non-Gaussianity of the CNN-compressed bins (Fig.~\ref{fig:gaussianity_compression}, bin~1) is sufficient to bias the ELI posterior width under the Gaussian likelihood assumption, even though the CNN summaries are well emulated by the GP (Appendix~\ref{app:gp_diagnostics}). Under ELI, the apparently broader $\Omega_{\rm m}$ posterior from MOPED-shear-2PCFs is instead consistent with the limited sensitivity of the second MOPED bin to cosmology, as discussed in Sect.~\ref{sec:eli_vs_lfi}.

For LFI, that goes beyond the emulator and Gaussian likelihood limitations, the two summary statistics are in good mutual consistency revealing a similar amount of cosmological information: both $\Omega_{\rm m}$ and $S_8$ posteriors are tight and well-centred, with NN-compressed shear-2PCFs and the CNN agreeing within $\lesssim 1\sigma$ in both parameters. In detail, we find $\Omega_{\rm m}^{\rm NN} = 0.305^{+0.026}_{-0.022}$ versus $\Omega_{\rm m}^{\rm CNN} = 0.279^{+0.018}_{-0.018}$, a $\approx 0.9\sigma$ difference on the combined uncertainty and a $\approx 1.3\times$ broader $1\sigma$ error for NN shear-2PCFs. While for the clustering parameter we report $S_8^{\rm NN} = 0.803^{+0.012}_{-0.013}$ versus $S_8^{\rm CNN} = 0.786^{+0.016}_{-0.016}$, a $\approx 0.8\sigma$ difference and a $\approx 1.3\times$ tighter $1\sigma$ error for NN shear-2PCFs.

Interestingly, the two probes exhibit a markedly asymmetric constraining power. The CNN achieves comparable uncertainties on $\Omega_{\rm m}$ and $S_8$ ($\sigma_{\Omega_{\rm m}}^{\rm CNN}\approx0.018$, $\sigma_{S_8}^{\rm CNN}\approx0.016$), whereas NN-compressed shear-2PCFs is $\approx30\%$ broader in $\Omega_{\rm m}$ ($\sigma_{\Omega_{\rm m}}^{\rm NN}\approx0.024$) but $\approx25$--$30\%$ tighter in $S_8$ ($\sigma_{S_8}^{\rm NN}\approx0.0125$). This behavior, consistently observed in ELI, reflects the different nature of the compressions: the CNN learns directly from the maps and extracts information on both parameters with the same weight, while the NN operates on shear-2PCFs, inheriting their intrinsic $\Omega_{\rm m}$--$\sigma_8$ degeneracy and preferentially preserving the degeneracy-breaking combination closest to $S_8$. Consequently, NN-compressed shear-2PCFs and the CNN trade constraining power between the two parameters, rather than one probe being uniformly more informative

Concluding, the proposed comparison confirms that ELI constraints are susceptible to the compounded limitations of MOPED compression and the Gaussian likelihood approximation differently affecting the two probes. Whereas LFI, by bypassing such caveats, is naturally immune to these failure modes, yielding consistent constraints between NN-compressed shear-2PCFs and the CNN across both cosmological parameters.

% ============================================================
% ============================================================
% ============================================================
\section{Conclusions}
\label{sec:conclusions}
% ============================================================

% --- Paragraph 1: Context & Goal ---
We presented a controlled comparison of ELI and LFI for WL cosmological inference in the context of Stage-IV surveys. Using a common simulation suite, matched summary statistics, and identical compression strategies, we assessed key points of agreement and divergence between the two frameworks and identified the numerical assumptions and caveats that drive their performance in our setup. Together with other relevant studies (e.g. \citealp{Alsing_2018, Lemos_2023, Zeghal_2025}), our results contribute to benchmarking, interpreting and jointly deploying ELI and LFI in future high-precision WL analyses.

% --- Paragraph 2: Summary of methodology ---
We built a simulation framework based on the \texttt{GLASS} forward model to compare ELI and LFI under controlled conditions using Euclid-like non-tomographic weak-lensing shear maps based on Gaussian random fields. We considered two representative summary statistics: shear-2PCFs, compressed either with the linear MOPED algorithm or a non-linear NN compressor, and a map-level CNN estimator. ELI combines a GP emulator, a simulation-based covariance matrix, and a Gaussian likelihood, while LFI learns the posterior directly from simulations using neural density estimators. Both approaches are applied to the same mock observation and MCMC setup, focusing on constraints on $(\Omega_{\rm m},S_8)$.

% --- Paragraph 3: Main finding 1: diagnostics and ELI--LFI agreement ---

Our first main finding is methodological: we show that TARP and the accompanying calibration diagnostics, until now used almost exclusively to validate LFI posteriors, are similarly informative when applied to ELI, and are markedly more sensitive to residual miscalibration than the goodness-of-fit tests (KS, $\chi^2$) traditionally used to check the Gaussianity of a compressed DV. This is best illustrated by MOPED-compressed shear-2PCFs, whose two bins individually pass or only jointly mildly fail the marginal Gaussianity test but clearly brings to miscalibrated posterior once probed with TARP. More generally, we find that ELI and LFI diverge whenever the ELI assumptions of an accurate emulator and a near-Gaussian compressed likelihood are not simultaneously met, and agree once they are. Under MOPED compression, the discrepancy is traceable to degraded GP emulation of the second bin, which carries most of the $\Omega_{\rm m}$ sensitivity, compounded by its mild failure of the Gaussianity test. For both NN-compressed shear-2PCFs and the CNN, by contrast, emulation is accurate and non-Gaussianity of the compressed likelihood is the sole driver. In all three cases, the cosmology-dependence of the covariance matrix has a negligible impact and cannot explain the discrepancy on its own. We recover agreement between the two frameworks once the emulator mapping and covariance noise are propagated into LFI through a mock Gaussian dataset, consistently across all configurations. For MOPED compression, this interpretation is further reinforced by restricting the analysis to the first bin alone, which is both accurately emulated and consistent with Gaussianity, and which likewise brings ELI and LFI into close agreement.

% --- Paragraph 4: Main finding 2: compression choice and probe dependence ---
Our second main finding concerns the type of compression. For ELI, this choice matters, since its Gaussian likelihood assumption is best matched by the MOPED linear transformation that better preserves near-Gaussianity of the compressed summaries, whereas NN, being non-linear, strongly violates it, broadening the $\Omega_{\rm m}$ error by a factor of $\approx 1.3$ and the $S_8$ error by a factor of $\approx 2.2$ compared to MOPED. LFI, by contrast, is robust to both approaches, recovering consistent posteriors with NN retaining slightly more cosmological information and improving $\Omega_{\rm m}$ precision by $\approx 30\%$.

% --- Paragraph 5: Main finding 3: two-point vs. map-level information ---
Our third main finding concerns the CNN--shear-2PCFs comparison. Even in this idealised Gaussian setting, where the two-point function is expected to be information-optimal, this result is not retrieved in ELI, due to its emulation and Gaussian likelihood approximation limitations: the CNN yields an $\Omega_{\rm m}$ marginal PDF $\approx 2\times$ better constrained than MOPED shear-2PCFs, driven by the emulation caveat of this compression scheme, but with a $\approx 1.6\times$ broader $S_8$ error, driven by the mild non-Gaussianity of the CNN bins. This expectation is instead met in LFI, which is free from both limitations: the CNN and NN shear-2PCFs constraints agree within $\approx 30\%$, though still trading constraining power differently between $\Omega_{\rm m}$ and $S_8$. These findings highlight the importance of validating the inference framework before drawing conclusions about the relative information content of different WL summary statistics.

% --- Paragraph 6: Limitations ---
In light of the outcomes presented in our analysis, we remark that several limitations apply. Our Gaussian random field setting makes shear-2PCFs sufficient by construction up to numerical effects; in realistic N-body WL fields, HOS would outperform two-point statistics and the ELI Gaussian likelihood assumption could be more severely violated. Also, our non-tomographic approach in a low resolution grid limits the conclusions as it differ from tomography with access to smaller scales that would increase constraining power but amplify ELI-specific caveats and impose stricter NDE stability requirements in LFI. Similarly, the absence of systematics such as baryonic feedback (e.g. \citealp{Semboloni_2011}), intrinsic alignment (e.g. \citealp{Troxel_2015}) and source-clustering (e.g. \citealp{Gatti_2023}) together with the reduced bi-dimensionality of the inference may hide significant red-flags in the two inference approaches, thus making necessary an increased realism of the forward model. 

% --- Paragraph 7: Outlook and recommendations ---
Within these constraints, we first recommend running ELI and LFI in parallel, using the posterior calibration diagnostics covered here to spot failure modes and improve the overall robustness of the inference. Given the recurring role of likelihood non-Gaussianity in driving the ELI--LFI discrepancy throughout this work, we further recommend the development of non-Gaussian likelihood modelling within ELI, which is strongly required to overcome potential limitations. This has a direct and relevant impact on the choice of compression scheme, for which we suggest further exploration in both inference frameworks, such as information-maximising neural networks (\citealp{Charnock_2018}) or mutual information maximization (\citealp{Makinen_2025}). As an extension of this work, the most natural next steps are: (i) replacing \texttt{GLASS} with N-body simulations to test the ELI--LFI comparison in the genuinely non-Gaussian regime, where HOS carry information beyond two-point functions; (ii) extending to a tomographic configuration to assess the interplay between increased DV dimensionality and the ELI and LFI caveats identified here; (iii) incorporating observational systematics and more realistic forward models; and (iv) expanding the parameter space to additional cosmological and nuisance parameters to test scalability and robustness in a higher-dimensional inference setting.

\begin{acknowledgements}

This research used resources of the National Energy Research
Scientific Computing Center, a DOE Office of Science User Facility
supported by the Office of Science of the U.S. Department of Energy
under Contract No. DE-AC02-05CH11231 using NERSC award
HEP-ERCAP0037257. This research has made use of computing facilities operated by Centre de données Astrophysiques de Marseille data center at Laboratoire Astrophysique de Marseille, Marseille, France. Simone Vinciguerra \& Nicolas Martinet acknowledge the funding of the French Agence Nationale de la Recherche for the PISCO project (grant ANR-22-CE31-0004) and of the Centre National d'Etudes Spatiales (CNES). Marco Gatti acknowledges support from the Ramón y Cajal program under grant RYC2024-049031-I, funded by MICIU/AEI/10.13039/501100011033 and by the European Social Fund Plus (FSE+).

\end{acknowledgements}

%%%%%%%%%%%%%%%%%%%%%%%%%%%%%%%%%%%%%%%%%%%%%%%%%%%%%%%%%%%%%%%%%%%%%%%%%%%%%%%%%%%%%%%

\bibliographystyle{aa}
\bibliography{PISCO_paper}

@article{Alsing_2018,
   title={Massive optimal data compression and density estimation for scalable, likelihood-free inference in cosmology},
   volume={477},
   ISSN={1365-2966},
   url={http://dx.doi.org/10.1093/mnras/sty819},
   DOI={10.1093/mnras/sty819},
   number={3},
   journal={Monthly Notices of the Royal Astronomical Society},
   publisher={Oxford University Press (OUP)},
   author={Alsing, Justin and Wandelt, Benjamin and Feeney, Stephen},
   year={2018},
   month=Mar, pages={2874--2885} }

@article{Armijo_2025,
   title={Cosmological constraints using Minkowski functionals from the first year data of the Hyper Suprime-Cam},
   volume={537},
   ISSN={1365-2966},
   url={http://dx.doi.org/10.1093/mnras/staf257},
   DOI={10.1093/mnras/staf257},
   number={4},
   journal={Monthly Notices of the Royal Astronomical Society},
   publisher={Oxford University Press (OUP)},
   author={Armijo, Joaquin and Marques, Gabriela A and Novaes, Camila P and Thiele, Leander and Cowell, Jessica A and Grand{\'o}n, Daniela and Shirasaki, Masato and Liu, Jia},
   year={2025},
   month=Feb, pages={3553--3560} }

@misc{Abbott_2026,
      title={Dark Energy Survey Year 6 Results: Cosmological Constraints from Galaxy Clustering and Weak Lensing}, 
      author={{DES Collaboration: Abbott} and M. Adamow and M. Aguena and A. Alarcon and S. S. Allam and O. Alves and A. Amon and D. Anbajagane and F. Andrade-Oliveira and S. Avila and D. Bacon and E. J. Baxter and J. Beas-Gonzalez and K. Bechtol and M. R. Becker and G. M. Bernstein and E. Bertin and J. Blazek and S. Bocquet and D. Brooks and D. Brout and H. Camacho and G. Camacho-Ciurana and R. Camilleri and G. Campailla and A. Campos and A. Carnero Rosell and M. Carrasco Kind and J. Carretero and P. Carrilho and F. J. Castander and R. Cawthon and C. Chang and A. Choi and J. M. Coloma-Nadal and M. Costanzi and M. Crocce and W. d'Assignies and L. N. da Costa and M. E. da Silva Pereira and T. M. Davis and J. De Vicente and J. DeRose and H. T. Diehl and S. Dodelson and P. Doel and C. Doux and A. Drlica-Wagner and T. F. Eifler and J. Elvin-Poole and J. Estrada and S. Everett and A. E. Evrard and J. Fang and A. Farahi and A. Fert{\'e} and B. Flaugher and P. Fosalba and J. Frieman and J. Garc{\'i}a-Bellido and M. Gatti and E. Gaztanaga and G. Giannini and P. Giles and K. Glazebrook and M. Gorsuch and D. Gruen and R. A. Gruendl and J. Gschwend and G. Gutierrez and I. Harrison and W. G. Hartley and E. Henning and K. Herner and S. R. Hinton and D. L. Hollowood and K. Honscheid and E. M. Huff and D. Huterer and B. Jain and D. J. James and M. Jarvis and N. Jeffrey and T. Jeltema and T. Kacprzak and S. Kent and A. Kovacs and E. Krause and R. Kron and K. Kuehn and O. Lahav and S. Lee and E. Legnani and C. Lidman and H. Lin and N. MacCrann and M. Manera and T. Manning and J. L. Marshall and S. Mau and J. McCullough and J. Mena-Fern{\'a}ndez and F. Menanteau and R. Miquel and J. J. Mohr and J. Muir and J. Myles and R. C. Nichol and B. Nord and J. H. O'Donnell and R. L. C. Ogando and A. Palmese and M. Paterno and J. Peoples and W. J. Percival and D. Petravick and A. Pieres and A. A. Plazas Malag{\'o}n and A. Porredon and A. Pourtsidou and J. Prat and C. Preston and M. Raveri and W. Riquelme and M. Rodriguez-Monroy and P. Rogozenski and A. K. Romer and A. Roodman and R. Rosenfeld and A. J. Ross and E. Rozo and E. S. Rykoff and S. Samuroff and C. S{\'a}nchez and E. Sanchez and D. Sanchez Cid and T. Schutt and I. Sevilla-Noarbe and E. Sheldon and N. Sherman and T. Shin and M. Smith and M. Soares-Santos and E. Suchyta and M. E. C. Swanson and M. Tabbutt and G. Tarle and D. Thomas and C. To and A. Tong and L. Toribio San Cipriano and M. A. Troxel and M. Tsedrik and D. L. Tucker and V. Vikram and A. R. Walker and N. Weaverdyck and R. H. Wechsler and D. H. Weinberg and J. Weller and V. Wetzell and A. Whyley and R. D. Wilkinson and P. Wiseman and H. -Y. Wu and M. Yamamoto and B. Yanny and B. Yin and G. Zacharegkas and Y. Zhang and J. Zuntz},
      year={2026},
      eprint={2601.14559},
      archivePrefix={arXiv},
      primaryClass={astro-ph.CO},
      url={https://arxiv.org/abs/2601.14559}, 
}

@article{Amendola_2018,
   title={Cosmology and fundamental physics with the Euclid satellite},
   volume={21},
   ISSN={1433-8351},
   url={http://dx.doi.org/10.1007/s41114-017-0010-3},
   DOI={10.1007/s41114-017-0010-3},
   number={1},
   journal={Living Reviews in Relativity},
   publisher={Springer Science and Business Media LLC},
   author={Amendola, Luca and Appleby, Stephen and Avgoustidis, Anastasios and Bacon, David and Baker, Tessa and Baldi, Marco and Bartolo, Nicola and Blanchard, Alain and Bonvin, Camille and Borgani, Stefano and Branchini, Enzo and Burrage, Clare and Camera, Stefano and Carbone, Carmelita and Casarini, Luciano and Cropper, Mark and de Rham, Claudia and Dietrich, J{\"o}rg P. and Di Porto, Cinzia and Durrer, Ruth and Ealet, Anne and Ferreira, Pedro G. and Finelli, Fabio and Garc{\'i}a-Bellido, Juan and Giannantonio, Tommaso and Guzzo, Luigi and Heavens, Alan and Heisenberg, Lavinia and Heymans, Catherine and Hoekstra, Henk and Hollenstein, Lukas and Holmes, Rory and Hwang, Zhiqi and Jahnke, Knud and Kitching, Thomas D. and Koivisto, Tomi and Kunz, Martin and La Vacca, Giuseppe and Linder, Eric and March, Marisa and Marra, Valerio and Martins, Carlos and Majerotto, Elisabetta and Markovic, Dida and Marsh, David and Marulli, Federico and Massey, Richard and Mellier, Yannick and Montanari, Francesco and Mota, David F. and Nunes, Nelson J. and Percival, Will and Pettorino, Valeria and Porciani, Cristiano and Quercellini, Claudia and Read, Justin and Rinaldi, Massimiliano and Sapone, Domenico and Sawicki, Ignacy and Scaramella, Roberto and Skordis, Constantinos and Simpson, Fergus and Taylor, Andy and Thomas, Shaun and Trotta, Roberto and Verde, Licia and Vernizzi, Filippo and Vollmer, Adrian and Wang, Yun and Weller, Jochen and Zlosnik, Tom},
   year={2018},
   month=Apr }

@misc{Abell_2009,
      title={LSST Science Book, Version 2.0}, 
      author={{LSST Science Collaboration: Abell} and Julius Allison and Scott F. Anderson and John R. Andrew and J. Roger P. Angel and Lee Armus and David Arnett and S. J. Asztalos and Tim S. Axelrod and Stephen Bailey and D. R. Ballantyne and Justin R. Bankert and Wayne A. Barkhouse and Jeffrey D. Barr and L. Felipe Barrientos and Aaron J. Barth and James G. Bartlett and Andrew C. Becker and Jacek Becla and Timothy C. Beers and Joseph P. Bernstein and Rahul Biswas and Michael R. Blanton and Joshua S. Bloom and John J. Bochanski and Pat Boeshaar and Kirk D. Borne and Marusa Bradac and W. N. Brandt and Carrie R. Bridge and Michael E. Brown and Robert J. Brunner and James S. Bullock and Adam J. Burgasser and James H. Burge and David L. Burke and Phillip A. Cargile and Srinivasan Chandrasekharan and George Chartas and Steven R. Chesley and You-Hua Chu and David Cinabro and Mark W. Claire and Charles F. Claver and Douglas Clowe and A. J. Connolly and Kem H. Cook and Jeff Cooke and Asantha Cooray and Kevin R. Covey and Christopher S. Culliton and Roelof de Jong and Willem H. de Vries and Victor P. Debattista and Francisco Delgado and Ian P. Dell'Antonio and Saurav Dhital and Rosanne Di Stefano and Mark Dickinson and Benjamin Dilday and S. G. Djorgovski and Gregory Dobler and Ciro Donalek and Gregory Dubois-Felsmann and Josef Durech and Ardis Eliasdottir and Michael Eracleous and Laurent Eyer and Emilio E. Falco and Xiaohui Fan and Christopher D. Fassnacht and Harry C. Ferguson and Yanga R. Fernandez and Brian D. Fields and Douglas Finkbeiner and Eduardo E. Figueroa and Derek B. Fox and Harold Francke and James S. Frank and Josh Frieman and Sebastien Fromenteau and Muhammad Furqan and Gaspar Galaz and A. Gal-Yam and Peter Garnavich and Eric Gawiser and John Geary and Perry Gee and Robert R. Gibson and Kirk Gilmore and Emily A. Grace and Richard F. Green and William J. Gressler and Carl J. Grillmair and Salman Habib and J. S. Haggerty and Mario Hamuy and Alan W. Harris and Suzanne L. Hawley and Alan F. Heavens and Leslie Hebb and Todd J. Henry and Edward Hileman and Eric J. Hilton and Keri Hoadley and J. B. Holberg and Matt J. Holman and Steve B. Howell and Leopoldo Infante and Zeljko Ivezic and Suzanne H. Jacoby and Bhuvnesh Jain and R and Jedicke and M. James Jee and J. Garrett Jernigan and Saurabh W. Jha and Kathryn V. Johnston and R. Lynne Jones and Mario Juric and Mikko Kaasalainen and Styliani and Kafka and Steven M. Kahn and Nathan A. Kaib and Jason Kalirai and Jeff Kantor and Mansi M. Kasliwal and Charles R. Keeton and Richard Kessler and Zoran Knezevic and Adam Kowalski and Victor L. Krabbendam and K. Simon Krughoff and Shrinivas Kulkarni and Stephen Kuhlman and Mark Lacy and Sebastien Lepine and Ming Liang and Amy Lien and Paulina Lira and Knox S. Long and Suzanne Lorenz and Jennifer M. Lotz and R. H. Lupton and Julie Lutz and Lucas M. Macri and Ashish A. Mahabal and Rachel Mandelbaum and Phil Marshall and Morgan May and Peregrine M. McGehee and Brian T. Meadows and Alan Meert and Andrea Milani and Christopher J. Miller and Michelle Miller and David Mills and Dante Minniti and David Monet and Anjum S. Mukadam and Ehud Nakar and Douglas R. Neill and Jeffrey A. Newman and Sergei Nikolaev and Martin Nordby and Paul O'Connor and Masamune Oguri and John Oliver and Scot S. Olivier and Julia K. Olsen and Knut Olsen and Edward W. Olszewski and Hakeem Oluseyi and Nelson D. Padilla and Alex Parker and Joshua Pepper and John R. Peterson and Catherine Petry and Philip A. Pinto and James L. Pizagno and Bogdan Popescu and Andrej Prsa and Veljko Radcka and M. Jordan Raddick and Andrew Rasmussen and Arne Rau and Jeonghee Rho and James E. Rhoads and Gordon T. Richards and Stephen T. Ridgway and Brant E. Robertson and Rok Roskar and Abhijit Saha and Ata Sarajedini and Evan Scannapieco and Terry Schalk and Rafe Schindler and Samuel Schmidt and Sarah Schmidt and Donald P. Schneider and German Schumacher and Ryan Scranton and Jacques Sebag and Lynn G. Seppala and Ohad Shemmer and Joshua D. Simon and M. Sivertz and Howard A. Smith and J. Allyn Smith and Nathan Smith and Anna H. Spitz and Adam Stanford and Keivan G. Stassun and Jay Strader and Michael A. Strauss and Christopher W. Stubbs and Donald W. Sweeney and Alex Szalay and Paula Szkody and Masahiro Takada and Paul Thorman and David E. Trilling and Virginia Trimble and Anthony Tyson and Richard Van Berg and Daniel Vanden Berk and Jake VanderPlas and Licia Verde and Bojan Vrsnak and Lucianne M. Walkowicz and Benjamin D. Wandelt and Sheng Wang and Yun Wang and Michael Warner and Risa H. Wechsler and Andrew A. West and Oliver Wiecha and Benjamin F. Williams and Beth Willman and David Wittman and Sidney C. Wolff and W. Michael Wood-Vasey and Przemek Wozniak and Patrick Young and Andrew Zentner and Hu Zhan},
      year={2009},
      eprint={0912.0201},
      archivePrefix={arXiv},
      primaryClass={astro-ph.IM},
      url={https://arxiv.org/abs/0912.0201}, 
}

@misc{akiba2019,
      title={Optuna: A Next-generation Hyperparameter Optimization Framework}, 
      author={Takuya Akiba and Shotaro Sano and Toshihiko Yanase and Takeru Ohta and Masanori Koyama},
      year={2019},
      eprint={1907.10902},
      archivePrefix={arXiv},
      primaryClass={cs.LG},
      url={https://arxiv.org/abs/1907.10902}, 
}

@article{Ajani2023,
   title={<i>Euclid</i>preparation: XXVIII. Forecasts for ten different higher-order weak lensing statistics},
   volume={675},
   ISSN={1432-0746},
   url={http://dx.doi.org/10.1051/0004-6361/202346017},
   DOI={10.1051/0004-6361/202346017},
   journal={Astronomy \& Astrophysics},
   publisher={EDP Sciences},
   author={{Euclid Collaboration: Ajani}, V. and Baldi, M. and Barthelemy, A. and Boyle, A. and Burger, P. and Cardone, V. F. and Cheng, S. and Codis, S. and Giocoli, C. and Harnois-D{\'e}raps, J. and Heydenreich, S. and Kansal, V. and Kilbinger, M. and Linke, L. and Llinares, C. and Martinet, N. and Parroni, C. and Peel, A. and Pires, S. and Porth, L. and Tereno, I. and Uhlemann, C. and Vicinanza, M. and Vinciguerra, S. and Aghanim, N. and Auricchio, N. and Bonino, D. and Branchini, E. and Brescia, M. and Brinchmann, J. and Camera, S. and Capobianco, V. and Carbone, C. and Carretero, J. and Castander, F. J. and Castellano, M. and Cavuoti, S. and Cimatti, A. and Cledassou, R. and Congedo, G. and Conselice, C. J. and Conversi, L. and Corcione, L. and Courbin, F. and Cropper, M. and Da Silva, A. and Degaudenzi, H. and Di Giorgio, A. M. and Dinis, J. and Douspis, M. and Dubath, F. and Dupac, X. and Farrens, S. and Ferriol, S. and Fosalba, P. and Frailis, M. and Franceschi, E. and Galeotta, S. and Garilli, B. and Gillis, B. and Grazian, A. and Grupp, F. and Hoekstra, H. and Holmes, W. and Hornstrup, A. and Hudelot, P. and Jahnke, K. and Jhabvala, M. and K{\"u}mmel, M. and Kitching, T. and Kunz, M. and Kurki-Suonio, H. and Lilje, P. B. and Lloro, I. and Maiorano, E. and Mansutti, O. and Marggraf, O. and Markovic, K. and Marulli, F. and Massey, R. and Mei, S. and Mellier, Y. and Meneghetti, M. and Moresco, M. and Moscardini, L. and Niemi, S.-M. and Nightingale, J. and Nutma, T. and Padilla, C. and Paltani, S. and Pedersen, K. and Pettorino, V. and Polenta, G. and Poncet, M. and Popa, L. A. and Raison, F. and Renzi, A. and Rhodes, J. and Riccio, G. and Romelli, E. and Roncarelli, M. and Rossetti, E. and Saglia, R. and Sapone, D. and Sartoris, B. and Schneider, P. and Schrabback, T. and Secroun, A. and Seidel, G. and Serrano, S. and Sirignano, C. and Stanco, L. and Starck, J.-L. and Tallada-Cresp{\'i}, P. and Taylor, A. N. and Toledo-Moreo, R. and Torradeflot, F. and Tutusaus, I. and Valentijn, E. A. and Valenziano, L. and Vassallo, T. and Wang, Y. and Weller, J. and Zamorani, G. and Zoubian, J. and Andreon, S. and Bardelli, S. and Boucaud, A. and Bozzo, E. and Colodro-Conde, C. and Di Ferdinando, D. and Fabbian, G. and Farina, M. and Graci{\'a}-Carpio, J. and Keih{\"a}nen, E. and Lindholm, V. and Maino, D. and Mauri, N. and Neissner, C. and Schirmer, M. and Scottez, V. and Zucca, E. and Akrami, Y. and Baccigalupi, C. and Balaguera-Antol{\'i}nez, A. and Ballardini, M. and Bernardeau, F. and Biviano, A. and Blanchard, A. and Borgani, S. and Borlaff, A. S. and Burigana, C. and Cabanac, R. and Cappi, A. and Carvalho, C. S. and Casas, S. and Castignani, G. and Castro, T. and Chambers, K. C. and Cooray, A. R. and Coupon, J. and Courtois, H. M. and Davini, S. and de la Torre, S. and De Lucia, G. and Desprez, G. and Dole, H. and Escartin, J. A. and Escoffier, S. and Ferrero, I. and Finelli, F. and Ganga, K. and Garcia-Bellido, J. and George, K. and Giacomini, F. and Gozaliasl, G. and Hildebrandt, H. and Jimenez Mu{\~n}oz, A. and Joachimi, B. and Kajava, J. J. E. and Kirkpatrick, C. C. and Legrand, L. and Loureiro, A. and Magliocchetti, M. and Maoli, R. and Marcin, S. and Martinelli, M. and Martins, C. J. A. P. and Matthew, S. and Maurin, L. and Metcalf, R. B. and Monaco, P. and Morgante, G. and Nadathur, S. and Nucita, A. A. and Popa, V. and Potter, D. and Pourtsidou, A. and P{\"o}ntinen, M. and Reimberg, P. and S{\'a}nchez, A. G. and Sakr, Z. and Schneider, A. and Sefusatti, E. and Sereno, M. and Shulevski, A. and Spurio Mancini, A. and Steinwagner, J. and Teyssier, R. and Valiviita, J. and Veropalumbo, A. and Viel, M. and Zinchenko, I. A.},
   year={2023},
   month=jul, pages={A120} }

@article{bishop1994mdn,
  author  = {Christopher M. Bishop},
  title   = {Mixture Density Networks},
  journal = {Technical Report NCRG/94/004},
  year    = {1994},
  institution = {Aston University}
}

@article{Czado_2009,
    author = {Czado, Claudia and Gneiting, Tilmann and Held, Leonhard},
    title = {Predictive Model Assessment for Count Data},
    journal = {Biometrics},
    volume = {65},
    number = {4},
    pages = {1254-1261},
    year = {2009},
    month = {12},
    issn = {0006-341X},
    doi = {10.1111/j.1541-0420.2009.01191.x},
    url = {https://doi.org/10.1111/j.1541-0420.2009.01191.x},
    eprint = {https://academic.oup.com/biometrics/article-pdf/65/4/1254/52742999/biometrics_65_4_1254.pdf},
}

@article{Charnock_2018,
   title={Automatic physical inference with information maximizing neural networks},
   volume={97},
   ISSN={2470-0029},
   url={http://dx.doi.org/10.1103/PhysRevD.97.083004},
   DOI={10.1103/physrevd.97.083004},
   number={8},
   journal={Physical Review D},
   publisher={American Physical Society (APS)},
   author={Charnock, Tom and Lavaux, Guilhem and Wandelt, Benjamin D.},
   year={2018},
   month=Apr }

@article{Cranmer_2020,
   title={The frontier of simulation-based inference},
   volume={117},
   ISSN={1091-6490},
   url={http://dx.doi.org/10.1073/pnas.1912789117},
   DOI={10.1073/pnas.1912789117},
   number={48},
   journal={Proceedings of the National Academy of Sciences},
   publisher={National Academy of Sciences},
   author={Cranmer, Kyle and Brehmer, Johann and Louppe, Gilles},
   year={2020},
   month=May, pages={30055--30062} }

@article{Carron_2013,
   title={On the assumption of Gaussianity for cosmological two-point statistics and parameter dependent covariance matrices},
   volume={551},
   ISSN={1432-0746},
   url={http://dx.doi.org/10.1051/0004-6361/201220538},
   DOI={10.1051/0004-6361/201220538},
   journal={Astronomy \& Astrophysics},
   publisher={EDP Sciences},
   author={Carron, J.},
   year={2013},
   month=Feb, pages={A88} }

@misc{Deraps_2024,
      title={KiDS-1000 and DES-Y1 combined: Cosmology from peak count statistics}, 
      author={Joachim Harnois-Deraps and Sven Heydenreich and Benjamin Giblin and Nicolas Martinet and Tilman Troester and Marika Asgari and Pierre Burger and Tiago Castro and Klaus Dolag and Catherine Heymans and Hendrik Hildebrandt and Benjamin Joachimi and Angus H. Wright},
      year={2024},
      eprint={2405.10312},
      archivePrefix={arXiv},
      primaryClass={astro-ph.CO},
      url={https://arxiv.org/abs/2405.10312}, 
}

@misc{durkan2019neuralsplineflows, title={Neural Spline Flows}, author={Conor Durkan and Artur Bekasov and Iain Murray and George Papamakarios}, year={2019}, eprint={1906.04032}, archivePrefix={arXiv}, primaryClass={stat.ML}, url={https://arxiv.org/abs/1906.04032}, }

@article{Deraps_2019,
   title={Cosmic shear covariance matrix in <i>w</i>CDM: Cosmology matters},
   volume={631},
   ISSN={1432-0746},
   url={http://dx.doi.org/10.1051/0004-6361/201935912},
   DOI={10.1051/0004-6361/201935912},
   journal={Astronomy \& Astrophysics},
   publisher={EDP Sciences},
   author={Harnois-D{\'e}raps, J. and Giblin, B. and Joachimi, B.},
   year={2019},
   month=Nov, pages={A160} }

@article{Eifler_2009,
   title={Dependence of cosmic shear covariances on cosmology: Impact on parameter estimation},
   volume={502},
   ISSN={1432-0746},
   url={http://dx.doi.org/10.1051/0004-6361/200811276},
   DOI={10.1051/0004-6361/200811276},
   number={3},
   journal={Astronomy \& Astrophysics},
   publisher={EDP Sciences},
   author={Eifler, T. and Schneider, P. and Hartlap, J.},
   year={2009},
   month=jun, pages={721--731} }

@article{Fu_2008,
  author  = {Fu, L. and Semboloni, E. and Hoekstra, H. and Kilbinger, M. and van Waerbeke, L. and Tereno, I. and Mellier, Y. and Heymans, C. and Coupon, J. and Benabed, K. and Benjamin, J. and Bertin, E. and Dor{\'e}, O. and Hudson, M. J. and Ilbert, O. and Maoli, R. and Marmo, C. and McCracken, H. J. and M{\'e}nard, B.},
  title   = {Very weak lensing in the CFHTLS wide: cosmology from cosmic shear in the linear regime},
  journal = {Astronomy \& Astrophysics},
  volume  = {479},
  number  = {1},
  pages   = {9--25},
  year    = {2008},
  doi     = {10.1051/0004-6361:20078522}
}

@article{Fluri_2022,
   title={Full 
<mml:math xmlns:mml=``http://www.w3.org/1998/Math/MathML'' display=``inline''><mml:mrow><mml:mi>w</mml:mi><mml:mi>CDM</mml:mi></mml:mrow></mml:math>
 analysis of KiDS-1000 weak lensing maps using deep learning},
   volume={105},
   ISSN={2470-0029},
   url={http://dx.doi.org/10.1103/PhysRevD.105.083518},
   DOI={10.1103/physrevd.105.083518},
   number={8},
   journal={Physical Review D},
   publisher={American Physical Society (APS)},
   author={Fluri, Janis and Kacprzak, Tomasz and Lucchi, Aurelien and Schneider, Aurel and Refregier, Alexandre and Hofmann, Thomas},
   year={2022},
   month=Apr }

@ARTICLE{ForemanMackey2013,
       author = {{Foreman-Mackey}, Daniel and {Hogg}, David W. and {Lang}, Dustin and {Goodman}, Jonathan},
        title = "{emcee: The MCMC Hammer}",
      journal = {\pasp},
         year = 2013,
        month = mar,
       volume = {125},
       number = {925},
        pages = {306},
          doi = {10.1086/670067},
archivePrefix = {arXiv},
       eprint = {1202.3665},
 primaryClass = {astro-ph.IM},
       adsurl = {https://ui.adsabs.harvard.edu/abs/2013PASP..125..306F}
}

@misc{Gatti_2022,
      title={Dark Energy Survey Year 3 results: cosmology with moments of weak lensing mass maps}, 
      author={M. Gatti and B. Jain and C. Chang and M. Raveri and D. Z{\"u}rcher and L. Secco and L. Whiteway and N. Jeffrey and C. Doux and T. Kacprzak and D. Bacon and P. Fosalba and A. Alarcon and A. Amon and K. Bechtol and M. Becker and G. Bernstein and J. Blazek and A. Campos and A. Choi and C. Davis and J. Derose and S. Dodelson and F. Elsner and J. Elvin-Poole and S. Everett and A. Ferte and D. Gruen and I. Harrison and D. Huterer and M. Jarvis and E. Krause and P. F. Leget and P. Lemos and N. Maccrann and J. Mccullough and J. Muir and J. Myles and A. Navarro and S. Pandey and J. Prat and R. P. Rollins and A. Roodman and C. Sanchez and E. Sheldon and T. Shin and M. Troxel and I. Tutusaus and B. Yin and M. Aguena and S. Allam and F. Andrade-Oliveira and J. Anni and E. Bertin and D. Brooks and D. L. Burke and A. Carnero Rosell and M. Carrasco Kind and J. Carretero and R. Cawthon and M. Costanzi and L. N. da Costa and M. E. S. Pereira and J. De Vicente and S. Desai and H. T. Diehl and J. P. Dietrich and P. Doel and A. Drlica-Wagner and K. Eckert and A. E. Evrard and I. Ferrero and J. Garc{\'i}a-Bellido and E. Gaztanaga and T. Giannantonio and R. A. Gruendl and J. Gschwend and G. Gutierrez and S. R. Hinton and D. L. Hollowood and K. Honscheid and D. J. James and K. Kuehn and N. Kuropatkin and O. Laha and C. Lidman and M. A. G. Maia and J. L. Marshall and P. Melchior and F. Menanteau and R. Miquel and R. Morgan and A. Palmese and F. Paz-Chinch{\'o}n and A. Pieres and A. A. Plazas Malag{\'o}n and K. Reil and M. Rodriguez-Monroyv and A. K. Romer and E. Sanchez and M. Schubnell and S. Serrano and I. Sevilla-Noarbe and M. Smith and M. Soares-Santos and E. Suchyta and G. Tarle and D. Thomas and C. To and T. N. Varga},
      year={2022},
      eprint={2110.10141},
      archivePrefix={arXiv},
      primaryClass={astro-ph.CO},
      url={https://arxiv.org/abs/2110.10141}, 
}

@misc{Gatti_2023,
      title={Detection of the significant impact of source clustering on higher-order statistics with DES Year 3 weak gravitational lensing data}, 
      author={M. Gatti and N. Jeffrey and L. Whiteway and V. Ajani and T. Kacprzak and D. Z{\"u}rcher and C. Chang and B. Jain and J. Blazek and E. Krause and A. Alarcon and A. Amon and K. Bechtol and M. Becker and G. Bernstein and A. Campos and R. Chen and A. Choi and C. Davis and J. Derose and H. T. Diehl and S. Dodelson and C. Doux and K. Eckert and J. Elvin-Poole and S. Everett and A. Ferte and D. Gruen and R. Gruendl and I. Harrison and W. G. Hartley and K. Herner and E. M. Huff and M. Jarvis and N. Kuropatkin and P. F. Leget and N. MacCrann and J. McCullough and J. Myles and A. Navarro-Alsina and S. Pandey and J. Prat and M. Raveri and R. P. Rollins and A. Roodman and C. Sanchez and L. F. Secco and I. Sevilla-Noarbe and E. Sheldon and T. Shin and M. Troxel and I. Tutusaus and T. N. Varga and B. Yanny and B. Yin and Y. Zhang and J. Zuntz and S. S. Allam and O. Alves and M. Aguena and D. Bacon and E. Bertin and D. Brooks and D. L. Burke and A. Carnero Rosell and J. Carretero and R. Cawthon and L. N. da Costa and T. M. Davis and J. De Vicente and S. Desai and P. Doel and J. Garc{\'i}a-Bellido and G. Giannini and G. Gutierrez and I. Ferrero and J. Frieman and S. R. Hinton and D. L. Hollowood and K. Honscheid and D. J. James and K. Kuehn and O. Lahav and J. L. Marshall and J. Mena-Fern{\'a}ndez and R. Miquel and R. L. C. Ogando and A. Palmese and M. E. S. Pereira and A. A. Plazas Malag{\'o}n and M. Rodriguez-Monroy and S. Samuroff and E. Sanchez and M. Schubnell and M. Smith and F. Sobreira and E. Suchyta and M. E. C. Swanson and G. Tarle and N. Weaverdyck and P. Wiseman},
      year={2023},
      eprint={2307.13860},
      archivePrefix={arXiv},
      primaryClass={astro-ph.CO},
      url={https://arxiv.org/abs/2307.13860}, 
}

@article{Gomes_2025,
   title={Dark Energy Survey Year 3 Results: Cosmological constraints from second- and third-order shear statistics},
   volume={112},
   ISSN={2470-0029},
   url={http://dx.doi.org/10.1103/sxlz-t9gb},
   DOI={10.1103/sxlz-t9gb},
   number={12},
   journal={Physical Review D},
   publisher={American Physical Society (APS)},
   author={Gomes, R. C. H. and Sugiyama, S. and Jain, B. and Jarvis, M. and Anbajagane, D. and Halder, A. and Marques, G. A. and Pandey, S. and Marshall, J. and Alarcon, A. and Amon, A. and Bechtol, K. and Becker, M. and Bernstein, G. and Campos, A. and Cawthon, R. and Chang, C. and Chen, R. and Choi, A. and Cordero, J. and Davis, C. and Derose, J. and Dodelson, S. and Doux, C. and Eckert, K. and Elsner, F. and Elvin-Poole, J. and Everett, S. and Fert{\'e}, A. and Gatti, M. and Giannini, G. and Gruen, D. and Harrison, I. and Herner, K. and Huff, E. M. and Huterer, D. and Kuropatkin, N. and Leget, P. F. and Maccrann, N. and Mccullough, J. and Muir, J. and Myles, J. and Navarro Alsina, A. and Prat, J. and Raveri, M. and Rollins, R. P. and Roodman, A. and Ross, A. J. and Rykoff, E. S. and S{\'a}nchez, C. and Secco, L. F. and Sheldon, E. and Shin, T. and Troxel, M. and Tutusaus, I. and Varga, T. N. and Yanny, B. and Yin, B. and Zhang, Y. and Zuntz, J. and Aguena, M. and Andrade-Oliveira, F. and Bacon, D. and Blazek, J. and Bocquet, S. and Brooks, D. and Carnero Rosell, A. and Carretero, J. and Costanzi, M. and da Costa, L. and da Silva Pereira, M. E. and Davis, T. M. and De Vicente, J. and Diehl, H. T. and Flaugher, B. and Frieman, J. and Gutierrez, G. and Hinton, S. R. and Hollowood, D. L. and Honscheid, K. and James, D. J. and Jeffrey, N. and Lee, S. and Mena-Fern{\'a}ndez, J. and Miquel, R. and Ogando, R. L. C. and Plazas Malag{\'o}n, A. A. and Porredon, A. and Sanchez, E. and Sanchez Cid, D. and Samuroff, S. and Smith, M. and Suchyta, E. and Swanson, M. E. C. and Thomas, D. and Vikram, V. and Weller, J. and Yamamoto, M. and },
   year={2025},
   month=Dec }

@article{Gelman_Rubin_1992,
author = {Andrew Gelman and Donald B. Rubin},
title = {{Inference from Iterative Simulation Using Multiple Sequences}},
volume = {7},
journal = {Statistical Science},
number = {4},
publisher = {Institute of Mathematical Statistics},
pages = {457 -- 472},
year = {1992},
doi = {10.1214/ss/1177011136},
URL = {https://doi.org/10.1214/ss/1177011136}
}

@book{Goodfellow_2016,
    title={Deep Learning},
    author={Ian Goodfellow and Yoshua Bengio and Aaron Courville},
    publisher={MIT Press},
    note={\url{http://www.deeplearningbook.org}},
    year={2016}
}

@ARTICLE{Gorski2005,
       author = {{G{\'o}rski}, K.~M. and {Hivon}, E. and {Banday}, A.~J. and {Wandelt}, B.~D. and {Hansen}, F.~K. and {Reinecke}, M. and {Bartelmann}, M.},
        title = "{HEALPix: A Framework for High-Resolution Discretization and Fast Analysis of Data Distributed on the Sphere}",
      journal = {\apj},
         year = 2005,
        month = apr,
       volume = {622},
       number = {2},
        pages = {759-771},
          doi = {10.1086/427976},
archivePrefix = {arXiv},
       eprint = {astro-ph/0409513},
 primaryClass = {astro-ph},
       adsurl = {https://ui.adsabs.harvard.edu/abs/2005ApJ...622..759G}
}

@article{Heavens_2017,
   title={Massive data compression for parameter-dependent covariance matrices},
   volume={472},
   ISSN={1365-2966},
   url={http://dx.doi.org/10.1093/mnras/stx2326},
   DOI={10.1093/mnras/stx2326},
   number={4},
   journal={Monthly Notices of the Royal Astronomical Society},
   publisher={Oxford University Press (OUP)},
   author={Heavens, Alan F. and Sellentin, Elena and de Mijolla, Damien and Vianello, Alvise},
   year={2017},
   month=sep, pages={4244--4250} }

@article{Hartlap_2006,
   title={Why your model parameter confidences might be too optimistic.  Unbiased estimation of the inverse covariance matrix},
   volume={464},
   ISSN={1432-0746},
   url={http://dx.doi.org/10.1051/0004-6361:20066170},
   DOI={10.1051/0004-6361:20066170},
   number={1},
   journal={Astronomy \& Astrophysics},
   publisher={EDP Sciences},
   author={Hartlap, J. and Simon, P. and Schneider, P.},
   year={2006},
   month=Dec, pages={399--404} }

@InProceedings{He_2016,
author = {He, Kaiming and Zhang, Xiangyu and Ren, Shaoqing and Sun, Jian},
title = {Deep Residual Learning for Image Recognition},
booktitle = {Proceedings of the IEEE Conference on Computer Vision and Pattern Recognition (CVPR)},
month = jun,
year = {2016}
}

@article{Heavens_2000,
   title={Massive lossless data compression and multiple parameter estimation from galaxy spectra},
   volume={317},
   ISSN={1365-2966},
   url={http://dx.doi.org/10.1046/j.1365-8711.2000.03692.x},
   DOI={10.1046/j.1365-8711.2000.03692.x},
   number={4},
   journal={Monthly Notices of the Royal Astronomical Society},
   publisher={Oxford University Press (OUP)},
   author={Heavens, A. F. and Jimenez, R. and Lahav, O.},
   year={2000},
   month=Oct, pages={965--972} }

@article{Ho_2024,
   title={LtU-ILI: An All-in-One Framework for Implicit Inference in Astrophysics and Cosmology},
   volume={7},
   ISSN={2565-6120},
   url={http://dx.doi.org/10.33232/001c.120559},
   DOI={10.33232/001c.120559},
   journal={The Open Journal of Astrophysics},
   publisher={Maynooth University},
   author={Ho, Matthew and Bartlett, Deaglan J. and Chartier, Nicolas and Cuesta-Lazaro, Carolina and Ding, Simon and Lapel, Axel and Lemos, Pablo and Lovell, Christopher C. and Makinen, T. Lucas and Modi, Chirag and Pandya, Viraj and Pandey, Shivam and Perez, Lucia A. and Wandelt, Benjamin and Bryan, Greg L.},
   year={2024},
   month=jul }

@misc{Jeffrey_2024,
      title={Dark Energy Survey Year 3 results: likelihood-free, simulation-based $w$CDM inference with neural compression of weak-lensing map statistics}, 
      author={N. Jeffrey and L. Whiteway and M. Gatti and J. Williamson and J. Alsing and A. Porredon and J. Prat and C. Doux and B. Jain and C. Chang and T. -Y. Cheng and T. Kacprzak and P. Lemos and A. Alarcon and A. Amon and K. Bechtol and M. R. Becker and G. M. Bernstein and A. Campos and A. Carnero Rosell and R. Chen and A. Choi and J. DeRose and A. Drlica-Wagner and K. Eckert and S. Everett and A. Fert{\'e} and D. Gruen and R. A. Gruendl and K. Herner and M. Jarvis and J. McCullough and J. Myles and A. Navarro-Alsina and S. Pandey and M. Raveri and R. P. Rollins and E. S. Rykoff and C. S{\'a}nchez and L. F. Secco and I. Sevilla-Noarbe and E. Sheldon and T. Shin and M. A. Troxel and I. Tutusaus and T. N. Varga and B. Yanny and B. Yin and J. Zuntz and M. Aguena and S. S. Allam and O. Alves and D. Bacon and S. Bocquet and D. Brooks and L. N. da Costa and T. M. Davis and J. De Vicente and S. Desai and H. T. Diehl and I. Ferrero and J. Frieman and J. Garc{\'i}a-Bellido and E. Gaztanaga and G. Giannini and G. Gutierrez and S. R. Hinton and D. L. Hollowood and K. Honscheid and D. Huterer and D. J. James and O. Lahav and S. Lee and J. L. Marshall and J. Mena-Fern{\'a}ndez and R. Miquel and A. Pieres and A. A. Plazas Malag{\'o}n and A. Roodman and M. Sako and E. Sanchez and D. Sanchez Cid and M. Smith and E. Suchyta and M. E. C. Swanson and G. Tarle and D. L. Tucker and N. Weaverdyck and J. Weller and P. Wiseman and M. Yamamoto},
      year={2024},
      eprint={2403.02314},
      archivePrefix={arXiv},
      primaryClass={astro-ph.CO},
      url={https://arxiv.org/abs/2403.02314}, 
}

@misc{jarvis2015,
       author = {{Jarvis}, Mike},
        title = "{TreeCorr: Two-point correlation functions}",
 howpublished = {Astrophysics Source Code Library, record ascl:1508.007},
         year = 2015,
        month = aug,
          eid = {ascl:1508.007},
archivePrefix = {ascl},
       eprint = {1508.007},
       adsurl = {https://ui.adsabs.harvard.edu/abs/2015ascl.soft08007J}
}

@article{Kilbinger_2015,
   title={Cosmology with cosmic shear observations: a review},
   volume={78},
   ISSN={1361-6633},
   url={http://dx.doi.org/10.1088/0034-4885/78/8/086901},
   DOI={10.1088/0034-4885/78/8/086901},
   number={8},
   journal={Reports on Progress in Physics},
   publisher={IOP Publishing},
   author={Kilbinger, Martin},
   year={2015},
   month=jul, pages={086901} }

@ARTICLE{Kaiser_1993,
       author = {{Kaiser}, Nick and {Squires}, Gordon},
        title = "{Mapping the Dark Matter with Weak Gravitational Lensing}",
      journal = {\apj},
         year = 1993,
        month = feb,
       volume = {404},
        pages = {441},
          doi = {10.1086/172297},
       adsurl = {https://ui.adsabs.harvard.edu/abs/1993ApJ...404..441K}
}

@ARTICLE{Lecun_1998,
  author={Lecun, Y. and Bottou, L. and Bengio, Y. and Haffner, P.},
  journal={Proceedings of the IEEE}, 
  title={Gradient-based learning applied to document recognition}, 
  year={1998},
  volume={86},
  number={11},
  pages={2278-2324},
  doi={10.1109/5.726791}}

@misc{Lemos_2023,
      title={Sampling-Based Accuracy Testing of Posterior Estimators for General Inference}, 
      author={Pablo Lemos and Adam Coogan and Yashar Hezaveh and Laurence Perreault-Levasseur},
      year={2023},
      eprint={2302.03026},
      archivePrefix={arXiv},
      primaryClass={stat.ML},
      url={https://arxiv.org/abs/2302.03026}, 
}

@misc{Li_2023,
      title={Hyper Suprime-Cam Year 3 Results: Cosmology from Cosmic Shear Two-point Correlation Functions}, 
      author={Xiangchong Li and Tianqing Zhang and Sunao Sugiyama and Roohi Dalal and Ryo Terasawa and Markus M. Rau and Rachel Mandelbaum and Masahiro Takada and Surhud More and Michael A. Strauss and Hironao Miyatake and Masato Shirasaki and Takashi Hamana and Masamune Oguri and Wentao Luo and Atsushi J. Nishizawa and Ryuichi Takahashi and Andrina Nicola and Ken Osato and Arun Kannawadi and Tomomi Sunayama and Robert Armstrong and James Bosch and Yutaka Komiyama and Robert H. Lupton and Nate B. Lust and Lauren A. MacArthur and Satoshi Miyazaki and Hitoshi Murayama and Takahiro Nishimichi and Yuki Okura and Paul A. Price and Philip J. Tait and Masayuki Tanaka and Shiang-Yu Wang},
      year={2023},
      eprint={2304.00702},
      archivePrefix={arXiv},
      primaryClass={astro-ph.CO},
      url={https://arxiv.org/abs/2304.00702}, 
}

@article{Linder_2003,
   title={Exploring the Expansion History of the Universe},
   volume={90},
   ISSN={1079-7114},
   url={http://dx.doi.org/10.1103/PhysRevLett.90.091301},
   DOI={10.1103/physrevlett.90.091301},
   number={9},
   journal={Physical Review Letters},
   publisher={American Physical Society (APS)},
   author={Linder, Eric V.},
   year={2003},
   month=Mar }

@misc{Laureijs_2011,
      title={Euclid Definition Study Report}, 
      author={R. Laureijs and J. Amiaux and S. Arduini and J. -L. Augu{\`e}res and J. Brinchmann and R. Cole and M. Cropper and C. Dabin and L. Duvet and A. Ealet and B. Garilli and P. Gondoin and L. Guzzo and J. Hoar and H. Hoekstra and R. Holmes and T. Kitching and T. Maciaszek and Y. Mellier and F. Pasian and W. Percival and J. Rhodes and G. Saavedra Criado and M. Sauvage and R. Scaramella and L. Valenziano and S. Warren and R. Bender and F. Castander and A. Cimatti and O. Le F{\`e}vre and H. Kurki-Suonio and M. Levi and P. Lilje and G. Meylan and R. Nichol and K. Pedersen and V. Popa and R. Rebolo Lopez and H. -W. Rix and H. Rottgering and W. Zeilinger and F. Grupp and P. Hudelot and R. Massey and M. Meneghetti and L. Miller and S. Paltani and S. Paulin-Henriksson and S. Pires and C. Saxton and T. Schrabback and G. Seidel and J. Walsh and N. Aghanim and L. Amendola and J. Bartlett and C. Baccigalupi and J. -P. Beaulieu and K. Benabed and J. -G. Cuby and D. Elbaz and P. Fosalba and G. Gavazzi and A. Helmi and I. Hook and M. Irwin and J. -P. Kneib and M. Kunz and F. Mannucci and L. Moscardini and C. Tao and R. Teyssier and J. Weller and G. Zamorani and M. R. Zapatero Osorio and O. Boulade and J. J. Foumond and A. Di Giorgio and P. Guttridge and A. James and M. Kemp and J. Martignac and A. Spencer and D. Walton and T. Bl{\"u}mchen and C. Bonoli and F. Bortoletto and C. Cerna and L. Corcione and C. Fabron and K. Jahnke and S. Ligori and F. Madrid and L. Martin and G. Morgante and T. Pamplona and E. Prieto and M. Riva and R. Toledo and M. Trifoglio and F. Zerbi and F. Abdalla and M. Douspis and C. Grenet and S. Borgani and R. Bouwens and F. Courbin and J. -M. Delouis and P. Dubath and A. Fontana and M. Frailis and A. Grazian and J. Koppenh{\"o}fer and O. Mansutti and M. Melchior and M. Mignoli and J. Mohr and C. Neissner and K. Noddle and M. Poncet and M. Scodeggio and S. Serrano and N. Shane and J. -L. Starck and C. Surace and A. Taylor and G. Verdoes-Kleijn and C. Vuerli and O. R. Williams and A. Zacchei and B. Altieri and I. Escudero Sanz and R. Kohley and T. Oosterbroek and P. Astier and D. Bacon and S. Bardelli and C. Baugh and F. Bellagamba and C. Benoist and D. Bianchi and A. Biviano and E. Branchini and C. Carbone and V. Cardone and D. Clements and S. Colombi and C. Conselice and G. Cresci and N. Deacon and J. Dunlop and C. Fedeli and F. Fontanot and P. Franzetti and C. Giocoli and J. Garcia-Bellido and J. Gow and A. Heavens and P. Hewett and C. Heymans and A. Holland and Z. Huang and O. Ilbert and B. Joachimi and E. Jennins and E. Kerins and A. Kiessling and D. Kirk and R. Kotak and O. Krause and O. Lahav and F. van Leeuwen and J. Lesgourgues and M. Lombardi and M. Magliocchetti and K. Maguire and E. Majerotto and R. Maoli and F. Marulli and S. Maurogordato and H. McCracken and R. McLure and A. Melchiorri and A. Merson and M. Moresco and M. Nonino and P. Norberg and J. Peacock and R. Pello and M. Penny and V. Pettorino and C. Di Porto and L. Pozzetti and C. Quercellini and M. Radovich and A. Rassat and N. Roche and S. Ronayette and E. Rossetti and B. Sartoris and P. Schneider and E. Semboloni and S. Serjeant and F. Simpson and C. Skordis and G. Smadja and S. Smartt and P. Spano and S. Spiro and M. Sullivan and A. Tilquin and R. Trotta and L. Verde and Y. Wang and G. Williger and G. Zhao and J. Zoubian and E. Zucca},
      year={2011},
      eprint={1110.3193},
      archivePrefix={arXiv},
      primaryClass={astro-ph.CO},
      url={https://arxiv.org/abs/1110.3193}, 
}

@article{LeCunEtAl1998,
  author       = {LeCun, Yann and Bottou, L{\'e}on and Bengio, Yoshua
                  and Haffner, Patrick},
  title        = {{Gradient-based learning applied to document recognition}},
  journal      = {Proceedings of the IEEE},
  year         = {1998},
  volume       = {86},
  pages        = {2278--2324},
  doi          = {10.1109/5.726791},
}

@article{Lewis_2000,
   title={Efficient Computation of Cosmic Microwave Background Anisotropies in Closed Friedmann-Robertson-Walker Models},
   volume={538},
   ISSN={1538-4357},
   url={http://dx.doi.org/10.1086/309179},
   DOI={10.1086/309179},
   number={2},
   journal={The Astrophysical Journal},
   publisher={American Astronomical Society},
   author={Lewis, Antony and Challinor, Anthony and Lasenby, Anthony},
   year={2000},
   month=Aug, pages={473--476} }

@misc{Loshchilov_2019,
      title={Decoupled Weight Decay Regularization}, 
      author={Ilya Loshchilov and Frank Hutter},
      year={2019},
      eprint={1711.05101},
      archivePrefix={arXiv},
      primaryClass={cs.LG},
      url={https://arxiv.org/abs/1711.05101}, 
}

@ARTICLE{Martinet_2018,
       author = {{Martinet}, Nicolas and {Schneider}, Peter and {Hildebrandt}, Hendrik and {Shan}, HuanYuan and {Asgari}, Marika and {Dietrich}, J{\"o}rg P. and {Harnois-D{\'e}raps}, Joachim and {Erben}, Thomas and {Grado}, Aniello and {Heymans}, Catherine and {Hoekstra}, Henk and {Klaes}, Dominik and {Kuijken}, Konrad and {Merten}, Julian and {Nakajima}, Reiko},
        title = "{KiDS-450: cosmological constraints from weak-lensing peak statistics - II: Inference from shear peaks using N-body simulations}",
      journal = {\mnras},
         year = 2018,
        month = feb,
       volume = {474},
       number = {1},
        pages = {712-730},
          doi = {10.1093/mnras/stx2793},
archivePrefix = {arXiv},
       eprint = {1709.07678},
 primaryClass = {astro-ph.CO},
       adsurl = {https://ui.adsabs.harvard.edu/abs/2018MNRAS.474..712M}
}

@article{Massey_2007b,
    author = {Massey, Richard and Heymans, Catherine and Berg{\'e}, Joel and Bernstein, Gary and Bridle, Sarah and Clowe, Douglas and Dahle, H{\aa}kon and Ellis, Richard and Erben, Thomas and Hetterscheidt, Marco and High, F. William and Hirata, Christopher and Hoekstra, Henk and Hudelot, Patrick and Jarvis, Mike and Johnston, David and Kuijken, Konrad and Margoniner, Vera and Mandelbaum, Rachel and Mellier, Yannick and Nakajima, Reiko and Paulin-Henriksson, Stephane and Peeples, Molly and Roat, Chris and Refregier, Alexandre and Rhodes, Jason and Schrabback, Tim and Schirmer, Mischa and Seljak, Uro{\v s} and Semboloni, Elisabetta and Van Waerbeke, Ludovic},
    title = {The Shear Testing Programme 2: Factors affecting high-precision weak-lensing analyses},
    journal = {Monthly Notices of the Royal Astronomical Society},
    volume = {376},
    number = {1},
    pages = {13-38},
    year = {2007},
    month = {03},
    issn = {0035-8711},
    doi = {10.1111/j.1365-2966.2006.11315.x},
    url = {https://doi.org/10.1111/j.1365-2966.2006.11315.x},
    eprint = {https://academic.oup.com/mnras/article-pdf/376/1/13/2947714/mnras0376-0013.pdf},
}

@misc{Makinen_2025,
      title={Hybrid Summary Statistics}, 
      author={T. Lucas Makinen and Ce Sui and Benjamin D. Wandelt and Natalia Porqueres and Alan Heavens},
      year={2025},
      eprint={2410.07548},
      archivePrefix={arXiv},
      primaryClass={stat.ML},
      url={https://arxiv.org/abs/2410.07548}, 
}

@article{Mellier_2025,
   title={<i>Euclid</i>: I. Overview of the <i>Euclid</i> mission},
   volume={697},
   ISSN={1432-0746},
   url={http://dx.doi.org/10.1051/0004-6361/202450810},
   DOI={10.1051/0004-6361/202450810},
   journal={Astronomy \& Astrophysics},
   publisher={EDP Sciences},
   author={{Euclid Collaboration: Mellier}, Y. and Abdurro'uf and Acevedo Barroso, J. A. and Ach{\'u}carro, A. and Adamek, J. and Adam, R. and Addison, G. E. and Aghanim, N. and Aguena, M. and Ajani, V. and Akrami, Y. and Al-Bahlawan, A. and Alavi, A. and Albuquerque, I. S. and Alestas, G. and Alguero, G. and Allaoui, A. and Allen, S. W. and Allevato, V. and Alonso-Tetilla, A. V. and Altieri, B. and Alvarez-Candal, A. and Alvi, S. and Amara, A. and Amendola, L. and Amiaux, J. and Andika, I. T. and Andreon, S. and Andrews, A. and Angora, G. and Angulo, R. E. and Annibali, F. and Anselmi, A. and Anselmi, S. and Arcari, S. and Archidiacono, M. and Aric{\`o}, G. and Arnaud, M. and Arnouts, S. and Asgari, M. and Asorey, J. and Atayde, L. and Atek, H. and Atrio-Barandela, F. and Aubert, M. and Aubourg, E. and Auphan, T. and Auricchio, N. and Aussel, B. and Aussel, H. and Avelino, P. P. and Avgoustidis, A. and Avila, S. and Awan, S. and Azzollini, R. and Baccigalupi, C. and Bachelet, E. and Bacon, D. and Baes, M. and Bagley, M. B. and Bahr-Kalus, B. and Balaguera-Antolinez, A. and Balbinot, E. and Balcells, M. and Baldi, M. and Baldry, I. and Balestra, A. and Ballardini, M. and Ballester, O. and Balogh, M. and Ba{\~n}ados, E. and Barbier, R. and Bardelli, S. and Baron, M. and Barreiro, T. and Barrena, R. and Barriere, J.-C. and Barros, B. J. and Barthelemy, A. and Bartolo, N. and Basset, A. and Battaglia, P. and Battisti, A. J. and Baugh, C. M. and Baumont, L. and Bazzanini, L. and Beaulieu, J.-P. and Beckmann, V. and Belikov, A. N. and Bel, J. and Bellagamba, F. and Bella, M. and Bellini, E. and Benabed, K. and Bender, R. and Benevento, G. and Bennett, C. L. and Benson, K. and Bergamini, P. and Bermejo-Climent, J. R. and Bernardeau, F. and Bertacca, D. and Berthe, M. and Berthier, J. and Bethermin, M. and Beutler, F. and Bevillon, C. and Bhargava, S. and Bhatawdekar, R. and Bianchi, D. and Bisigello, L. and Biviano, A. and Blake, R. P. and Blanchard, A. and Blazek, J. and Blot, L. and Bosco, A. and Bodendorf, C. and Boenke, T. and B{\"o}hringer, H. and Boldrini, P. and Bolzonella, M. and Bonchi, A. and Bonici, M. and Bonino, D. and Bonino, L. and Bonvin, C. and Bon, W. and Booth, J. T. and Borgani, S. and Borlaff, A. S. and Borsato, E. and Bosco, A. and Bose, B. and Botticella, M. T. and Boucaud, A. and Bouche, F. and Boucher, J. S. and Boutigny, D. and Bouvard, T. and Bouwens, R. and Bouy, H. and Bowler, R. A. A. and Bozza, V. and Bozzo, E. and Branchini, E. and Brando, G. and Brau-Nogue, S. and Brekke, P. and Bremer, M. N. and Brescia, M. and Breton, M.-A. and Brinchmann, J. and Brinckmann, T. and Brockley-Blatt, C. and Brodwin, M. and Brouard, L. and Brown, M. L. and Bruton, S. and Bucko, J. and Buddelmeijer, H. and Buenadicha, G. and Buitrago, F. and Burger, P. and Burigana, C. and Busillo, V. and Busonero, D. and Cabanac, R. and Cabayol-Garcia, L. and Cagliari, M. S. and Caillat, A. and Caillat, L. and Calabrese, M. and Calabro, A. and Calderone, G. and Calura, F. and Camacho Quevedo, B. and Camera, S. and Campos, L. and Ca{\~n}as-Herrera, G. and Candini, G. P. and Cantiello, M. and Capobianco, V. and Cappellaro, E. and Cappelluti, N. and Cappi, A. and Caputi, K. I. and Cara, C. and Carbone, C. and Cardone, V. F. and Carella, E. and Carlberg, R. G. and Carle, M. and Carminati, L. and Caro, F. and Carrasco, J. M. and Carretero, J. and Carrilho, P. and Carron Duque, J. and Carry, B. and Carvalho, A. and Carvalho, C. S. and Casas, R. and Casas, S. and Casenove, P. and Casey, C. M. and Cassata, P. and Castander, F. J. and Castelao, D. and Castellano, M. and Castiblanco, L. and Castignani, G. and Castro, T. and Cavet, C. and Cavuoti, S. and Chabaud, P.-Y. and Chambers, K. C. and Charles, Y. and Charlot, S. and Chartab, N. and Chary, R. and Chaumeil, F. and Cho, H. and Chon, G. and Ciancetta, E. and Ciliegi, P. and Cimatti, A. and Cimino, M. and Cioni, M.-R. L. and Claydon, R. and Cleland, C. and Cl{\'e}ment, B. and Clements, D. L. and Clerc, N. and Clesse, S. and Codis, S. and Cogato, F. and Colbert, J. and Cole, R. E. and Coles, P. and Collett, T. E. and Collins, R. S. and Colodro-Conde, C. and Colombo, C. and Combes, F. and Conforti, V. and Congedo, G. and Conseil, S. and Conselice, C. J. and Contarini, S. and Contini, T. and Conversi, L. and Cooray, A. R. and Copin, Y. and Corasaniti, P.-S. and Corcho-Caballero, P. and Corcione, L. and Cordes, O. and Corpace, O. and Correnti, M. and Costanzi, M. and Costille, A. and Courbin, F. and Courcoult Mifsud, L. and Courtois, H. M. and Cousinou, M.-C. and Covone, G. and Cowell, T. and Cragg, C. and Cresci, G. and Cristiani, S. and Crocce, M. and Cropper, M. and Crouzet, P. E. and Csizi, B. and Cuby, J.-G. and Cucchetti, E. and Cucciati, O. and Cuillandre, J.-C. and Cunha, P. A. C. and Cuozzo, V. and Daddi, E. and D'Addona, M. and Dafonte, C. and Dagoneau, N. and Dalessandro, E. and Dalton, G. B. and D'Amico, G. and Dannerbauer, H. and Danto, P. and Das, I. and Da Silva, A. and da Silva, R. and d'Assignies Doumerg, W. and Daste, G. and Davies, J. E. and Davini, S. and Dayal, P. and de Boer, T. and Decarli, R. and De Caro, B. and Degaudenzi, H. and Degni, G. and de Jong, J. T. A. and de la Bella, L. F. and de la Torre, S. and Delhaise, F. and Delley, D. and Delucchi, G. and De Lucia, G. and Denniston, J. and De Paolis, F. and De Petris, M. and Derosa, A. and Desai, S. and Desjacques, V. and Despali, G. and Desprez, G. and De Vicente-Albendea, J. and Deville, Y. and Dias, J. D. F. and D{\'i}az-S{\'a}nchez, A. and Diaz, J. J. and Di Domizio, S. and Diego, J. M. and Di Ferdinando, D. and Di Giorgio, A. M. and Dimauro, P. and Dinis, J. and Dolag, K. and Dolding, C. and Dole, H. and Dom{\'i}nguez S{\'a}nchez, H. and Dor{\'e}, O. and Dournac, F. and Douspis, M. and Dreihahn, H. and Droge, B. and Dryer, B. and Dubath, F. and Duc, P.-A. and Ducret, F. and Duffy, C. and Dufresne, F. and Duncan, C. A. J. and Dupac, X. and Duret, V. and Durrer, R. and Durret, F. and Dusini, S. and Ealet, A. and Eggemeier, A. and Eisenhardt, P. R. M. and Elbaz, D. and Elkhashab, M. Y. and Ellien, A. and Endicott, J. and Enia, A. and Erben, T. and Escartin Vigo, J. A. and Escoffier, S. and Escudero Sanz, I. and Essert, J. and Ettori, S. and Ezziati, M. and Fabbian, G. and Fabricius, M. and Fang, Y. and Farina, A. and Farina, M. and Farinelli, R. and Farrens, S. and Faustini, F. and Feltre, A. and Ferguson, A. M. N. and Ferrando, P. and Ferrari, A. G. and Ferr{\'e}-Mateu, A. and Ferreira, P. G. and Ferreras, I. and Ferrero, I. and Ferriol, S. and Ferruit, P. and Filleul, D. and Finelli, F. and Finkelstein, S. L. and Finoguenov, A. and Fiorini, B. and Flentge, F. and Focardi, P. and Fonseca, J. and Fontana, A. and Fontanot, F. and Fornari, F. and Fosalba, P. and Fossati, M. and Fotopoulou, S. and Fouchez, D. and Fourmanoit, N. and Frailis, M. and Fraix-Burnet, D. and Franceschi, E. and Franco, A. and Franzetti, P. and Freihoefer, J. and Frenk, C. S. and Frittoli, G. and Frugier, P.-A. and Frusciante, N. and Fumagalli, A. and Fumagalli, M. and Fumana, M. and Fu, Y. and Gabarra, L. and Galeotta, S. and Galluccio, L. and Ganga, K. and Gao, H. and Garc{\'i}a-Bellido, J. and Garcia, K. and Gardner, J. P. and Garilli, B. and Gaspar-Venancio, L.-M. and Gasparetto, T. and Gautard, V. and Gavazzi, R. and Gaztanaga, E. and Genolet, L. and Genova Santos, R. and Gentile, F. and George, K. and Gerbino, M. and Ghaffari, Z. and Giacomini, F. and Gianotti, F. and Gibb, G. P. S. and Gillard, W. and Gillis, B. and Ginolfi, M. and Giocoli, C. and Girardi, M. and Giri, S. K. and Goh, L. W. K. and G{\'o}mez-Alvarez, P. and Gonzalez-Perez, V. and Gonzalez, A. H. and Gonzalez, E. J. and Gonzalez, J. C. and Gouyou Beauchamps, S. and Gozaliasl, G. and Gracia-Carpio, J. and Grandis, S. and Granett, B. R. and Granvik, M. and Grazian, A. and Gregorio, A. and Grenet, C. and Grillo, C. and Grupp, F. and Gruppioni, C. and Gruppuso, A. and Guerbuez, C. and Guerrini, S. and Guidi, M. and Guillard, P. and Gutierrez, C. M. and Guttridge, P. and Guzzo, L. and Gwyn, S. and Haapala, J. and Haase, J. and Haddow, C. R. and Hailey, M. and Hall, A. and Hall, D. and Hamaus, N. and Haridasu, B. S. and Harnois-D{\'e}raps, J. and Harper, C. and Hartley, W. G. and Hasinger, G. and Hassani, F. and Hatch, N. A. and Haugan, S. V. H. and H{\"a}u{\ss}ler, B. and Heavens, A. and Heisenberg, L. and Helmi, A. and Helou, G. and Hemmati, S. and Henares, K. and Herent, O. and Hern{\'a}ndez-Monteagudo, C. and Heuberger, T. and Hewett, P. C. and Heydenreich, S. and Hildebrandt, H. and Hirschmann, M. and Hjorth, J. and Hoar, J. and Hoekstra, H. and Holland, A. D. and Holliman, M. S. and Holmes, W. and Hook, I. and Horeau, B. and Hormuth, F. and Hornstrup, A. and Hosseini, S. and Hu, D. and Hudelot, P. and Hudson, M. J. and Huertas-Company, M. and Huff, E. M. and Hughes, A. C. N. and Humphrey, A. and Hunt, L. K. and Huynh, D. D. and Ibata, R. and Ichikawa, K. and Iglesias-Groth, S. and Ilbert, O. and Ili{\'c}, S. and Ingoglia, L. and Iodice, E. and Israel, H. and Israelsson, U. E. and Izzo, L. and Jablonka, P. and Jackson, N. and Jacobson, J. and Jafariyazani, M. and Jahnke, K. and Jain, B. and Jansen, H. and Jarvis, M. J. and Jasche, J. and Jauzac, M. and Jeffrey, N. and Jhabvala, M. and Jimenez-Teja, Y. and Jimenez Mu{\~n}oz, A. and Joachimi, B. and Johansson, P. H. and Joudaki, S. and Jullo, E. and Kajava, J. J. E. and Kang, Y. and Kannawadi, A. and Kansal, V. and Karagiannis, D. and K{\"a}rcher, M. and Kashlinsky, A. and Kazandjian, M. V. and Keck, F. and Keih{\"a}nen, E. and Kerins, E. and Kermiche, S. and Khalil, A. and Kiessling, A. and Kiiveri, K. and Kilbinger, M. and Kim, J. and King, R. and Kirkpatrick, C. C. and Kitching, T. and Kluge, M. and Knabenhans, M. and Knapen, J. H. and Knebe, A. and Kneib, J.-P. and Kohley, R. and Koopmans, L. V. E. and Koskinen, H. and Koulouridis, E. and Kou, R. and Kov{\'a}cs, A. and Kova{\v c}i{\'c}, I. and Kowalczyk, A. and Koyama, K. and Kraljic, K. and Krause, O. and Kruk, S. and Kubik, B. and Kuchner, U. and Kuijken, K. and K{\"u}mmel, M. and Kunz, M. and Kurki-Suonio, H. and Lacasa, F. and Lacey, C. G. and La Franca, F. and Lagarde, N. and Lahav, O. and Laigle, C. and La Marca, A. and La Marle, O. and Lamine, B. and Lam, M. C. and Lan{\c c}on, A. and Landt, H. and Langer, M. and Lapi, A. and Larcheveque, C. and Larsen, S. S. and Lattanzi, M. and Laudisio, F. and Laugier, D. and Laureijs, R. and Laurent, V. and Lavaux, G. and Lawrenson, A. and Lazanu, A. and Lazeyras, T. and Le Boulc'h, Q. and Le Brun, A. M. C. and Le Brun, V. and Leclercq, F. and Lee, S. and Le Graet, J. and Legrand, L. and Leirvik, K. N. and Le Jeune, M. and Lembo, M. and Le Mignant, D. and Lepinzan, M. D. and Lepori, F. and Le Reun, A. and Leroy, G. and Lesci, G. F. and Lesgourgues, J. and Leuzzi, L. and Levi, M. E. and Liaudat, T. I. and Libet, G. and Liebing, P. and Ligori, S. and Lilje, P. B. and Lin, C.-C. and Linde, D. and Linder, E. and Lindholm, V. and Linke, L. and Li, S.-S. and Liu, S. J. and Lloro, I. and Lobo, F. S. N. and Lodieu, N. and Lombardi, M. and Lombriser, L. and Lonare, P. and Longo, G. and L{\'o}pez-Caniego, M. and Lopez Lopez, X. and Lorenzo Alvarez, J. and Loureiro, A. and Loveday, J. and Lusso, E. and Macias-Perez, J. and Maciaszek, T. and Maggio, G. and Magliocchetti, M. and Magnard, F. and Magnier, E. A. and Magro, A. and Mahler, G. and Mainetti, G. and Maino, D. and Maiorano, E. and Maiorano, E. and Malavasi, N. and Mamon, G. A. and Mancini, C. and Mandelbaum, R. and Manera, M. and Manj{\'o}n-Garc{\'i}a, A. and Mannucci, F. and Mansutti, O. and Manteiga Outeiro, M. and Maoli, R. and Maraston, C. and Marcin, S. and Marcos-Arenal, P. and Margalef-Bentabol, B. and Marggraf, O. and Marinucci, D. and Marinucci, M. and Markovic, K. and Marleau, F. R. and Marpaud, J. and Martignac, J. and Mart{\'i}n-Fleitas, J. and Martin-Moruno, P. and Martin, E. L. and Martinelli, M. and Martinet, N. and Martin, H. and Martins, C. J. A. P. and Marulli, F. and Massari, D. and Massey, R. and Masters, D. C. and Matarrese, S. and Matsuoka, Y. and Matthew, S. and Maughan, B. J. and Mauri, N. and Maurin, L. and Maurogordato, S. and McCarthy, K. and McConnachie, A. W. and McCracken, H. J. and McDonald, I. and McEwen, J. D. and McPartland, C. J. R. and Medinaceli, E. and Mehta, V. and Mei, S. and Melchior, M. and Melin, J.-B. and M{\'e}nard, B. and Mendes, J. and Mendez-Abreu, J. and Meneghetti, M. and Mercurio, A. and Merlin, E. and Metcalf, R. B. and Meylan, G. and Migliaccio, M. and Mignoli, M. and Miller, L. and Miluzio, M. and Milvang-Jensen, B. and Mimoso, J. P. and Miquel, R. and Miyatake, H. and Mobasher, B. and Mohr, J. J. and Monaco, P. and Mongui{\'o}, M. and Montoro, A. and Mora, A. and Moradinezhad Dizgah, A. and Moresco, M. and Moretti, C. and Morgante, G. and Morisset, N. and Moriya, T. J. and Morris, P. W. and Mortlock, D. J. and Moscardini, L. and Mota, D. F. and Mottet, S. and Moustakas, L. A. and Moutard, T. and M{\"u}ller, T. and Munari, E. and Murphree, G. and Murray, C. and Murray, N. and Musi, P. and Nadathur, S. and Nagam, B. C. and Nagao, T. and Naidoo, K. and Nakajima, R. and Nally, C. and Natoli, P. and Navarro-Alsina, A. and Navarro Girones, D. and Neissner, C. and Nersesian, A. and Nesseris, S. and Nguyen-Kim, H. N. and Nicastro, L. and Nichol, R. C. and Nielbock, M. and Niemi, S.-M. and Nieto, S. and Nilsson, K. and Noller, J. and Norberg, P. and Nouri-Zonoz, A. and Ntelis, P. and Nucita, A. A. and Nugent, P. and Nunes, N. J. and Nutma, T. and Ocampo, I. and Odier, J. and Oesch, P. A. and Oguri, M. and Magalhaes Oliveira, D. and Onoue, M. and Oosterbroek, T. and Oppizzi, F. and Ordenovic, C. and Osato, K. and Pacaud, F. and Pace, F. and Padilla, C. and Paech, K. and Pagano, L. and Page, M. J. and Palazzi, E. and Paltani, S. and Pamuk, S. and Pandolfi, S. and Paoletti, D. and Paolillo, M. and Papaderos, P. and Pardede, K. and Parimbelli, G. and Parmar, A. and Partmann, C. and Pasian, F. and Passalacqua, F. and Paterson, K. and Patrizii, L. and Pattison, C. and Paulino-Afonso, A. and Paviot, R. and Peacock, J. A. and Pearce, F. R. and Pedersen, K. and Peel, A. and Peletier, R. F. and Pellejero Ibanez, M. and Pello, R. and Penny, M. T. and Percival, W. J. and Perez-Garrido, A. and Perotto, L. and Pettorino, V. and Pezzotta, A. and Pezzuto, S. and Philippon, A. and Pierre, M. and Piersanti, O. and Pietroni, M. and Piga, L. and Pilo, L. and Pires, S. and Pisani, A. and Pizzella, A. and Pizzuti, L. and Plana, C. and Polenta, G. and Pollack, J. E. and Poncet, M. and P{\"o}ntinen, M. and Pool, P. and Popa, L. A. and Popa, V. and Popp, J. and Porciani, C. and Porth, L. and Potter, D. and Poulain, M. and Pourtsidou, A. and Pozzetti, L. and Prandoni, I. and Pratt, G. W. and Prezelus, S. and Prieto, E. and Pugno, A. and Quai, S. and Quilley, L. and Racca, G. D. and Raccanelli, A. and R{\'a}cz, G. and Radinovi{\'c}, S. and Radovich, M. and Ragagnin, A. and Ragnit, U. and Raison, F. and Ramos-Chernenko, N. and Ranc, C. and Rasera, Y. and Raylet, N. and Rebolo, R. and Refregier, A. and Reimberg, P. and Reiprich, T. H. and Renk, F. and Renzi, A. and Retre, J. and Revaz, Y. and Reyl{\'e}, C. and Reynolds, L. and Rhodes, J. and Ricci, F. and Ricci, M. and Riccio, G. and Ricken, S. O. and Rissanen, S. and Risso, I. and Rix, H.-W. and Robin, A. C. and Rocca-Volmerange, B. and Rocci, P.-F. and Rodenhuis, M. and Rodighiero, G. and Rodriguez Monroy, M. and Rollins, R. P. and Romanello, M. and Roman, J. and Romelli, E. and Romero-Gomez, M. and Roncarelli, M. and Rosati, P. and Rosset, C. and Rossetti, E. and Roster, W. and Rottgering, H. J. A. and Rozas-Fern{\'a}ndez, A. and Ruane, K. and Rubino-Martin, J. A. and Rudolph, A. and Ruppin, F. and Rusholme, B. and Sacquegna, S. and S{\'a}ez-Casares, I. and Saga, S. and Saglia, R. and Sahl{\'e}n, M. and Saifollahi, T. and Sakr, Z. and Salvalaggio, J. and Salvaterra, R. and Salvati, L. and Salvato, M. and Salvignol, J.-C. and S{\'a}nchez, A. G. and Sanchez, E. and Sanders, D. B. and Sapone, D. and Saponara, M. and Sarpa, E. and Sarron, F. and Sartori, S. and Sartoris, B. and Sassolas, B. and Sauniere, L. and Sauvage, M. and Sawicki, M. and Scaramella, R. and Scarlata, C. and Scharr{\'e}, L. and Schaye, J. and Schewtschenko, J. A. and Schindler, J.-T. and Schinnerer, E. and Schirmer, M. and Schmidt, F. and Schmidt, F. and Schmidt, M. and Schneider, A. and Schneider, M. and Schneider, P. and Sch{\"o}neberg, N. and Schrabback, T. and Schultheis, M. and Schulz, S. and Schuster, N. and Schwartz, J. and Sciotti, D. and Scodeggio, M. and Scognamiglio, D. and Scott, D. and Scottez, V. and Secroun, A. and Sefusatti, E. and Seidel, G. and Seiffert, M. and Sellentin, E. and Selwood, M. and Semboloni, E. and Sereno, M. and Serjeant, S. and Serrano, S. and Setnikar, G. and Shankar, F. and Sharples, R. M. and Short, A. and Shulevski, A. and Shuntov, M. and Sias, M. and Sikkema, G. and Silvestri, A. and Simon, P. and Sirignano, C. and Sirri, G. and Skottfelt, J. and Slezak, E. and Sluse, D. and Smith, G. P. and Smith, L. C. and Smith, R. E. and Smit, S. J. A. and Soldano, F. and Solheim, B. G. B. and Sorce, J. G. and Sorrenti, F. and Soubrie, E. and Spinoglio, L. and Spurio Mancini, A. and Stadel, J. and Stagnaro, L. and Stanco, L. and Stanford, S. A. and Starck, J.-L. and Stassi, P. and Steinwagner, J. and Stern, D. and Stone, C. and Strada, P. and Strafella, F. and Stramaccioni, D. and Surace, C. and Sureau, F. and Suyu, S. H. and Swindells, I. and Szafraniec, M. and Szapudi, I. and Taamoli, S. and Talia, M. and Tallada-Cresp{\'i}, P. and Tanidis, K. and Tao, C. and Tarr{\'i}o, P. and Tavagnacco, D. and Taylor, A. N. and Taylor, J. E. and Taylor, P. L. and Teixeira, E. M. and Tenti, M. and Teodoro Idiago, P. and Teplitz, H. I. and Tereno, I. and Tessore, N. and Testa, V. and Testera, G. and Tewes, M. and Teyssier, R. and Theret, N. and Thizy, C. and Thomas, P. D. and Toba, Y. and Toft, S. and Toledo-Moreo, R. and Tolstoy, E. and Tommasi, E. and Torbaniuk, O. and Torradeflot, F. and Tortora, C. and Tosi, S. and Tosti, S. and Trifoglio, M. and Troja, A. and Trombetti, T. and Tronconi, A. and Tsedrik, M. and Tsyganov, A. and Tucci, M. and Tutusaus, I. and Uhlemann, C. and Ulivi, L. and Urbano, M. and Vacher, L. and Vaillon, L. and Valageas, P. and Valdes, I. and Valentijn, E. A. and Valenziano, L. and Valieri, C. and Valiviita, J. and Van den Broeck, M. and Vassallo, T. and Vavrek, R. and Vega-Ferrero, J. and Venemans, B. and Venhola, A. and Ventura, S. and Verdoes Kleijn, G. and Vergani, D. and Verma, A. and Vernizzi, F. and Veropalumbo, A. and Verza, G. and Vescovi, C. and Vibert, D. and Viel, M. and Vielzeuf, P. and Viglione, C. and Viitanen, A. and Villaescusa-Navarro, F. and Vinciguerra, S. and Visticot, F. and Voggel, K. and von Wietersheim-Kramsta, M. and Vriend, W. J. and Wachter, S. and Walmsley, M. and Walth, G. and Walton, D. M. and Walton, N. A. and Wander, M. and Wang, L. and Wang, Y. and Weaver, J. R. and Weller, J. and Wetzstein, M. and Whalen, D. J. and Whittam, I. H. and Widmer, A. and Wiesmann, M. and Wilde, J. and Williams, O. R. and Winther, H.-A. and Wittje, A. and Wong, J. H. W. and Wright, A. H. and Yankelevich, V. and Yeung, H. W. and Yoon, M. and Youles, S. and Yung, L. Y. A. and Zacchei, A. and Zalesky, L. and Zamorani, G. and Zamorano Vitorelli, A. and Zanoni Marc, M. and Zennaro, M. and Zerbi, F. M. and Zinchenko, I. A. and Zoubian, J. and Zucca, E. and Zumalacarregui, M.},
   year={2025},
   month=Apr, pages={A1} }

@article{Martinet2019,
   title={<i>Euclid</i> preparation: IV. Impact of undetected galaxies on weak-lensing shear measurements},
   volume={627},
   ISSN={1432-0746},
   url={http://dx.doi.org/10.1051/0004-6361/201935187},
   DOI={10.1051/0004-6361/201935187},
   journal={Astronomy \& Astrophysics},
   publisher={EDP Sciences},
   author={ {Euclid Collaboration: Martinet}, N. and Schrabback, T. and Hoekstra, H. and Tewes, M. and Herbonnet, R. and Schneider, P. and Hernandez-Martin, B. and Taylor, A. N. and Brinchmann, J. and Carvalho, C. S. and Castellano, M. and Congedo, G. and Gillis, B. R. and Jullo, E. and K{\"u}mmel, M. and Ligori, S. and Lilje, P. B. and Padilla, C. and Paris, D. and Peacock, J. A. and Pilo, S. and Pujol, A. and Scott, D. and Toledo-Moreo, R.},
   year={2019},
   month=jul, pages={A59} }

@misc{Novaes_2024,
      title={Cosmology from HSC Y1 Weak Lensing with Combined Higher-Order Statistics and Simulation-based Inference}, 
      author={Camila P. Novaes and Leander Thiele and Joaquin Armijo and Sihao Cheng and Jessica A. Cowell and Gabriela A. Marques and Elisa G. M. Ferreira and Masato Shirasaki and Ken Osato and Jia Liu},
      year={2024},
      eprint={2409.01301},
      archivePrefix={arXiv},
      primaryClass={astro-ph.CO},
      url={https://arxiv.org/abs/2409.01301}, 
}

@article{Oehl_2026,
   title={The Non-Gaussian Weak-Lensing Likelihood: A Multivariate Copula Construction and Impact on Cosmological Constraints},
   volume={9},
   ISSN={2565-6120},
   url={http://dx.doi.org/10.33232/001c.163550},
   DOI={10.33232/001c.163550},
   journal={The Open Journal of Astrophysics},
   publisher={Maynooth University},
   author={Oehl, Veronika and Tr{\"o}ster, Tilman},
   year={2026},
   month=jun }

@article{Percival_2021,
   title={Matching Bayesian and frequentist coverage probabilities when using an approximate data covariance matrix},
   volume={510},
   ISSN={1365-2966},
   url={http://dx.doi.org/10.1093/mnras/stab3540},
   DOI={10.1093/mnras/stab3540},
   number={3},
   journal={Monthly Notices of the Royal Astronomical Society},
   publisher={Oxford University Press (OUP)},
   author={Percival, Will J and Friedrich, Oliver and Sellentin, Elena and Heavens, Alan},
   year={2021},
   month=Dec, pages={3207--3221} }

@article{Porqueres_2021,
   title={Lifting weak lensing degeneracies with a field-based likelihood},
   volume={509},
   ISSN={1365-2966},
   url={http://dx.doi.org/10.1093/mnras/stab3234},
   DOI={10.1093/mnras/stab3234},
   number={3},
   journal={Monthly Notices of the Royal Astronomical Society},
   publisher={Oxford University Press (OUP)},
   author={Porqueres, Natalia and Heavens, Alan and Mortlock, Daniel and Lavaux, Guilhem},
   year={2021},
   month=Nov, pages={3194--3202} }

@article{Perraudin_2019,
   title={DeepSphere: Efficient spherical convolutional neural network with HEALPix sampling for cosmological applications},
   volume={27},
   ISSN={2213-1337},
   url={http://dx.doi.org/10.1016/j.ascom.2019.03.004},
   DOI={10.1016/j.ascom.2019.03.004},
   journal={Astronomy and Computing},
   publisher={Elsevier BV},
   author={Perraudin, N. and Defferrard, M. and Kacprzak, T. and Sgier, R.},
   year={2019},
   month=Apr, pages={130--146} }

@misc{Papamakarios_2019,
      title={Neural Density Estimation and Likelihood-free Inference}, 
      author={George Papamakarios},
      year={2019},
      eprint={1910.13233},
      archivePrefix={arXiv},
      primaryClass={stat.ML},
      url={https://arxiv.org/abs/1910.13233}, 
}

@misc{papamakarios2018maskedautoregressiveflowdensity, title={Masked Autoregressive Flow for Density Estimation}, author={George Papamakarios and Theo Pavlakou and Iain Murray}, year={2018}, eprint={1705.07057}, archivePrefix={arXiv}, primaryClass={stat.ML}, url={https://arxiv.org/abs/1705.07057}, }

@misc{Pedregosa_2011,
      title={Scikit-learn: Machine Learning in Python}, 
      author={Fabian Pedregosa and Ga{\"e}l Varoquaux and Alexandre Gramfort and Vincent Michel and Bertrand Thirion and Olivier Grisel and Mathieu Blondel and Andreas M{\"u}ller and Joel Nothman and Gilles Louppe and Peter Prettenhofer and Ron Weiss and Vincent Dubourg and Jake Vanderplas and Alexandre Passos and David Cournapeau and Matthieu Brucher and Matthieu Perrot and {\'E}douard Duchesnay},
      year={2018},
      eprint={1201.0490},
      archivePrefix={arXiv},
      primaryClass={cs.LG},
      url={https://arxiv.org/abs/1201.0490}, 
}

@article{Rumelhart_1986,
  author  = {Rumelhart, David E. and Hinton, Geoffrey E. and Williams, Ronald J.},
  title   = {Learning representations by back-propagating errors},
  journal = {Nature},
  volume  = {323},
  number  = {6088},
  pages   = {533--536},
  year    = {1986},
  month   = {October},
  doi     = {10.1038/323533a0},
  url     = {https://doi.org/10.1038/323533a0},
  issn    = {1476-4687}
}

@article{Ribli_2019,
   title={Weak lensing cosmology with convolutional neural networks on noisy data},
   volume={490},
   ISSN={1365-2966},
   url={http://dx.doi.org/10.1093/mnras/stz2610},
   DOI={10.1093/mnras/stz2610},
   number={2},
   journal={Monthly Notices of the Royal Astronomical Society},
   publisher={Oxford University Press (OUP)},
   author={Ribli, Dezs{\H{o}} and Pataki, B{\'a}lint {\'A}rmin and Zorrilla Matilla, Jos{\'e} Manuel and Hsu, Daniel and Haiman, Zolt{\'a}n and Csabai, Istv{\'a}n},
   year={2019},
   month=sep, pages={1843--1860} }

@article{Semboloni_2011,
   title={Quantifying the effect of baryon physics on weak lensing tomography: Baryon physics and weak lensing tomography},
   volume={417},
   ISSN={0035-8711},
   url={http://dx.doi.org/10.1111/j.1365-2966.2011.19385.x},
   DOI={10.1111/j.1365-2966.2011.19385.x},
   number={3},
   journal={Monthly Notices of the Royal Astronomical Society},
   publisher={Oxford University Press (OUP)},
   author={Semboloni, Elisabetta and Hoekstra, Henk and Schaye, Joop and van Daalen, Marcel P. and McCarthy, Ian G.},
   year={2011},
   month=sep, pages={2020--2035} }

@article{Schneider_1996,
   title={Detection of (dark) matter concentrations via weak gravitational lensing},
   volume={283},
   ISSN={1365-2966},
   url={http://dx.doi.org/10.1093/mnras/283.3.837},
   DOI={10.1093/mnras/283.3.837},
   number={3},
   journal={Monthly Notices of the Royal Astronomical Society},
   publisher={Oxford University Press (OUP)},
   author={Schneider, P.},
   year={1996},
   month=Dec, pages={837--853} }

@article{Troxel_2015,
   title={The intrinsic alignment of galaxies and its impact on weak gravitational lensing in an era of precision cosmology},
   volume={558},
   ISSN={0370-1573},
   url={http://dx.doi.org/10.1016/j.physrep.2014.11.001},
   DOI={10.1016/j.physrep.2014.11.001},
   journal={Physics Reports},
   publisher={Elsevier BV},
   author={Troxel, M.A. and Ishak, Mustapha},
   year={2015},
   month=Feb, pages={1--59} }

@misc{Talts_2020,
      title={Validating Bayesian Inference Algorithms with Simulation-Based Calibration}, 
      author={Sean Talts and Michael Betancourt and Daniel Simpson and Aki Vehtari and Andrew Gelman},
      year={2020},
      eprint={1804.06788},
      archivePrefix={arXiv},
      primaryClass={stat.ME},
      url={https://arxiv.org/abs/1804.06788}, 
}

@article{TejeroCantero_2020, doi = {10.21105/joss.02505}, url = {https://doi.org/10.21105/joss.02505}, year = {2020}, publisher = {The Open Journal}, volume = {5}, number = {52}, pages = {2505}, author = {Tejero-Cantero, Alvaro and Boelts, Jan and Deistler, Michael and Lueckmann, Jan-Matthis and Durkan, Conor and Gon{\c c}alves, Pedro J. and Greenberg, David S. and Macke, Jakob H.}, title = {sbi: A toolkit for simulation-based inference}, journal = {Journal of Open Source Software} }

@article{Tessore_2023,
   title={GLASS: Generator for Large Scale Structure},
   volume={6},
   ISSN={2565-6120},
   url={http://dx.doi.org/10.21105/astro.2302.01942},
   DOI={10.21105/astro.2302.01942},
   journal={The Open Journal of Astrophysics},
   publisher={Maynooth University},
   author={Tessore, Nicolas and Loureiro, Arthur and Joachimi, Benjamin and von Wietersheim-Kramsta, Maximilian and Jeffrey, Niall},
   year={2023},
   month=Mar }

@article{Vinciguerra_2026,
   title={<i>Euclid</i>
                    preparation: LXXXV. Toward a DR1 application of higher-order weak lensing statistics},
   volume={707},
   ISSN={1432-0746},
   url={http://dx.doi.org/10.1051/0004-6361/202557573},
   DOI={10.1051/0004-6361/202557573},
   journal={Astronomy \& Astrophysics},
   publisher={EDP Sciences},
   author={{Euclid Collaboration: Vinciguerra}, S. and Bouch{\`e}, F. and Martinet, N. and Castiblanco, L. and Uhlemann, C. and Pires, S. and Harnois-D{\'e}raps, J. and Giocoli, C. and Baldi, M. and Cardone, V. F. and Vadal{\`a}, A. and Dagoneau, N. and Linke, L. and Sellentin, E. and Taylor, P. L. and Broxterman, J. C. and Heydenreich, S. and Tinnaneri Sreekanth, V. and Porqueres, N. and Porth, L. and Gatti, M. and Grand{\'o}n, D. and Barthelemy, A. and Bernardeau, F. and Tersenov, A. and Hoekstra, H. and Starck, J.-L. and Cheng, S. and Burger, P. A. and Tereno, I. and Scaramella, R. and Altieri, B. and Andreon, S. and Auricchio, N. and Baccigalupi, C. and Bardelli, S. and Biviano, A. and Branchini, E. and Brescia, M. and Camera, S. and Ca{\~n}as-Herrera, G. and Capobianco, V. and Carbone, C. and Carretero, J. and Castellano, M. and Castignani, G. and Cavuoti, S. and Chambers, K. C. and Cimatti, A. and Colodro-Conde, C. and Congedo, G. and Conversi, L. and Copin, Y. and Courbin, F. and Courtois, H. M. and Cropper, M. and Da Silva, A. and Degaudenzi, H. and de la Torre, S. and De Lucia, G. and Dole, H. and Dubath, F. and Dupac, X. and Dusini, S. and Escoffier, S. and Farina, M. and Farinelli, R. and Farrens, S. and Faustini, F. and Ferriol, S. and Finelli, F. and Frailis, M. and Franceschi, E. and Fumana, M. and Galeotta, S. and George, K. and Gillis, B. and Gracia-Carpio, J. and Grazian, A. and Grupp, F. and Haugan, S. V. H. and Holmes, W. and Hormuth, F. and Hornstrup, A. and Hudelot, P. and Jahnke, K. and Jhabvala, M. and Joachimi, B. and Keih{\"a}nen, E. and Kermiche, S. and Kiessling, A. and Kilbinger, M. and Kubik, B. and Kunz, M. and Kurki-Suonio, H. and Le Brun, A. M. C. and Ligori, S. and Lilje, P. B. and Lindholm, V. and Lloro, I. and Mainetti, G. and Maino, D. and Mansutti, O. and Marggraf, O. and Martinelli, M. and Marulli, F. and Massey, R. J. and Medinaceli, E. and Mei, S. and Melchior, M. and Mellier, Y. and Meneghetti, M. and Meylan, G. and Mora, A. and Moresco, M. and Moscardini, L. and Neissner, C. and Niemi, S. -M. and Padilla, C. and Paltani, S. and Pasian, F. and Pedersen, K. and Pettorino, V. and Polenta, G. and Poncet, M. and Popa, L. A. and Raison, F. and Renzi, A. and Rhodes, J. and Riccio, G. and Romelli, E. and Roncarelli, M. and Saglia, R. and Sakr, Z. and S{\'a}nchez, A. G. and Sapone, D. and Sartoris, B. and Schneider, P. and Schrabback, T. and Secroun, A. and Seidel, G. and Serrano, S. and Sirignano, C. and Sirri, G. and Spurio Mancini, A. and Stanco, L. and Steinwagner, J. and Tallada-Cresp{\'i}, P. and Taylor, A. N. and Tessore, N. and Toft, S. and Toledo-Moreo, R. and Torradeflot, F. and Tutusaus, I. and Valiviita, J. and Vassallo, T. and Wang, Y. and Weller, J. and Zacchei, A. and Zamorani, G. and Zerbi, F. M. and Zucca, E. and Ballardini, M. and Bolzonella, M. and Boucaud, A. and Bozzo, E. and Burigana, C. and Cabanac, R. and Calabrese, M. and Cappi, A. and Escartin Vigo, J. A. and Gabarra, L. and Hartley, W. G. and Maoli, R. and Mart{\'i}n-Fleitas, J. and Matthew, S. and Mauri, N. and Metcalf, R. B. and Pezzotta, A. and P{\"o}ntinen, M. and Risso, I. and Scottez, V. and Sereno, M. and Tenti, M. and Viel, M. and Wiesmann, M. and Akrami, Y. and Andika, I. T. and Angulo, R. E. and Anselmi, S. and Archidiacono, M. and Atrio-Barandela, F. and Aubourg, E. and Bertacca, D. and Bethermin, M. and Blanchard, A. and Blot, L. and Bonici, M. and Borgani, S. and Brown, M. L. and Bruton, S. and Calabro, A. and Camacho Quevedo, B. and Caro, F. and Carvalho, C. S. and Castro, T. and Cogato, F. and Conseil, S. and Cooray, A. R. and Desprez, G. and D{\'i}az-S{\'a}nchez, A. and Diaz, J. J. and Di Domizio, S. and Diego, J. M. and Elkhashab, M. Y. and Fang, Y. and Ferreira, P. G. and Finoguenov, A. and Franco, A. and Ganga, K. and Garc{\'i}a-Bellido, J. and Gasparetto, T. and Gautard, V. and Gavazzi, R. and Gaztanaga, E. and Giacomini, F. and Gianotti, F. and Gozaliasl, G. and Guidi, M. and Gutierrez, C. M. and Hall, A. and Hemmati, S. and Hildebrandt, H. and Hjorth, J. and Kajava, J. J. E. and Kang, Y. and Karagiannis, D. and Kiiveri, K. and Kim, J. and Kirkpatrick, C. C. and Kruk, S. and Legrand, L. and Lembo, M. and Lepori, F. and Leroy, G. and Lesci, G. F. and Lesgourgues, J. and Liaudat, T. I. and Macias-Perez, J. and Magliocchetti, M. and Mannucci, F. and Martins, C. J. A. P. and Maurin, L. and Miluzio, M. and Monaco, P. and Moretti, C. and Morgante, G. and Nadathur, S. and Naidoo, K. and Navarro-Alsina, A. and Nesseris, S. and Paoletti, D. and Passalacqua, F. and Paterson, K. and Patrizii, L. and Pisani, A. and Potter, D. and Quai, S. and Radovich, M. and Sacquegna, S. and Sahl{\'e}n, M. and Sanders, D. B. and Sarpa, E. and Schneider, A. and Sciotti, D. and Smith, L. C. and Tanidis, K. and Tao, C. and Testera, G. and Teyssier, R. and Tosi, S. and Troja, A. and Tucci, M. and Vergani, D. and Verza, G. and Walton, N. A.},
   year={2026},
   month=Mar, pages={A235} }

@article{Wright_2025,
   title={KiDS-Legacy: Cosmological constraints from cosmic shear with the complete Kilo-Degree Survey},
   volume={703},
   ISSN={1432-0746},
   url={http://dx.doi.org/10.1051/0004-6361/202554908},
   DOI={10.1051/0004-6361/202554908},
   journal={Astronomy \& Astrophysics},
   publisher={EDP Sciences},
   author={Wright, Angus H. and St{\"o}lzner, Benjamin and Asgari, Marika and Bilicki, Maciej and Giblin, Benjamin and Heymans, Catherine and Hildebrandt, Hendrik and Hoekstra, Henk and Joachimi, Benjamin and Kuijken, Konrad and Li, Shun-Sheng and Reischke, Robert and von Wietersheim-Kramsta, Maximilian and Yoon, Mijin and Burger, Pierre and Chisari, Nora Elisa and de Jong, Jelte and Dvornik, Andrej and Georgiou, Christos and Harnois-D{\'e}raps, Joachim and Jalan, Priyanka and William, Anjitha John and Joudaki, Shahab and Lesci, Giorgio Francesco and Linke, Laila and Loureiro, Arthur and Mahony, Constance and Maturi, Matteo and Miller, Lance and Moscardini, Lauro and Napolitano, Nicola R. and Porth, Lucas and Radovich, Mario and Schneider, Peter and Tr{\"o}ster, Tilman and Valentijn, Edwin and Wittje, Anna and Yan, Ziang and Zhang, Yun-Hao},
   year={2025},
   month=Nov, pages={A158} }

@ARTICLE{Zeghal_2025,
       author = {{Zeghal}, Justine and {Lanzieri}, Denise and {Lanusse}, Fran{\c{c}}ois and {Boucaud}, Alexandre and {Louppe}, Gilles and {Aubourg}, Eric and {Bayer}, Adrian E. and {LSST Dark Energy Science Collaboration}},
        title = "{Simulation-based inference benchmark for weak lensing cosmology}",
      journal = {\aap},
         year = 2025,
        month = jul,
       volume = {699},
          eid = {A327},
        pages = {A327},
          doi = {10.1051/0004-6361/202452410},
archivePrefix = {arXiv},
       eprint = {2409.17975},
 primaryClass = {astro-ph.CO},
       adsurl = {https://ui.adsabs.harvard.edu/abs/2025A&A...699A.327Z}
}

%%%%%%%%%%%%%%%%%%%%%%%%%%%%%%%%%%%%%%%%%%%%%%%%%%%%%%%%%%%%%%%%%%%%%%%%%%%%%%%%%%%%%%%

\clearpage

\begin{appendix}

% ============================================================
% ============================================================
\section{GP emulator diagnostic plots}
\label{app:gp_diagnostics}
% ============================================================

\begin{figure*}[t]
    \centering
    \includegraphics[width=0.7\textwidth, trim=0 50 0 55, clip]{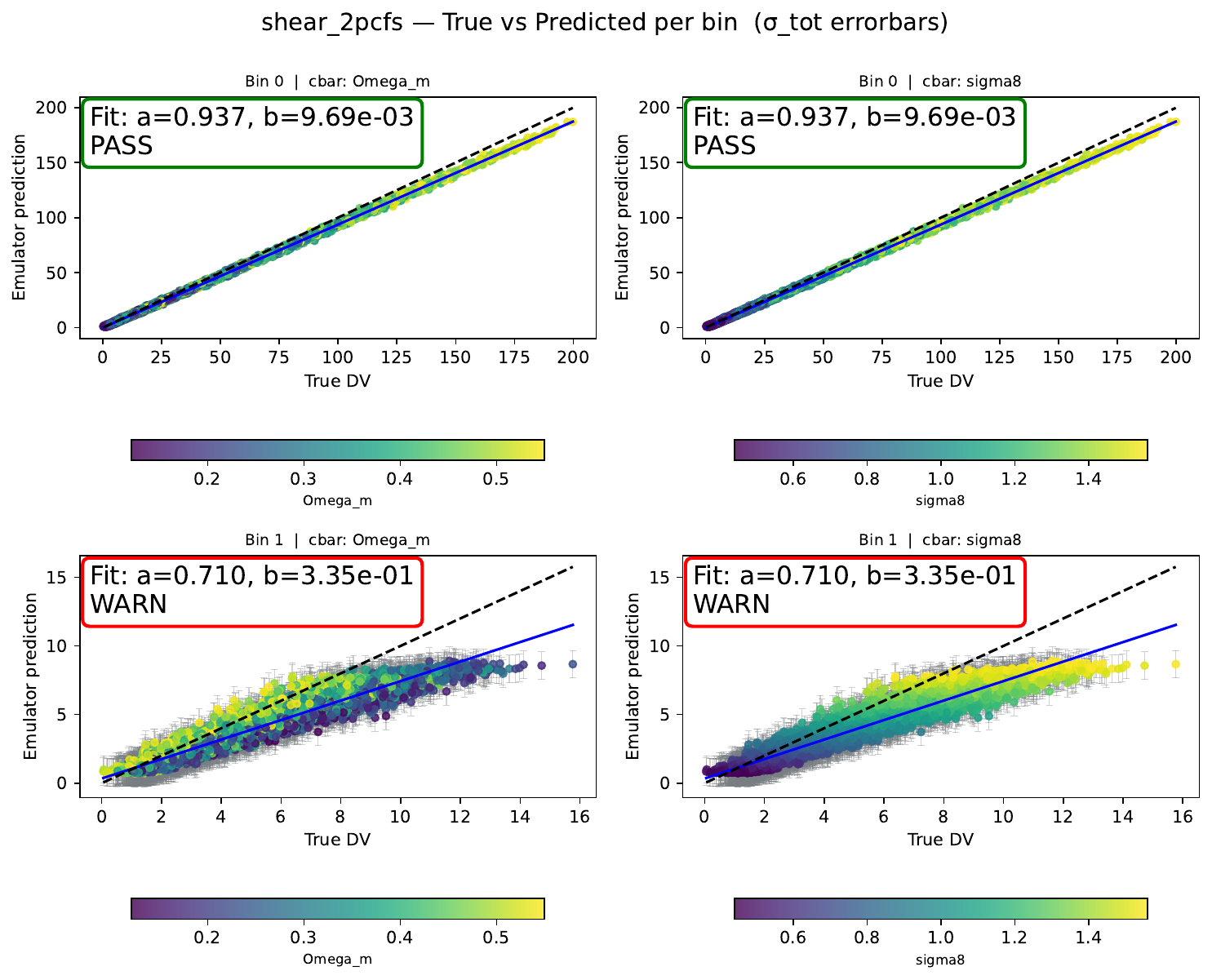}
    \caption{GP emulator true-vs-predicted calibration for the two MOPED-compressed bins (bin~0: first row, bin~1: second row) of shear-2PCFs. Each panel shows the GP prediction against the true DV value for all test nodes of $\mathrm{LHS}_1$, coloured by $\Omega_{\rm m}$ (left column) and $\sigma_8$ (right column), with $\sigma_{\rm tot}$ error bars. The dashed line marks the identity; the solid blue line shows the fitted linear regression with slope $a$ and intercept $b$. Bins with $|a - 1| > 0.10$ or $|b| > 0.10$ are flagged as WARN.}
    \label{fig:gp_true_vs_pred}
\end{figure*}

\begin{figure*}[t]
    \centering
    \includegraphics[width=0.7\textwidth, trim=0 50 0 55, clip]{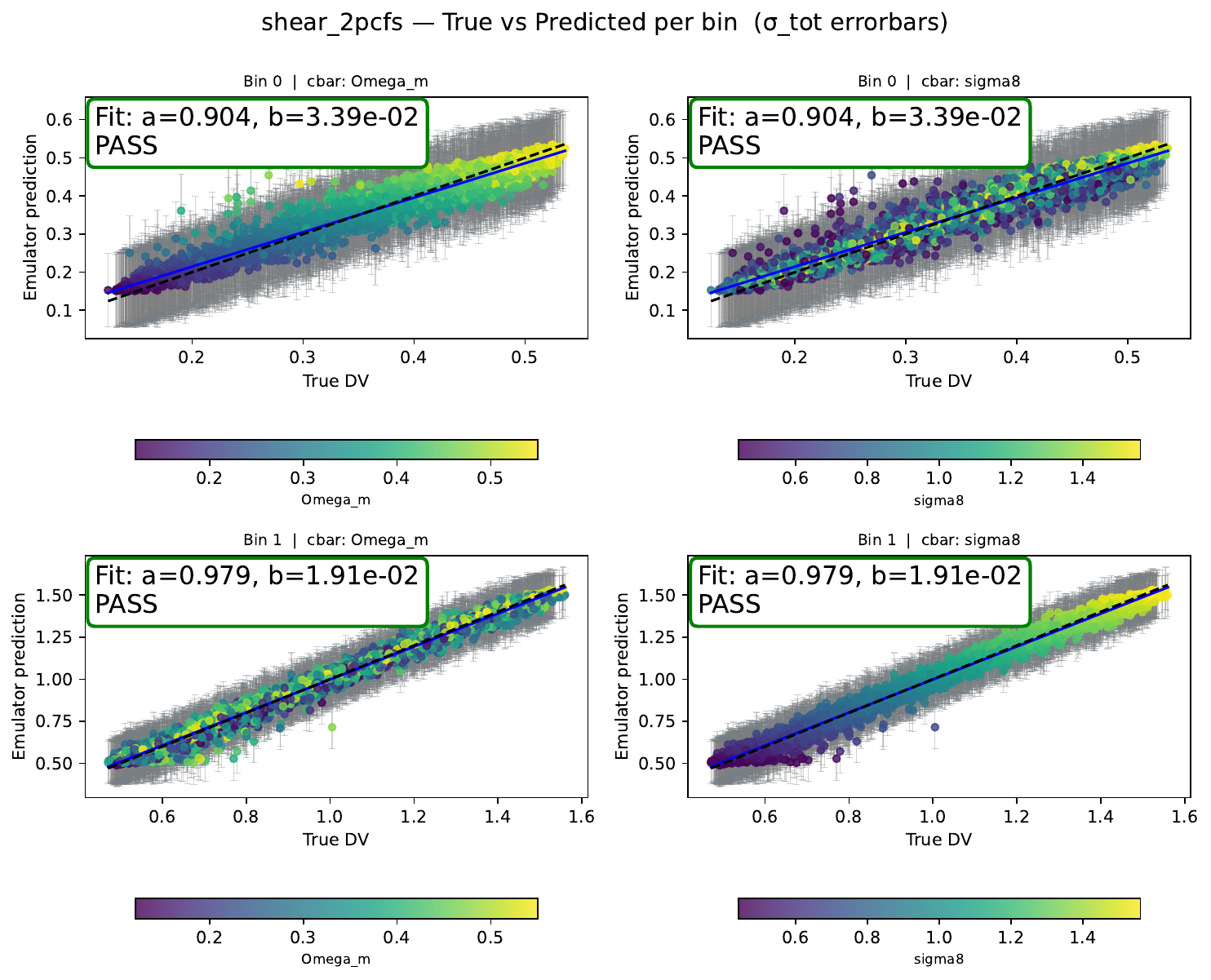}
    \caption{As Fig.~\ref{fig:gp_true_vs_pred}, for the two NN-compressed bins of shear-2PCFs.}
    \label{fig:gp_true_vs_pred_nn}
\end{figure*}

\begin{figure*}[t]
    \centering
    \includegraphics[width=0.7\textwidth, trim=0 0 0 45, clip]{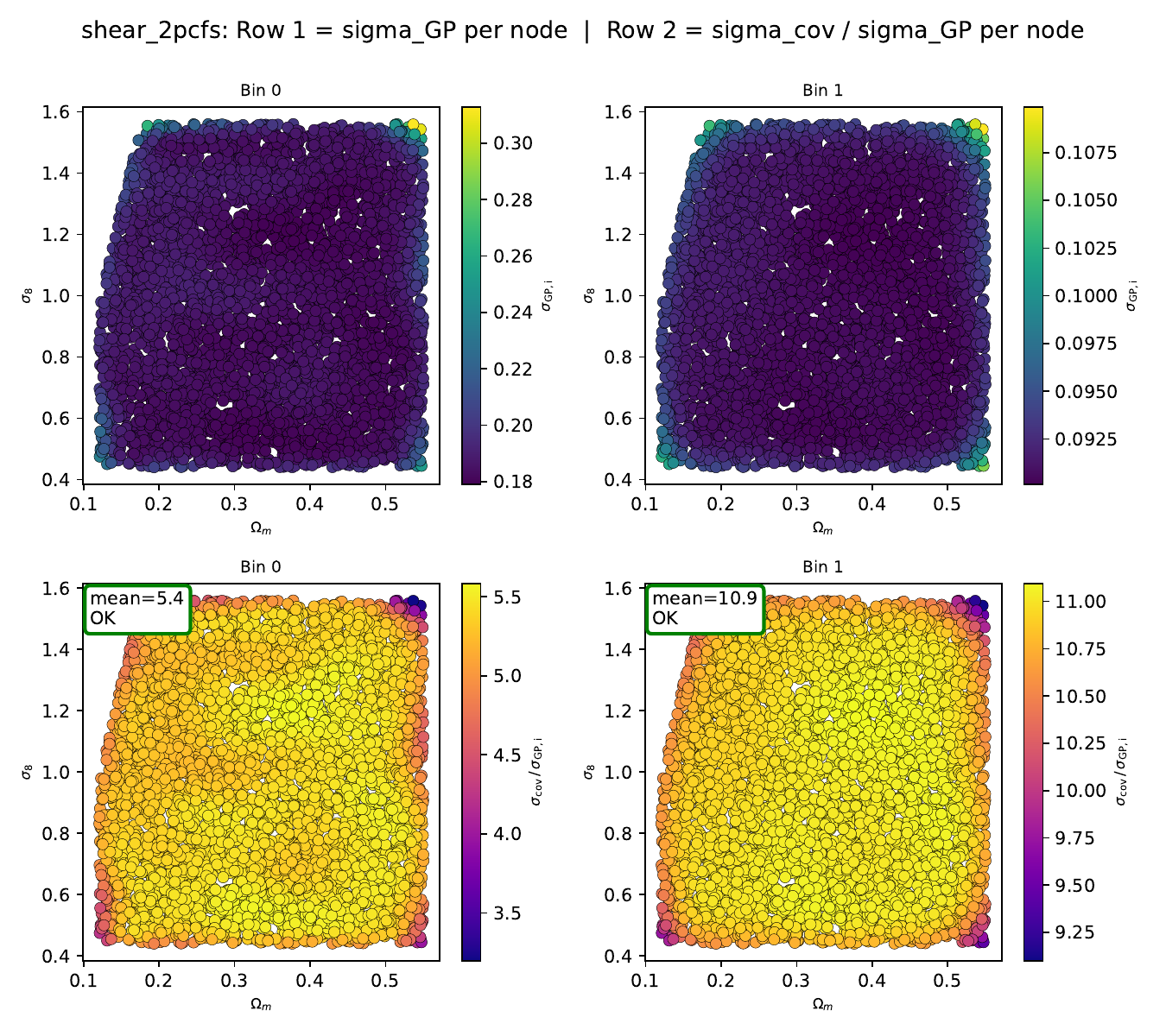}
    \caption{GP emulation uncertainty maps for the two MOPED-compressed bins of shear-2PCFs. Top row: GP interpolation uncertainty $\sigma_{{\rm GP},i}$ per test node in the $(\Omega_{\rm m}, \sigma_8)$ plane. Bottom row: ratio $\sigma_{{\rm cov},i} / \sigma_{{\rm GP},i}$, where $\sigma_{{\rm cov},i} = \sqrt{\hat{\mathsf{C}}_{ii}}$ is the DV noise per bin; the mean ratio is reported in the legend and flagged as OK when it exceeds 5.}
    \label{fig:gp_sigma_plane}
\end{figure*}

We present GP diagnostic figures illustrating the emulation quality of each compressed bin. True-vs-predicted calibration plots are shown for shear-2PCFs under both MOPED and NN compression, while the emulation uncertainty maps are shown for shear-2PCFs under MOPED compression only since those are consistent for the other configurations.

\emph{True-vs-predicted calibration plots} (Figs.~\ref{fig:gp_true_vs_pred}- \ref{fig:gp_true_vs_pred_nn}) show the GP prediction against the true DV value for each compressed bin across all test nodes of $\mathrm{LHS}_1$, with $\sigma_{\rm tot}$ error bars combining the GP interpolation variance and the DV covariance diagonal element per bin. A linear regression is fitted to each scatter; a well-calibrated emulator yields slope $a \approx 1$ and intercept $b \approx 0$. Bins are flagged as WARN when $|a - 1| > 0.10$ or $|b| > 0.10$. For shear-2PCFs under MOPED compression (Fig.~\ref{fig:gp_true_vs_pred}), and as discussed in Sect.~\ref{sec:eli_vs_lfi}, the first bin (bin~0) achieves a good calibration ($a = 0.937$, $b \approx 0$, PASS), while the second bin (bin~1) shows a more significant deviation ($a = 0.710$, $b = 0.335$, WARN), indicating that the GP struggles to accurately emulate the cosmological dependence of this specific compressed summary, which concentrates the sensitivity to $\Omega_{\rm m}$. We note that the emulation accuracy degrades towards large values of $\Omega_{\rm m}$ and $\sigma_8$, where the signal amplitude is highest and the GP interpolation is more challenging; this has a limited impact on the inferred posteriors, which are concentrated in the region of interest around the fiducial cosmology. Under NN compression (Fig.~\ref{fig:gp_true_vs_pred_nn}) both bins pass the calibration test ($a = 0.904$, $b = 0.034$ for bin~0; $a = 0.979$, $b = 0.019$ for bin~1), and the CNN estimator, not reported here, is found to perform similarly.

\emph{Emulation uncertainty maps} (Fig.~\ref{fig:gp_sigma_plane}) show, for each compressed bin, the GP interpolation uncertainty $\sigma_{{\rm GP},i}$ per test node (top row) and the ratio $\sigma_{{\rm cov},i} / \sigma_{{\rm GP},i}$ (bottom row), both displayed in the $(\Omega_{\rm m}, \sigma_8)$ plane, where $\sigma_{{\rm cov},i} = \sqrt{\hat{\mathsf{C}}_{ii}}$ denotes the DV noise per bin given by the square root of the covariance diagonal element. A ratio $\sigma_{{\rm cov},i} / \sigma_{{\rm GP},i} \gg 1$ (flagged as OK when the mean ratio over the full prior exceeds 5) indicates that the GP interpolation error is negligible compared to the intrinsic DV noise, so that the Gaussian likelihood of Eq.~(\ref{eq:log_likelihood}) is not artificially broadened by emulation uncertainty. Even if found to be mostly negligible, the GP variance is nevertheless propagated in the covariance matrix as a linear addition to its diagonal elements. For bin~0 the mean ratio is $5.4$ (OK), and for bin~1 it reaches $10.9$ (OK), confirming that despite the calibration warning in bin~1, the GP uncertainty is small relative to the DV noise at the scale of each compressed summary; the ELI degradation in bin~1 is therefore driven by the inaccurate calibration ($a = 0.710$) rather than by emulation noise inflating the likelihood. We do not show the corresponding emulation uncertainty maps for shear-2PCFs under NN compression nor for the CNN estimator: the results are qualitatively similar to the MOPED case but with a lower GP interpolation uncertainty, making it even more negligible relative to the DV noise.

% ============================================================

% ------
\section{LFI grid-search optimization}
\label{app:lfi_nde_grid}
% ------

\begin{figure*}[t]
    \centering
    \includegraphics[width=\textwidth, trim=0 0 0 25, clip]{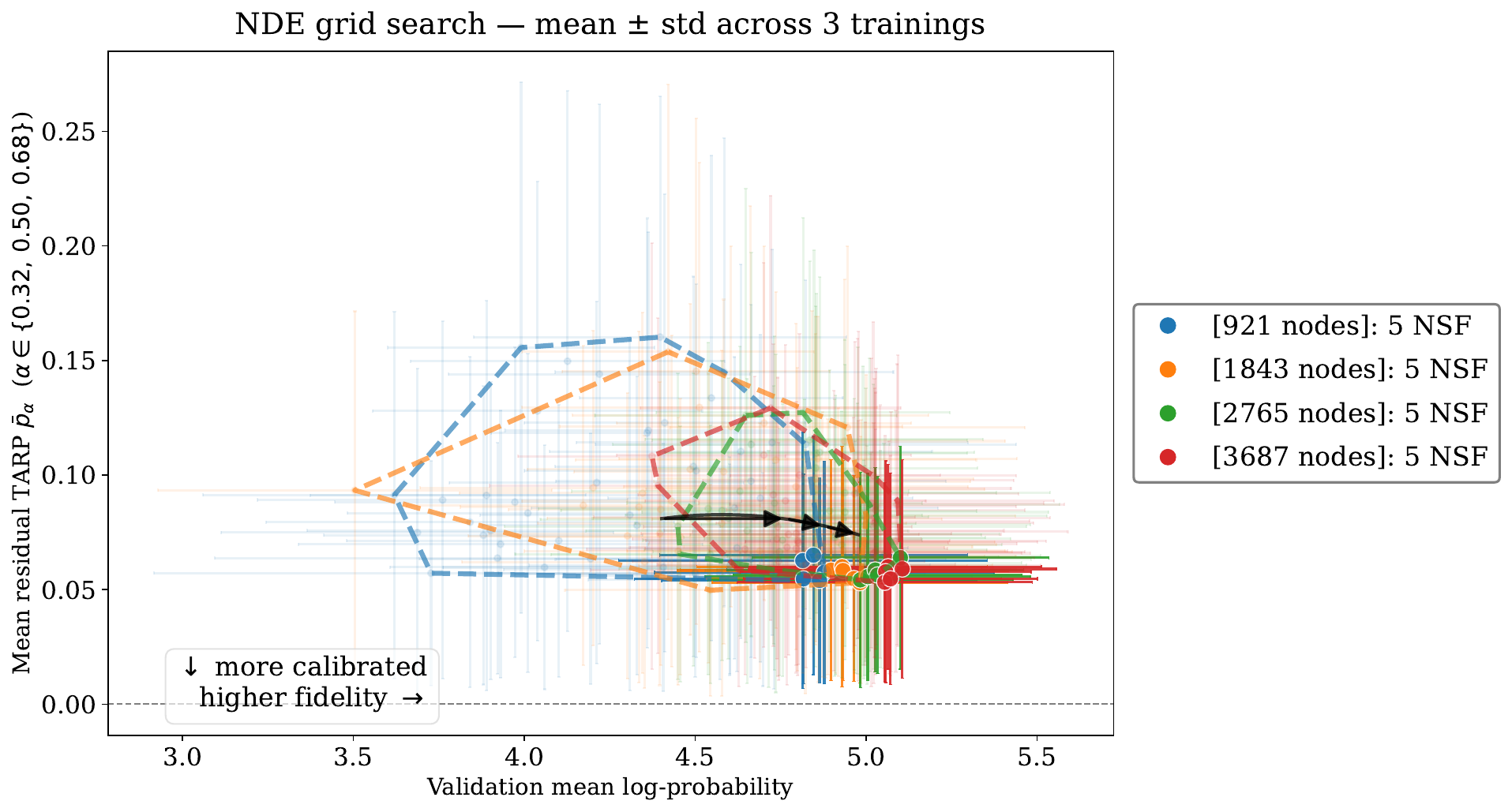}
    \caption{NDE grid-search quality plane for LFI with MOPED-compressed shear-2PCFs. Each point represents one of the $n_{\rm trials}=50$ trials, with marker shape indicating the NDE family (circles: NSF, squares: MAF, triangles: MDN). The four colours correspond to four fractions of the 3,687 training nodes of $\mathrm{LHS}_1$: 25\% (921 nodes, blue), 50\% (1\,843 nodes, orange), 75\% (2\,765 nodes, green), and 100\% (3\,687 nodes, red). Top-ranked trials (large opaque markers, white edge) are distinguished from the remaining trials (low-opacity markers); the legend reports the NDE families found as the 5 top-ranked for each training fraction. The dotted convex hull of each colour group delimits the extent of the solution cloud for that training fraction. The black curve with arrows connects the barycentres of successive clouds, tracing the convergence trend as training set size grows.}
    \label{fig:lfi_grid_search}
\end{figure*}

The quality of an NDE posterior depends non-trivially on the choice of architecture and hyperparameters, and the optimal configuration is not known a priori: it depends on the statistical properties of the input DVs, including the probe type, the noise level, and the compression method applied. Rather than making arbitrary choices, we adopt an automated grid search that systematically trains and evaluates a large number of NDE configurations drawn at random from a predefined hyperparameter space, ranks them according to a combined quality metric, and selects the best-performing configuration for each (estimator, compression) combination independently, ensuring that the inference is not limited by a suboptimal architecture choice. Since the grid search requires training and evaluating $n_{\rm trials}$ independent NDE per configuration, running it in NLE mode would demand an equally large number of MCMC chains to evaluate the posterior at each trial, making the computational cost prohibitive. We therefore use NPE mode during the grid search, which allows direct posterior sampling without MCMC and reduces the per-trial evaluation cost by orders of magnitude, while still providing a reliable ranking of architectures. The final best-net is then retrained in NLE mode for the final constraints.

This automated grid-search and selection procedure is fully general and can be applied to any NDE-based inference problem without modification, making it a versatile tool for simulation-based inference beyond the specific analyses presented here.

For each trial, the NDE ensemble is trained on a subset of the 3,687 training nodes from $\mathrm{LHS}_1$ (80\% of the 4,609 nodes available for inference). Within each trial, the \texttt{sbi} framework internally reserves 10\% of the loaded training data for early stopping, so the effective gradient-update data is 90\% of the loaded subset. Once trained, the NDE is applied to the 230 held-out validation nodes (5\% of the 4,609 nodes available for inference) to evaluate the two ranking metrics: the mean log-probability $\bar{\ell}$, computed by evaluating the learned posterior log-probability on 10,000 samples drawn per validation node and averaging across all 230 nodes; and the TARP residual bias $\bar{p}_\alpha = \langle |\mathrm{ECP}(\alpha) - \alpha| / \alpha \rangle_\alpha$, evaluated on up to 50 of these validation nodes via NPE direct sampling of 10,000 posterior samples per node and averaged over three confidence levels $\alpha \in \{0.32, 0.50, 0.68\}$ to anchor the ECP curve at representative points spanning the credible interval range, with the 50-node cap limiting the total computational cost of repeating this evaluation across all $n_{\rm trials}=50$ trials. The 692 test nodes (15\%) are never accessed during the grid search and are reserved exclusively for the final best-net diagnostics described in Sect.~\ref{sec:lfi_posterior_validation}. The grid search spans three NDE families (NSF, MAF, MDN) over the following hyperparameter space: hidden features per layer $\in \{32, 64\}$; number of flow transforms for NSF and MAF $\in \{4, 6\}$; number of Gaussian mixture components for MDN $\in \{3, 5\}$; learning rate $\in \{10^{-4}, 2\times10^{-4}, 5\times10^{-4}\}$; batch size $\in \{32, 64\}$. We perform $n_{\rm trials}=50$ randomly sampled trials; each trial trains an ensemble of $n_{\rm repeats}=3$ NPE networks sharing the same architecture but initialised with different random seeds, and the ensemble posterior is obtained by averaging the individual log-posteriors, thereby reducing sensitivity to the stochasticity of individual training runs. Trials are ranked by jointly maximising the validation log-probability and minimising the TARP residual, defining a two-dimensional quality plane where the best NDE can be identified unambiguously.

To probe convergence with training set size, the full 50-trial search is run at four fractions of the available training nodes: 25\%, 50\%, 75\%, and 100\%, corresponding to 921, 1\,843, 2\,765, and 3\,687 training nodes respectively. Fig.~\ref{fig:lfi_grid_search} shows the resulting quality-plane diagram for MOPED-compressed shear-2PCFs, while consistent behaviours are found for the other statistics and compression methods. The solution clouds shift monotonically towards higher log-probability and lower TARP residual as the training set grows. Importantly, the perimeter of each cloud, delimited by the dashed convex hull, shrinks as the number of training nodes increases: with more training data the performance of all trials converges, meaning that the final result becomes progressively less sensitive to the specific choice of hyperparameters. This reflects the fact that, when the training set is sufficiently large relative to the complexity of the inference problem, most reasonable architectures recover the likelihood comparably well. Improvement saturates between 2\,765 and 3\,687 nodes, and NSF architectures dominate the top-ranked trials across all four training fractions. It is worth noting that the lowest fraction of 921 training nodes already produces top-ranked NDE with a level of calibration and fidelity comparable to the higher training fractions, confirming that the low dimensionality of the inference problem allows well-calibrated posteriors to be obtained even with a relatively small training set. We nevertheless use the full set of 3,687 available training nodes for the rest of the analysis.

The top-ranked trials can be stacked together to define an optimized \emph{best-net} ensemble posterior. In our setup, we verified that combining up to 5 top-ranked trials produces posteriors fully consistent with those from the single top-ranked trial alone, and therefore select the individual best-ranked NDE to minimise computational cost without loss of accuracy for each statistic and compression method. The reconstructed posterior from the best-net is then validated with the full set of diagnostics described in Sect.~\ref{sec:lfi_posterior_validation}.

% ============================================================
\section{ELI and LFI posterior validation diagnostics}
\label{app:lfi_validation}
% ============================================================

\begin{figure}[h]
    \centering
    \begin{minipage}{0.49\columnwidth}
        \centering
        \text{ELI MOPED}\\[0.2em]
        \includegraphics[width=0.95\linewidth, height=5cm, keepaspectratio, trim=0 45 0 0, clip]{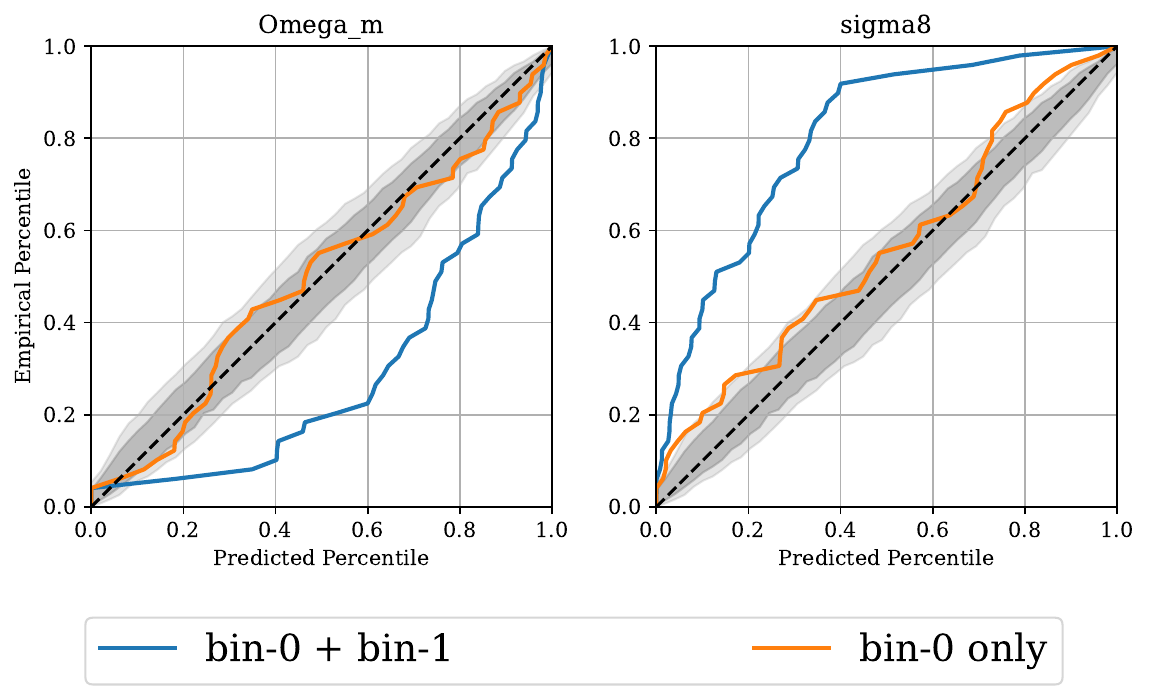}
    \end{minipage}
    \hfill
    \begin{minipage}{0.49\columnwidth}
        \centering
        \text{ELI NN}\\[0.6em]
        \includegraphics[width=0.87\linewidth, height=5cm, keepaspectratio]{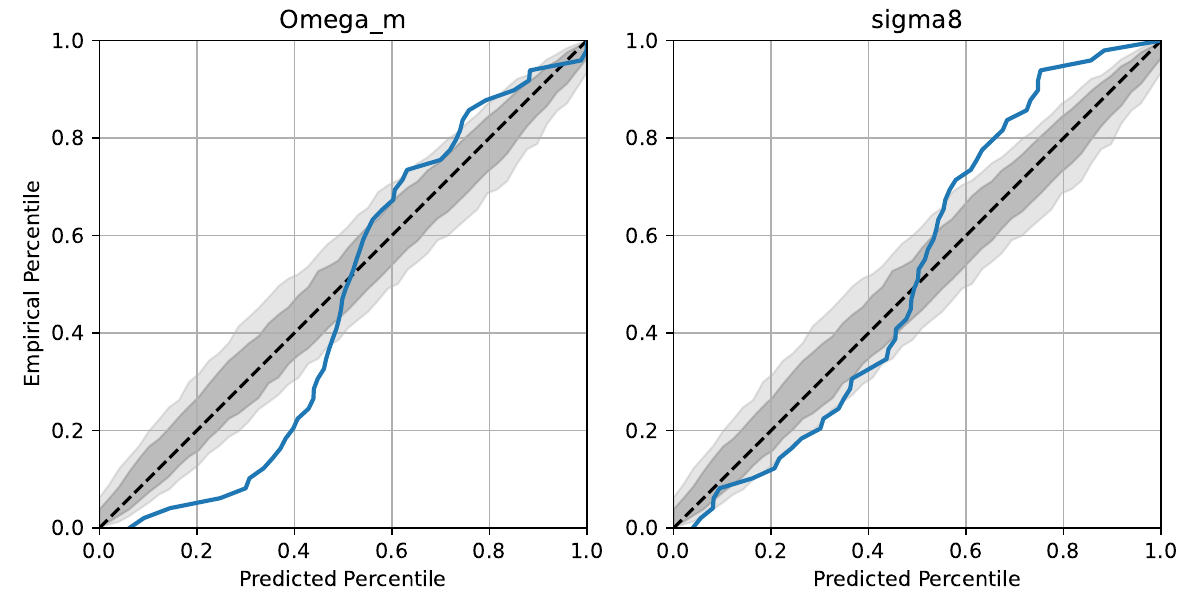}
    \end{minipage}
    \\[0.8em]
    \begin{minipage}{0.49\columnwidth}
        \centering
        \text{LFI MOPED}\\[-0.1em]
        \includegraphics[width=\linewidth, height=5cm, keepaspectratio]
        {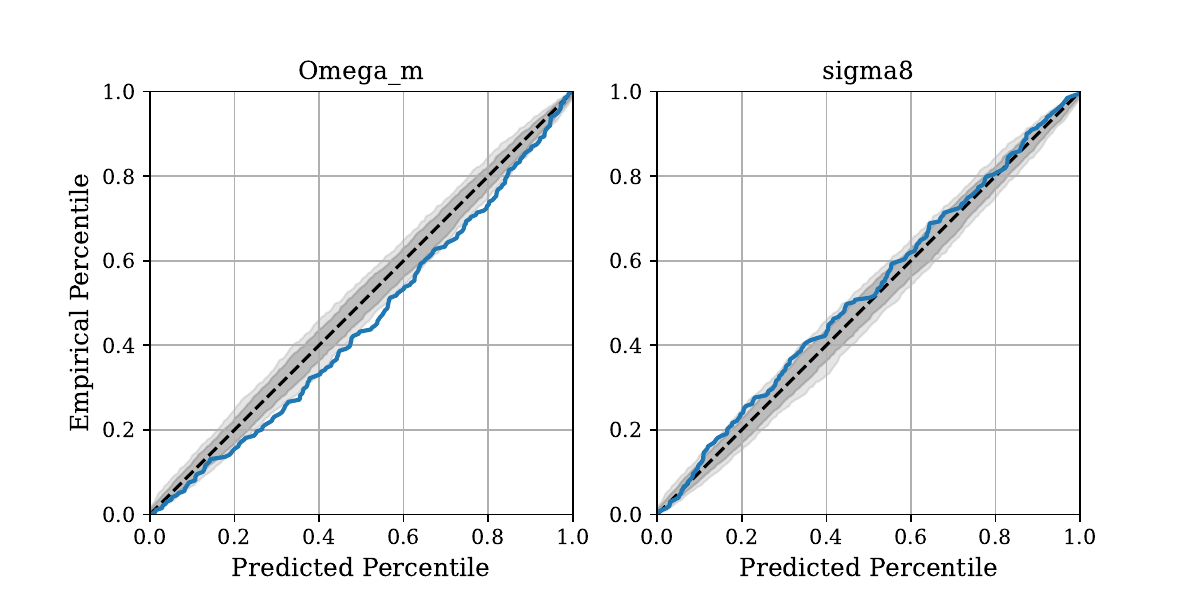}
    \end{minipage}
    \hfill
    \begin{minipage}{0.49\columnwidth}
        \centering
        \text{LFI NN}\\[-0.1em]
        \includegraphics[width=\linewidth, height=5cm, keepaspectratio]
        {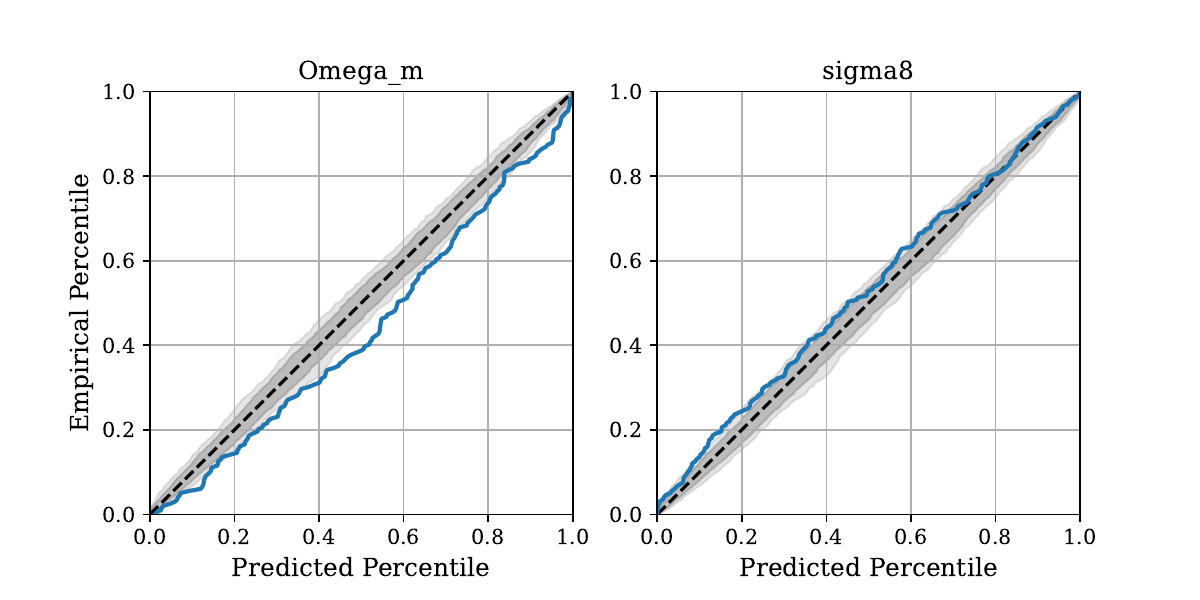}
    \end{minipage}
    \caption{Marginal coverage test comparing ELI (top row) and LFI best-net (bottom row) applied to shear-2PCFs, under MOPED (left column) and NN (right column) compression. For ELI MOPED, blue corresponds to the both-bins posterior and orange to the bin-0-only posterior. The empirical coverage probability $\hat{C}_d(\alpha)$ is plotted against the nominal level $\alpha$ for $\Omega_{\rm m}$ and $\sigma_8$; the dashed diagonal indicates perfect calibration, and shaded bands reflect the $1\sigma$ and $2\sigma$ error-bar obtained via bootstrap resamples.}
    \label{fig:lfi_coverage_compression}
\end{figure}

\begin{figure}[h]
    \centering
    \begin{minipage}{0.49\columnwidth}
        \centering
        \text{ELI MOPED}\\[0.4em]
        \includegraphics[width=\linewidth, height=5cm, keepaspectratio, trim=0 45 0 0, clip]{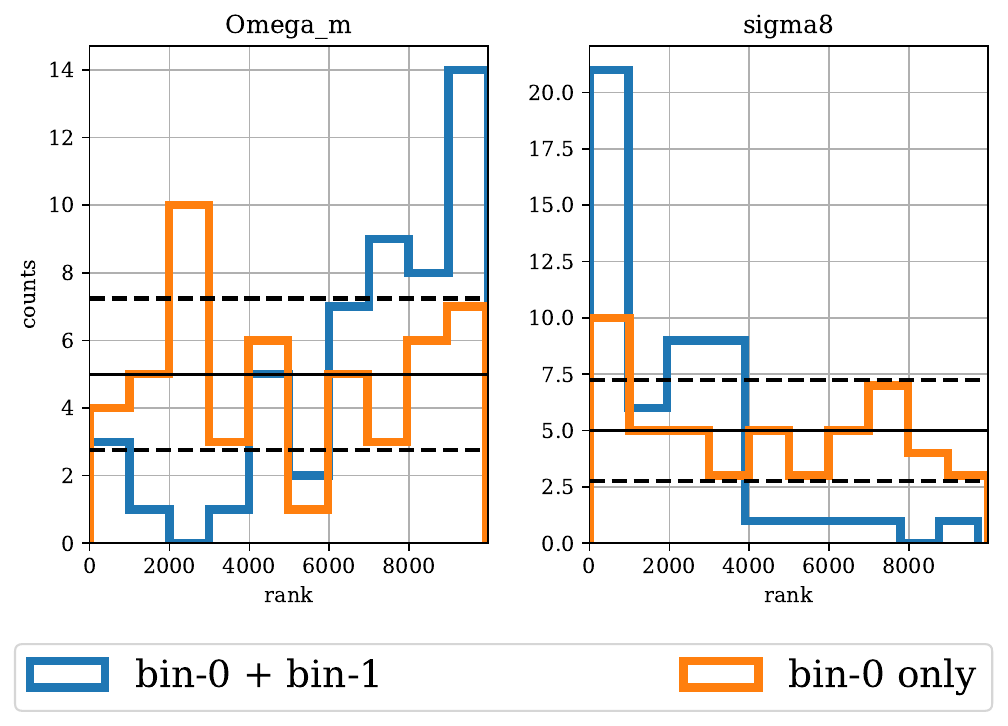}
    \end{minipage}
    \hfill
    \begin{minipage}{0.49\columnwidth}
        \centering
        \text{ELI NN}\\[0.2em]
        \includegraphics[width=\linewidth, height=5cm, keepaspectratio]{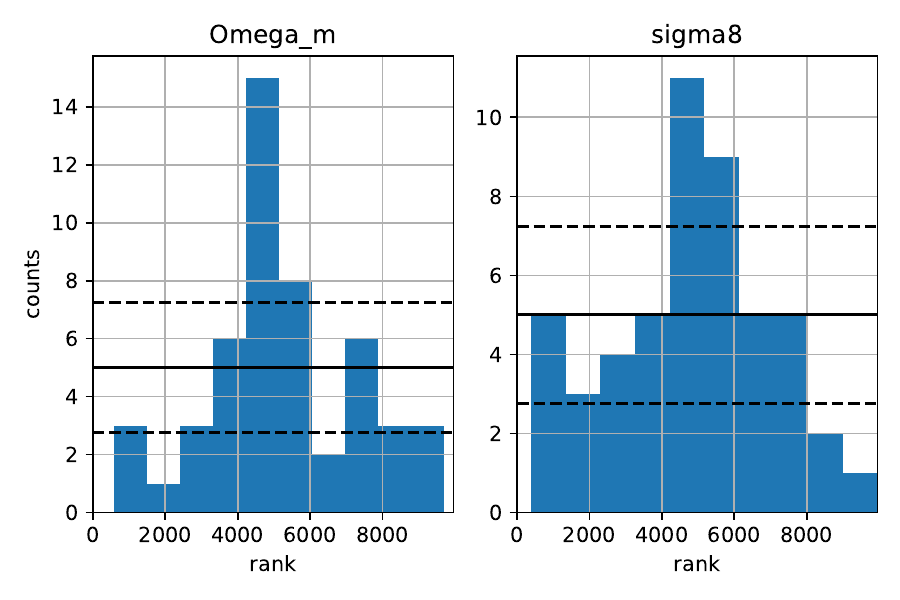}
    \end{minipage}
    \\[0.8em]
    \begin{minipage}{0.49\columnwidth}
        \centering
        \text{LFI MOPED}\\[-0.1em]
        \includegraphics[width=\linewidth, height=5cm, keepaspectratio]
        {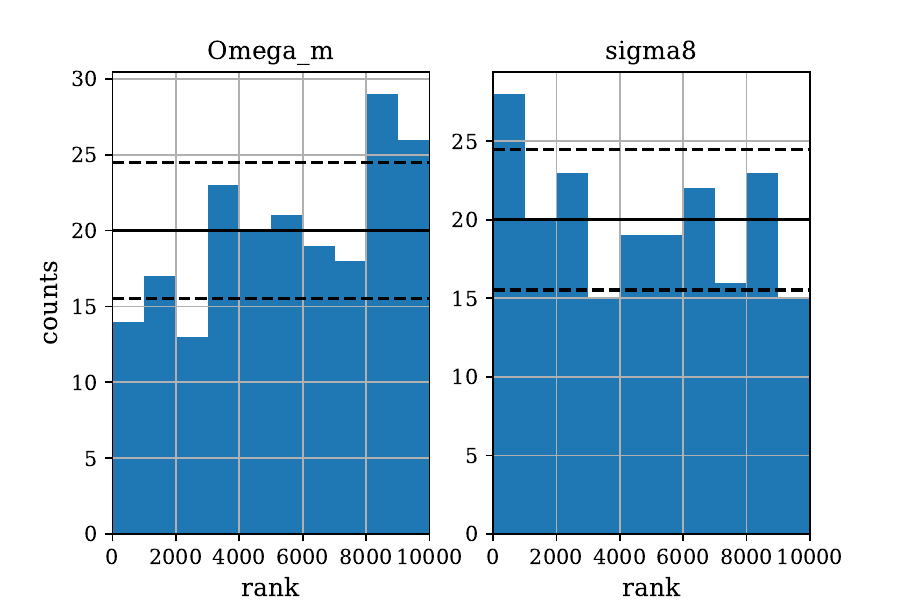}
    \end{minipage}
    \hfill
    \begin{minipage}{0.49\columnwidth}
        \centering
        \text{LFI NN}\\[-0.1em]
        \includegraphics[width=\linewidth, height=5cm, keepaspectratio]
        {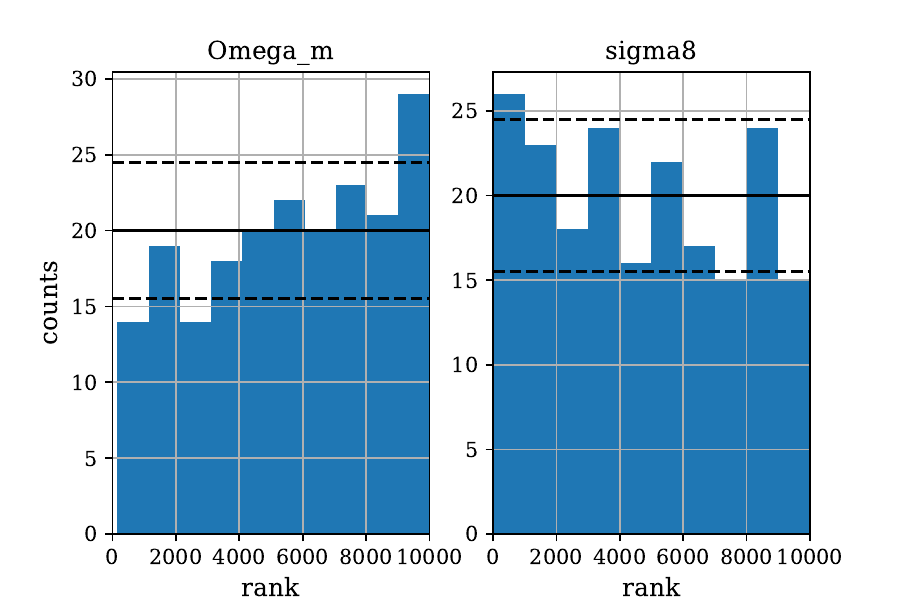}
    \end{minipage}
    \caption{Same as Fig.~\ref{fig:lfi_coverage_compression} but for the PIT/rank histogram test.}
    \label{fig:lfi_histogram_compression}
\end{figure}

\begin{figure}[h]
    \centering
    \begin{minipage}{0.49\columnwidth}
        \centering
        \text{ELI MOPED}\\[0.2em]
        \includegraphics[width=0.9\linewidth, height=5cm, keepaspectratio, trim=0 45 0 0, clip]{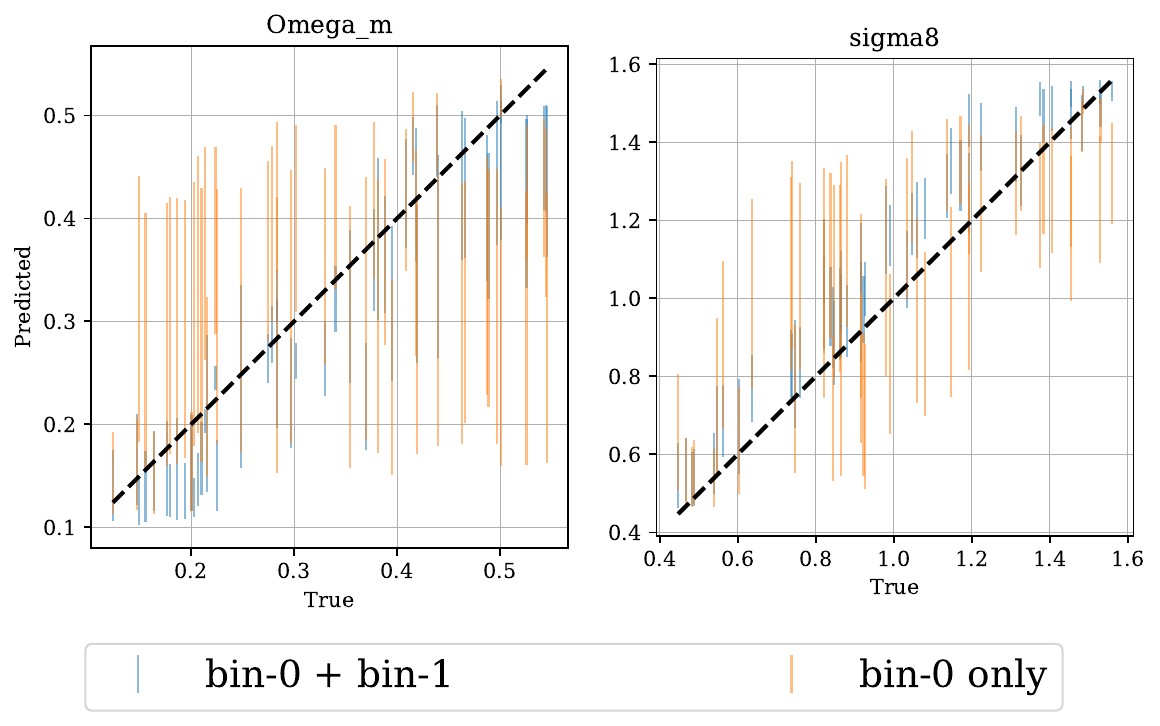}
    \end{minipage}
    \hfill
    \begin{minipage}{0.49\columnwidth}
        \centering
        \text{ELI NN}\\[0.2em]
        \includegraphics[width=0.9\linewidth, height=5cm, keepaspectratio]{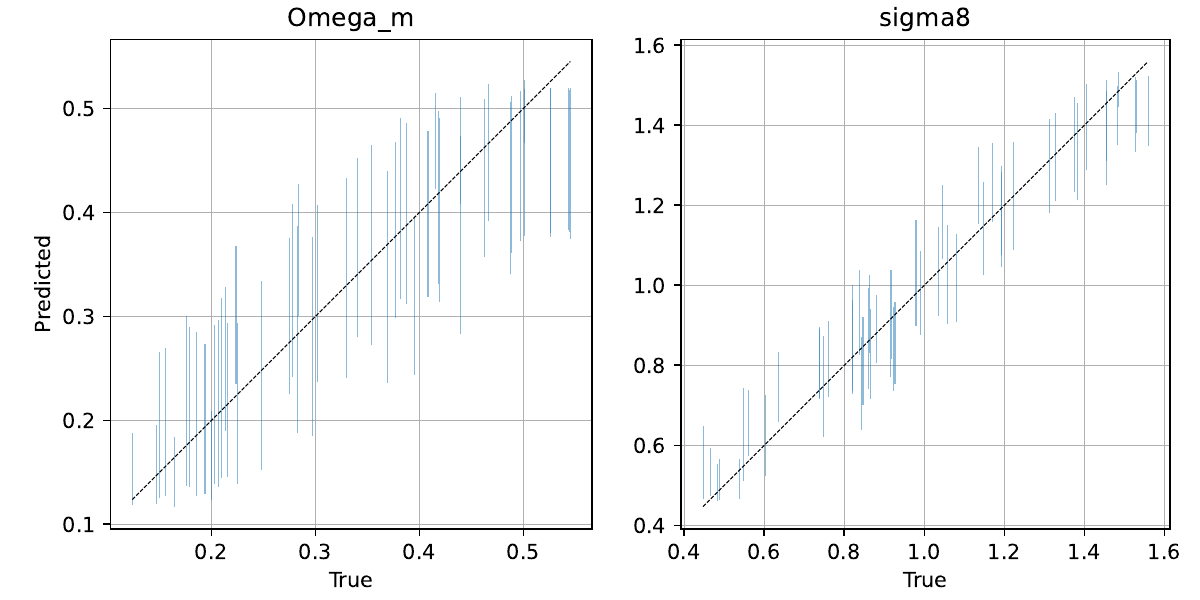}
    \end{minipage}
    \\[0.8em]
    \begin{minipage}{0.49\columnwidth}
        \centering
        \text{LFI MOPED}\\[-0.1em]
        \includegraphics[width=\linewidth, height=5cm, keepaspectratio]
        {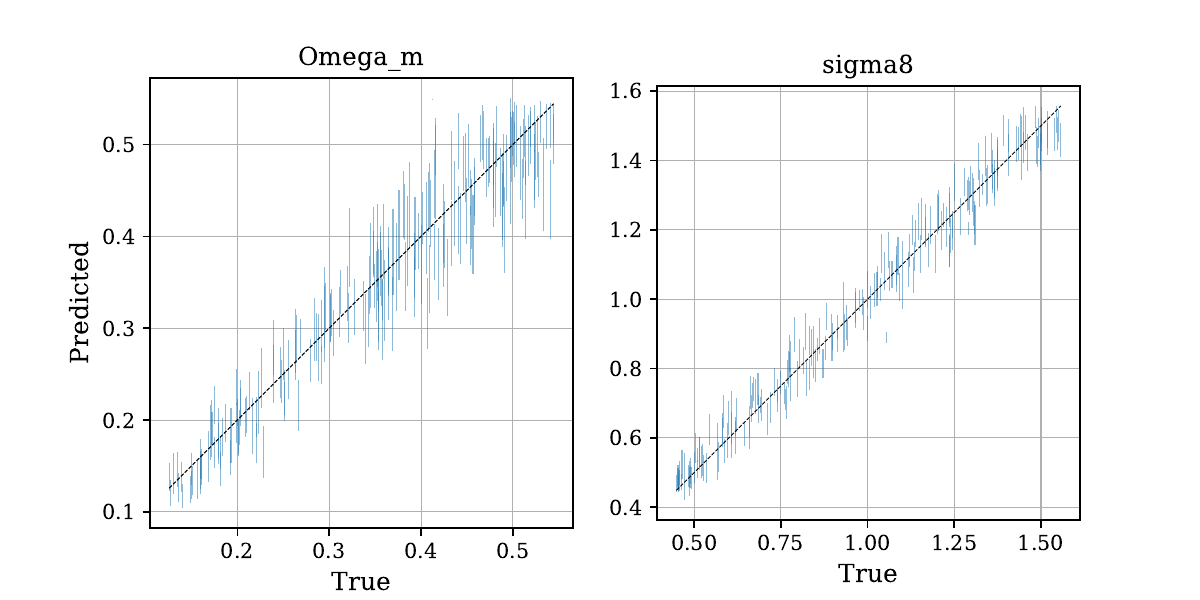}
    \end{minipage}
    \hfill
    \begin{minipage}{0.49\columnwidth}
        \centering
        \text{LFI NN}\\[-0.1em]
        \includegraphics[width=\linewidth, height=5cm, keepaspectratio]
        {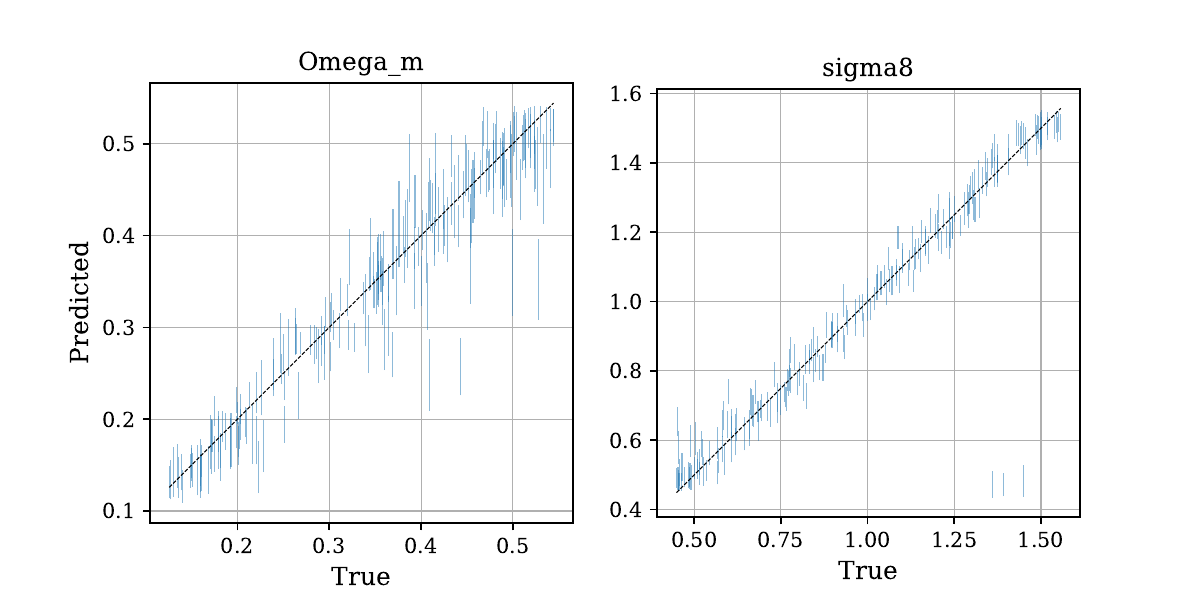}
    \end{minipage}
    \caption{Same as Fig.~\ref{fig:lfi_coverage_compression} but for the posterior predictive check test.}
    \label{fig:lfi_predictions_compression}
\end{figure}

We present the complementary set of posterior calibration diagnostics for shear-2PCFs, comparing MOPED and NN compression, for both the LFI best-net and the ELI tests introduced in Sect.~\ref{sec:lfi_posterior_validation}. For ELI under MOPED compression, we additionally overlay the nominal both-bins posterior with the bin-0-only posterior discussed in Sect.~\ref{sec:eli_vs_lfi}. These diagnostics complement the TARP joint calibration test shown in Fig.~\ref{fig:tarp_compression} and together constitute the full set of validation checks described in Sect.~\ref{sec:lfi_posterior_validation}.

\emph{Marginal coverage test} (Fig.~\ref{fig:lfi_coverage_compression}): for each parameter independently, the empirical coverage probability $\hat{C}_d(\alpha)$ is compared to the nominal level $\alpha$ across 692 test nodes together with $1\sigma$ and $2\sigma$ error-bar computed from bootstrap technique. A perfectly calibrated marginal posterior traces the diagonal $\hat{C}_d(\alpha) = \alpha$. Deviations above (below) the diagonal indicate under-confidence (overconfidence) in the marginal distribution of that parameter. Both MOPED and NN compression produce marginal coverage curves close to the diagonal for both $\Omega_{\rm m}$ and $\sigma_8$ for LFI, with only a residual bias of up to $20 \%$ at low confidence levels, which would be almost entirely explained by a $3 \, \sigma$ uncertainty and is not significant for the coverage levels on which our results are based (i.e., beyond $68 \%$). For ELI under MOPED compression, the both-bins posterior shows a marked under-confidence at low-to-intermediate percentiles for both $\Omega_{\rm m}$ and $\sigma_8$, lying well outside the bootstrap bands; restricting to bin-0-only removes this bias and yields a curve consistent with the diagonal, mirroring the same emulation and Gaussianity limitations of bin~1 already discussed in Sect.~\ref{sec:eli_vs_lfi}. For ELI under NN compression, the coverage curves for both parameters remain closer to the diagonal but with non-negligible order of miscalibration at low (high) percentile region for $\Omega_{\rm m}$ ($\sigma_8$).

\emph{PIT/rank histogram} (Fig.~\ref{fig:lfi_histogram_compression}): for each test node, $10^4$ posterior samples are drawn and the true parameter value is ranked among them; the ranks are then collected into a histogram across all 692 test nodes. A flat histogram indicates perfect calibration: a U-shaped distribution signals overconfidence (posterior too narrow, true value too often near the tails), while an inverted-U shape signals under-confidence (posterior too wide). Both compression schemes yield approximately uniform rank distributions for both $\Omega_{\rm m}$ and $\sigma_8$ for LFI, with no significant systematic deviations from uniformity, except at extreme ranks, which nevertheless do not produce a cosmological bias beyond $1 \, \sigma$. For ELI under MOPED compression, the both-bins posterior shows a strongly non-uniform histogram, most notably a sharp excess at low ranks for $\sigma_8$, indicating overconfidence; the bin-0-only histogram is comparatively closer to uniform, consistent with the improved calibration seen in the marginal coverage test. For ELI under NN compression, the histograms are also non-uniform resembling an inverted U-shape peaking at intermediate ranks for both parameters. These results confirm ELI uncalibrated posteriors because of the caveats explained in the main text (Sect.~\ref{sec:compression_comparison})

\emph{Posterior predictive checks} (Fig.~\ref{fig:lfi_predictions_compression}): the posterior median and width is compared against the true parameter value across all 692 test nodes. A well-calibrated posterior should produce scatter consistent with the posterior width, with no systematic offset. Both MOPED and NN compression show tight, unbiased scatter around the identity line for both parameters for LFI, with no evidence of systematic median shift across the parameter space. For ELI under MOPED compression, the both-bins posterior shows substantially smaller scatter and narrower error bars around the identity line, especially for $\Omega_{\rm m}$, than the bin-0-only posterior, reflecting the reduced constraining power once bin~1 is excluded. For ELI under NN compression, the scatter is comparably tight and well-centred on the identity line for both parameters, similar to the both bins MOPED case.

\end{appendix}

\end{document}